\documentclass{article}

\usepackage{arxiv}
\usepackage{longtable}

\usepackage[utf8]{inputenc} % allow utf-8 input
\usepackage[T1]{fontenc}    % use 8-bit T1 fonts
\usepackage{hyperref}       % hyperlinks
\usepackage{url}            % simple URL typesetting
\usepackage{booktabs}       % professional-quality tables
\usepackage{amsfonts}       % blackboard math symbols
\usepackage{nicefrac}       % compact symbols for 1/2, etc.
\usepackage{microtype}      % microtypography
\usepackage{lipsum}
\usepackage{graphicx}
\usepackage{multicol}
\usepackage{placeins}
\usepackage{multirow}
\usepackage{pdflscape}
\usepackage{geometry}
\usepackage{float}
\graphicspath{ {./media/} }
\usepackage{amsmath}

\title{Poly-Vector Retrieval: Reference and Content Embeddings for Legal Documents

}

\author{
 João Alberto de Oliveira Lima \\
  University of Brasília; Federal Senate of Brazil\\
  \texttt{joaolima@senado.leg.br} 
}

\begin{document}
\maketitle

\begin{abstract}
Retrieval-Augmented Generation (RAG) has emerged as an effective paradigm for generating contextually accurate answers by integrating Large Language Models (LLMs) with retrieval mechanisms. However, in legal contexts, users frequently reference norms by their labels or nicknames (e.g., \emph{Article 5 of the Constitution} or \emph{Consumer Defense Code (CDC)}), rather than by their content, posing challenges for traditional RAG approaches that rely solely on semantic embeddings of text. Furthermore, legal texts themselves heavily rely on explicit cross-references (e.g., `pursuant to Article 34') that function as pointers. Both scenarios pose challenges for traditional RAG approaches that rely solely on semantic embeddings of text, often failing to retrieve the necessary referenced content. This paper introduces \textbf{Poly-Vector Retrieval}, a method assigning multiple distinct embeddings to each legal provision: one embedding captures the \emph{content} (the full text), another captures the \emph{label} (the identifier or proper name), and optionally additional embeddings capture alternative denominations. Inspired by Frege's distinction between \emph{Sense} and \emph{Reference}, this poly-vector retrieval approach treats labels, identifiers and reference markers as rigid designators and content embeddings as carriers of semantic substance. Experiments on the Brazilian Federal Constitution demonstrate that Poly-Vector Retrieval significantly improves retrieval accuracy for label-centric queries  and potential to resolve internal and external cross-references, without compromising performance on purely semantic queries. The study discusses philosophical and practical implications of explicitly separating reference from content in vector embeddings and proposes future research directions for applying this approach to broader legal datasets and other domains characterized by explicit reference identifiers.
\end{abstract}


\begin{multicols}{2}\raggedcolumns

\section{Introduction}
\label{sec:intro}
Systems that combine information retrieval with large language models (i.e., Retrieval-Augmented Generation or RAG) \cite{lewis2020retrieval} have demonstrated effectiveness in knowledge-intensive tasks, including legal question-answering. These systems mitigate hallucinations by grounding model outputs in retrieved text passages from authoritative legal sources (e.g., statutes, court rulings, scholarly articles). 

Despite recent advances, one persistent challenge in \emph{legal} RAG arises when users formulate queries referencing legal provisions by \emph{label} or \emph{identifier}, as is common in legal practice. For instance, a lawyer might ask, ``What does Article 20 of the Constitution say?'' or ``Explain the main aspects of Article 5 of the Brazilian Constitution.'' From an embedding standpoint, such queries often appear lexically sparse (just ``Art.\ 20'' or ``Art.\ 5''), and hence may not be closely aligned with the more elaborate vocabulary of the actual text content (which might discuss ``assets of the Union,'' ``fundamental rights,'' etc.). 

Furthermore, legal texts themselves are replete with explicit cross-references (e.g., 'in accordance with Article 5' or 'pursuant to Law XYZ'). Standard semantic retrieval may fetch the text containing such a reference, but often fails to retrieve the content of the referenced provision if its text is not directly relevant to the user's original query, leading to incomplete context for the LLM. An effective legal RAG system must navigate both user-provided labels and these internal textual pointers.

\begin{figure*}[t]
\centering
\includegraphics[width=0.9\textwidth]{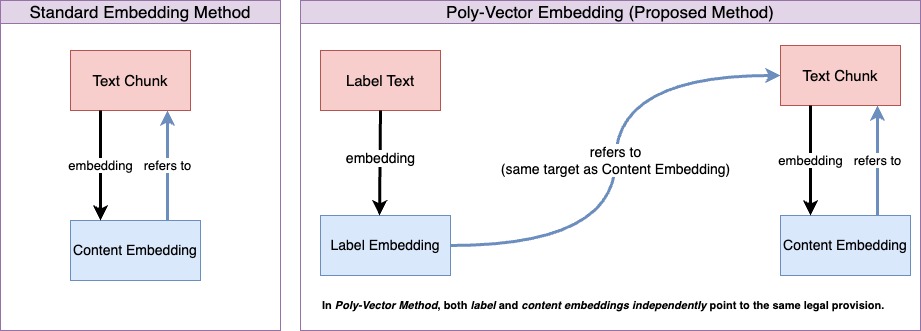}
\caption{Comparison between standard embedding-based retrieval and the Poly-Vector Retrieval Method }
\label{fig:poliv}
\end{figure*}

\textbf{Conventional embedding-based retrieval} typically excels at matching semantic similarity between a query and the text of a document chunk, but fails in purely \emph{referential} queries that rely on numeric or named identifiers. One might solve this partially with hybrid solutions (e.g., using a regex or BM25 to match ``Art.\ 20'' literally). However, these approaches risk missing non-literal references such as ``the twentieth article,'' or references in different languages, abbreviations (``Art.\ 5 CF/88''), or subtle variations that do not exactly match the text in the chunk. Furthermore, standard embeddings do not explicitly encode the notion that \emph{labels} can behave like \emph{names} or \emph{rigid designators} pointing to certain text passages.

To address this, we introduce \textbf{Poly-Vector Retrieval}. This strategy represents each legal provision (or text chunk) using \emph{multiple} vector embeddings stored within the index. Specifically, for a given provision, we generate:
\begin{itemize}
    \item \textbf{Content Embedding(s):} One or more dense vectors derived from the full text of the provision. These capture the semantic meaning, the substance, or the \emph{Sense} (in Frege's terms) of the legal rule.
    \item \textbf{Label Embedding(s):} One or more dense vectors derived from the provision's formal identifier(s) or common name(s). These capture the referential aspect, acting like pointers or \emph{References} to the content.
\end{itemize}

\begin{figure*}[t]
\centering
\includegraphics[width=0.9\textwidth]{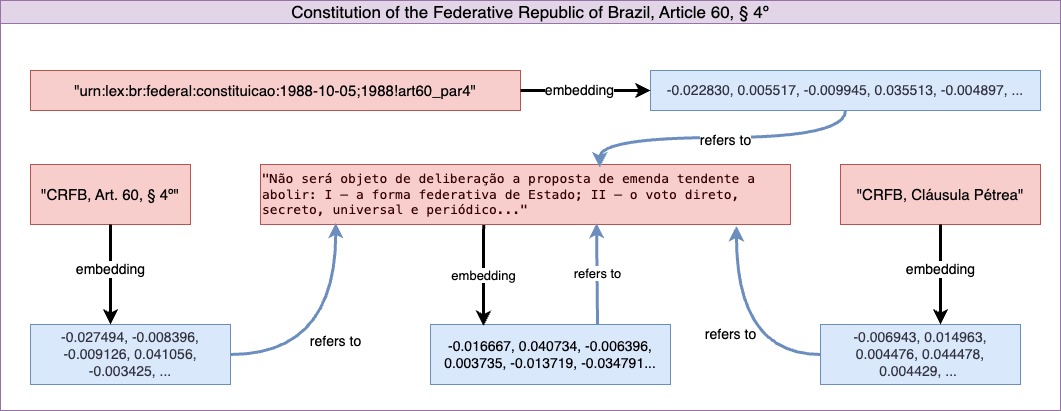}
\caption{Example of the Poly-Vector Retrieval Method for Article 60, Paragraph 4 of the Brazilian Constitution }
\label{fig:poliv-ex}
\end{figure*}

To illustrate the conceptual distinction between traditional and Poly-Vector retrieval approaches, Figure~\ref{fig:poliv} presents a comparative diagram. On the left, the standard embedding method assigns a single vector representation to each legal text chunk, which serves both semantic and referential purposes. On the right, the Poly-Vector method assigns two distinct vectors per provision: one embedding the semantic content of the legal text, and another embedding its label or identifier. Both embeddings independently point to the same legal provision, thereby aligning with the philosophical distinction between sense and reference. This dual-vector setup ensures that referential queries (e.g., ``What does Brazilian Constitution's Article 20 say?'') and semantic queries (e.g., ``Which provision discusses federal property?'') can both resolve accurately to the correct text chunk.

Consider, for instance, Article 60, Paragraph 4 of the Brazilian Constitution (see Figure~\ref{fig:poliv-ex}). Its formal \textbf{label} is ``Art. 60, § 4º'', and its \textbf{content} consists of the text listing matters that cannot be abolished by constitutional amendment (``§ 4º Não será objeto de deliberação a proposta de emenda tendente a abolir: I – a forma federativa de Estado; II – o voto direto, secreto, universal e periódico; III – a separação dos Poderes; IV – os direitos e garantias individuais.'').\footnote{Text of Art. 60, § 4º: "A proposed amendment intended to abolish: I – the federative form of State; II – direct, secret, universal, and periodic suffrage; III – the separation of Powers; IV – individual rights and guarantees, shall not be subject to deliberation."} Additionally, the provision is associated with a standardized identifier, the URN \texttt{"urn:lex:br:federal:constituicao:1988-10-05;\allowbreak 1988!art60\_par4"}, which uniquely identifies this normative device. This provision is also widely recognized by the \textbf{nickname} ``Cláusula Pétrea'' (Immutable Clause). Poly-Vector Retrieval creates distinct embeddings for the content text, the formal label ``Art. 60, § 4º'', the URN identifier, and potentially the nickname ``Cláusula Pétrea'', all linked to the same underlying textual provision. Thus, the system effectively retrieves the correct content whether a user queries ``What does Art. 60, § 4º say?'', ``Explain the Cláusulas Pétreas'', references the URN directly, or semantically describes the content (e.g., ``What are the limits on constitutional amendments?'').

We situate Poly-Vector Retrieval within philosophical discussions on names, reference, and meaning, starting with Frege's distinction between \emph{Sense (Sinn)} and \emph{Reference (Bedeutung)} \cite{frege1892sense}, through the debates between descriptivist and causal-historical theories of proper names (Searle \cite{searle1958proper}, Kripke \cite{kripke1980naming}), and more recent contributions by Kaplan, Evans, and Recanati. This theoretical lens clarifies why a single embedding may fail to capture both the referential function and the semantic content, and how distinct embeddings can remedy this by making reference an explicit dimension.

We validate Poly-Vector Retrieval on the 1988 Brazilian Federal Constitution, comparing:
\begin{enumerate}
    \item A \emph{standard} RAG approach with content-only embeddings,
    \item A \emph{multi-layered} approach \cite{lima2024unlocking} which indexes multiple textual granularities (articles, paragraphs), and
    \item Our proposed \emph{Poly-Vector} approach, which adds label embeddings for each article.
\end{enumerate}
We run queries that directly cite articles by their label (e.g., ``What does Art. 20 say?''), as well as semantically descriptive queries (e.g., ``Which provisions govern the distribution of Union property?''). The results show that Poly-Vector allow the retrieval of chunks with higher similarity for label-centric queries, while preserving strong performance on purely semantic searches.

This paper is structured as follows. Section \ref{sec:background} overviews the evolution of RAG and the philosophical foundations of labeling, reference, and sense in language. Section \ref{sec:poli-methodology} details the Poly-Vector approach. Sections \ref{sec:methodology}, \ref{sec:experiments} and \ref{sec:expected_results} describe our experimental setup on the Brazilian Constitution corpus, including baselines, queries, evaluation methods and expected results. Section \ref{sec:results} reports results and discusses key findings, highlighting how explicitly modeling both reference (label) and sense (content) yields substantial benefits. Finally, Section \ref{sec:conclusion} concludes with future directions, such as extending Poly-Vector to multiple legal documents, layering reference embeddings at multiple levels (articles, chapters, codes), and applying the approach to other domains.
\section{Theoretical Background and Related Work}
\label{sec:background}

\subsection{Evolution of RAG Systems}
RAG systems \cite{lewis2020retrieval} augment LLMs by retrieving relevant documents before generation. Early implementations often used sparse retrieval methods like TF-IDF or BM25. The advent of powerful sentence and document embedding models (e.g., Sentence-BERT \cite{reimers2019sentence}, OpenAI Ada models \cite{neelakantan2022text}) has led to the dominance of dense retrieval, where queries and documents are mapped to a shared vector space, and relevance is determined by vector similarity (e.g., cosine similarity).

Recognizing that fixed-size chunking might miss relevant information or provide insufficient context, techniques like the multi-layered retrieval  \cite{lima2024unlocking} that creates embeddings for different textual granularities (e.g., chapter, section, paragraphs, etc.), improving the chances of matching a query's semantic scope. However, these still primarily embed variations of the \textit{content}. Our Poly-Vector approach differs by creating embeddings specifically for the \textit{labels} that name the content, addressing a different aspect of information access.

This reliance solely on content semantics also poses challenges when encountering explicit cross-references common in legal and technical documents. A phrase like 'as defined in Art. 156' functions as a pointer, akin to a name or label, whose semantic content is minimal. Standard embedding methods struggle to bridge the gap between the reference marker and the substantive content it points to, potentially omitting crucial context during retrieval. Having a dedicated label embedding for such reference markers becomes especially useful, as it enables the RAG system to perform targeted reference expansion—proactively retrieving the content of 'Art. 156', for instance—potentially for the most relevant retrieved chunks, thus ensuring a more complete context.

\subsection{Philosophical Foundations: Sense, Reference, and Names}
The core idea of Poly-Vector Retrieval resonates strongly with long-standing philosophical discussions about meaning, reference, and the nature of names.

\textbf{Frege's Distinction between Sense and Reference:}\footnote{%
In a strictly Fregean perspective, the “reference” might be the normative device itself as an entity 
(and the written text would be its “sense” expressing that norm). 
However, from an information retrieval engineering standpoint, “label as \emph{Sinn}” 
and “text as \emph{Bedeutung}” works well enough—and is an instructive analogy 
for motivating the idea of \emph{Poly-Vector Retrieval}.
} Gottlob Frege \cite{frege1892sense} famously distinguished between the \textit{Sense} (Sinn) and \textit{Reference} (Bedeutung) of a term. The Reference is the object or entity the term designates in the world. The Sense is the ``mode of presentation'' of the reference, the cognitive content or way we grasp the object. Frege's classic example involves ``the Morning Star'' and ``the Evening Star''. Both expressions have the same Reference (the planet Venus) but different Senses, conveying different information and ways of identifying the object.

In the context of Poly-Vector RAG, a label like "Artigo 5º da Constituição Federal" can be seen as analogous to a Fregean Sense. It's a specific linguistic expression, a ``mode of presentation,'' that users employ to think about and request specific information. The Reference is the actual textual content of Article 5. Standard RAG primarily embeds the Reference (content). Poly-Vector RAG explicitly models both: it creates an embedding for the label (capturing aspects of its Sense or its role as a designator) and links it to the content (the Reference) which is used for generation.

\textbf{Proper Names and Descriptions:} The debate between descriptivist theories of proper names (like John Searle's initial view \cite{searle1958proper}, later refined into a cluster theory \cite{searle1969speech}) and direct reference theories (like Saul Kripke's causal-historical theory \cite{kripke1980naming}) is also relevant. Searle argued that names function like shorthand for a cluster of descriptions associated with the referent. Kripke argued against this, proposing that names refer directly via an initial ``baptism'' and a subsequent causal chain of transmission, without needing to satisfy specific descriptions.

Labels like ``Lei nº 8.078/1990'' or ``Artigo 5º da Constituição'' function similarly to proper names within the legal domain. They rigidly designate specific legal texts or provisions. While Kripke's causal theory emphasizes the historical chain, the Poly-Vector approach computationally leverages the \textit{referential function}  of these labels. The label embedding $E_l$ acts as the computational trigger associated with the name/label. The system doesn't necessarily rely on the label being a \textit{description}  of the content (aligning more with Kripke's intuition that names are not descriptions), but it recognizes that the label is the term users employ to access the reference. The link between $E_l$ and the content $C$ computationally establishes this reference relation. Even if some descriptive information is implicitly captured by the label embedding, its primary function in Poly-Vector is as a pointer.

The Poly-Vector approach, by using distinct representations for labels and content and leveraging them strategically during retrieval, implicitly acknowledges this complexity. It uses the label (linguistic form, often contextually salient) as an access key while relying on the content (the reference) for substantive grounding, reflecting a pragmatic separation of concerns in information retrieval.

\section{The Poly-Vector Retrieval Methodology}
\label{sec:poli-methodology}

The Poly-Vector Retrieval approach extends existing RAG frameworks by introducing a distinct representation layer for document labels. Let $\mathcal{D} = \{C_1, C_2, ..., C_N\}$ be a corpus of text chunks. Let $\mathcal{L}_i$ be the set of one or more labels associated with chunk $C_i$. A label $L \in \mathcal{L}_i$ can be a formal identifier (e.g., "Art. 1, Sec. II"), a common name (e.g., "Consumer Protection Code"), or any other designation used to refer to $C_i$.

The core components of the Poly-Vector approach are:

\begin{enumerate}
    \item \textbf{Content Embedding:} For each chunk $C_i \in \mathcal{D}$, compute a content embedding $E_c(C_i) \in \mathbb{R}^d$ using a suitable embedding model $\mathcal{M}_{emb}$.
    \[ E_c(C_i) = \mathcal{M}_{emb}(C_i) \]

    \item \textbf{Label Embedding:} For each label $L \in \mathcal{L}_i$ associated with chunk $C_i$, compute a label embedding $E_l(L) \in \mathbb{R}^d$ using the same or a potentially different embedding model $\mathcal{M}_{emb}$.
    \[ E_l(L) = \mathcal{M}_{emb}(L) \]

 Specifically, the label string $L$ used for embedding is constructed based on the type of legal provision it represents, always prepended with the full name of the norm: 
 
\begin{itemize}
    \item \emph{For fundamental textual units like Articles (or their specific parts like caput, paragraphs, items/incisos):} The label $L$ consists of the full norm name followed by the unit's canonical identifier within the document structure. For example, for the nineteenth item (Inciso XIX) of Article 5 of the Brazilian Constitution, the label string embedded would be \texttt{"Constituição da República Federativa do Brasil de 1988, Artigo 5º, Inciso XIX"}. For the article itself, it would be \texttt{"Constituição da República Federativa do Brasil de 1988, Artigo 5º"}.
    
    \item \emph{For higher-level structural groupings (e.g., Titles, Chapters, Sections):} The label $L$ is constructed by starting with the full norm name, followed by concatenating the names or identifiers of the grouping itself and all its parent groupings, forming a complete path. For example, the label string for Section II within Chapter II of Title VI of the Brazilian Constitution would be constructed as \texttt{"Constituição da República Federativa do Brasil de 1988, TÍTULO VI, CAPÍTULO II, Seção II"}. 
\end{itemize}
This constructed label string $L$ is then provided as input to the embedding model $\mathcal{M}_{emb}$ to generate the label vector $E_l(L)$.

    A single chunk $C_i$ might have multiple labels, leading to multiple label vectors $E_l(L)$ all pointing to the same content $C_i$ (See Figure~\ref{fig:poliv-ex}).

    \item \textbf{Indexing:} Store both sets of embeddings in a vector database or index. Crucially, maintain a mapping that links each label embedding $E_l(L)$ back to its corresponding content chunk $C_i$. This can be done by storing the ID or the full content of $C_i$ as metadata associated with $E_l(L)$. Content embeddings $E_c(C_i)$ can also be indexed for hybrid retrieval strategies.

\item \textbf{Retrieval Process:} Given a user query $Q$:
        \begin{enumerate}
            \item[(a)] Compute the query embedding $E_q = \mathcal{M}_{emb}(Q)$.

            \item[(b)] Define the unified search space $\mathcal{E}$ containing all embeddings associated with the corpus: 
            \[ \mathcal{E} = \bigcup_{C \in \mathcal{D}} \left( \{E_c(C)\} \cup \{ E_l(L) \mid L \in \mathcal{L}(C) \} \right) \]
            Ensure that each embedding in $\mathcal{E}$ maintains a reference (e.g., via metadata) to the original content chunk $C_i$ it represents (either directly as $E_c(C_i)$ or indirectly via the mapping from $E_l(L)$ to $C_i$).
            \item[(c)] Perform a \emph{single} similarity search (e.g., k-Nearest Neighbors based on cosine similarity) for the query embedding $E_q$ against the entire unified embedding space $\mathcal{E}$. This search retrieves the top-$k'$ most similar embeddings, which may include both content embeddings ($E_c^*$) and label embeddings ($E_l^*$).
            \item[(d)] Identify the corresponding content chunks for the top-$k'$ retrieved embeddings. If the retrieved embedding is a content embedding $E_c^*(C_j)$, the corresponding chunk is $C_j$. If the retrieved embedding is a label embedding $E_l^*(L)$ associated with chunk $C_i$, the corresponding chunk is $C_i$.
            \item[(e)] Aggregate and deduplicate the retrieved content chunks. Since multiple embeddings (e.g., a content embedding and one or more label embeddings) might point to the same chunk $C_i$, collect the unique set of content chunks $\{C_1^*, C_2^*, ..., C_m^*\}$, where $m \le k'$.
            \item[(f)] Rank the unique content chunks. A common strategy is to rank them based on the highest similarity score among the embeddings that retrieved them. For instance, if chunk $C_i$ was retrieved via $E_c(C_i)$ with score $s_c$ and via $E_l(L)$ with score $s_l$, its final score could be $\max(s_c, s_l)$. Select the top-$k$ ranked unique content chunks.
        \end{enumerate}

        This unified search space allows matching both user queries containing labels and potentially resolving explicit cross-references encountered within retrieved text passages during a more advanced retrieval strategy (though the latter is beyond the scope of the current experimental setup).

    \item \textbf{Context Augmentation and Generation:} Concatenate the retrieved content chunks $C^*_{1}, ..., C^*_{k}$ with the original query $Q$ to form the prompt for the LLM.
    \[ \text{Prompt} = \text{format}(Q, C^*_{1}, ..., C^*_{k}) \]
    The LLM then generates the final response based on this augmented context.
    \[ \text{Response} = \text{LLM}(\text{Prompt}) \]
\end{enumerate}

This process ensures that if a query like "What does Article 5 say?" is posed, its embedding $E_q$ is likely to be close to the label embedding $E_l(\text{"Article 5"})$. The system then retrieves the \textit{content} of Article 5, $C_{\text{Art.5}}$, providing the LLM with the relevant substance needed to answer the query accurately. This contrasts with standard RAG where $E_q$ might have low similarity to the content embedding $E_c(C_{\text{Art.5}})$ if the query phrasing is very different from the article's text.

A further refinement in our Poly-Vector approach was the introduction of a combined \emph{Identifier plus Label} embedding, denoted \texttt{I+L}. While we already generate separate embeddings for the label (e.g., \texttt{"CFRB, Artigo 5º"}) and the URN (e.g., \texttt{"urn:lex:br:federal:constituicao:1988-10-05;\allowbreak 1988!art5"}), we also test a hybrid embedding that concatenates both strings into one. This design was motivated by John Searle’s discussion of proper names as ``clusters'' of descriptions~\cite{searle1969speech}, suggesting that a name may be associated with a bundle of properties or references. In our legal context, we hypothesized that merging the formal label with its unique URN would yield a richer representation that more robustly captures the same referential object via different textual cues.

Our specific exploratory hypothesis was that for certain queries—especially those mixing partial references to both the article label and its URN—the \texttt{I+L} embedding might score higher in the similarity search than the label-only (\texttt{LBL}) or URN-only (\texttt{URN}) vectors. Conceptually, \texttt{I+L} is designed to function as a ``cluster'' capturing the joint identity of a provision, rather than relying on the user query matching just one of those references. In principle, this could make retrieval more resilient to partial or approximate matches, potentially increasing top-1 precision for referential queries.

\section{Methodology: Comparative Evaluation}
\label{sec:methodology}

We design a comparative evaluation of eight embedding-based retrieval methods on the same corpus, aiming to quantify the impact of hierarchical chunking, query normalization, and “poly-vector” label embeddings on retrieval performance. The corpus for all methods is the full text of the Brazilian Federal Constitution of 1988, chosen for its rich hierarchical structure (Titles, Chapters, Articles, paragraphs, clauses) and well-defined reference labels. All methods use the same underlying embedding model (\texttt{text-embedding-3-large} from OpenAI with $d=256$\footnote{The original embeddings generated by \texttt{text-embedding-3-large} have 3,072 dimensions. In this study, embeddings were truncated to 256 dimensions using the Matryoshka Representation Learning technique~\cite{kusupati2024matryoshkarepresentationlearning}.}) and cosine similarity measure for retrieval; they differ only in how the text is segmented/indexed and in whether queries are preprocessed. The eight retrieval strategies compared are:

\begin{enumerate}
    \item[(a)] \textbf{Blind Segmentation Baseline:} This method divides the constitution text into fixed-length overlapping chunks (800 tokens each, with a 400-token overlap) without regard to semantic or structural boundaries. Each chunk is encoded into a single embedding. This ``blind'' chunking approach, similar to how systems like ChatGPT (with retrieval plugins) handle long documents by sliding window segmentation\footnote{OpenAI documents this default strategy for its File Search tool, using 800-token chunks with a 400-token overlap. See \url{https://platform.openai.com/docs/assistants/tools/file-search}}, results in an index of \textbf{284 embeddings}. It serves as a baseline representing no special use of the document structure. It is important to note that this 'blind' chunking approach, common in general-purpose systems, carries significant risks in the legal domain, as arbitrary segmentation might split a provision mid-sentence or separate a crucial condition from its consequence, potentially altering the legal meaning.

    \item[(b)] \textbf{Flat Per-Article Baseline:} In this approach, each Article of the Constitution is treated as one document chunk. We generate one embedding per Article (including all its paragraphs and items). This flat indexing respects the natural top-level division and results in \textbf{276 embeddings}. It provides a baseline that uses coarse granularity (article-level) representations.

    \item[(c)] \textbf{Multi-layer Hierarchical Embeddings:} This method follows the approach proposed by \cite{lima2024unlocking}, creating embeddings not only for each Article but also for finer-grained components (each paragraph, clause, or item within the articles) as well as for higher-level groupings (such as Sections, Chapters, and Titles). This multi-layer embedding scheme generates a total of \textbf{2973 embeddings}, leveraging the inherent hierarchy of the legal text to improve retrieval of both specific and general information.

    \item[(d)] \textbf{Multi-layer + Query Normalization:} This extends the multi-layer approach above (using the same \textbf{2973 embeddings}) by incorporating the query preprocessing strategy described by \cite{lima2025improvingragretrievalpropositional}. Before embedding the user’s question, we automatically transform the query into its core propositional content by stripping away polite or interrogative phrasing (speech-act markers). This tests whether query normalization can improve alignment.

    \item[(e)] \textbf{Poly-Vector + Blind:} 
    This method augments the blind segmentation baseline (a) by creating \emph{three distinct label embeddings} (the unit’s label, its URN, and a combined label+URN string) for each structural provision in the Constitution. These label embeddings are then added to the baseline index of overlapping content chunks. Consequently, the unified index grows from the blind baseline’s 284 content embeddings to a total of \textbf{9203 embeddings}, allowing the system to match user queries that reference provisions by number, by URN, or by both.

    \item[(f)] \textbf{Poly-Vector + Flat:} 
    Similar to (e), this method supplements the flat per-article index (b) with the three label embeddings (label, URN, label+URN) for each article or higher-level section. The original flat strategy had 276 content embeddings; adding the triple label embeddings for the entire document yields \textbf{9195 embeddings} in total. This enables direct matches for queries that mention article identifiers or URNs, while still providing article-level semantic retrieval.

    \item[(g)] \textbf{Poly-Vector + Multi-layer:} 
    Extending the multi-layer hierarchical index (c), this approach adds three label embeddings per structural unit (e.g., clause, paragraph, article, section, chapter, title). Since the multi-layer index contains 2973 content embeddings, introducing an additional 3\,\(\times\)\,2973 label-based embeddings increases the index size to \textbf{11892 embeddings}. By indexing the label, URN, and combined label+URN for each provision, the system can retrieve specific parts of the Constitution whether a query references them by content, by formal identifier, or both.

    \item[(h)] \textbf{Poly-Vector + Multi-layer + Query Normalization:} 
    This most comprehensive strategy applies the same enhanced Poly-Vector + Multi-layer indexing (g), giving an index of \textbf{11892 embeddings}, and also includes the query normalization step described in method (d). Thus, it benefits from hierarchical chunking, explicit label embeddings (label, URN, label+URN), and the removal of extraneous question phrasing, offering robust performance across both semantic and referential queries.

\end{enumerate}

All eight methods are evaluated under identical conditions. For each method, we use the same embedding model (\texttt{text-embedding-3-large}, $d=256$) to encode the Constitution text (in Portuguese) and the user’s queries. Cosine similarity is used to compare query embeddings against the document embeddings in the vector index. By comparing these methods, we can isolate the effects of using multi-scale representations, query normalization, and additional label-based vectors on retrieval performance in a legal document context.

\begin{figure*}[t]
    \centering
    \includegraphics[width=0.9\textwidth]{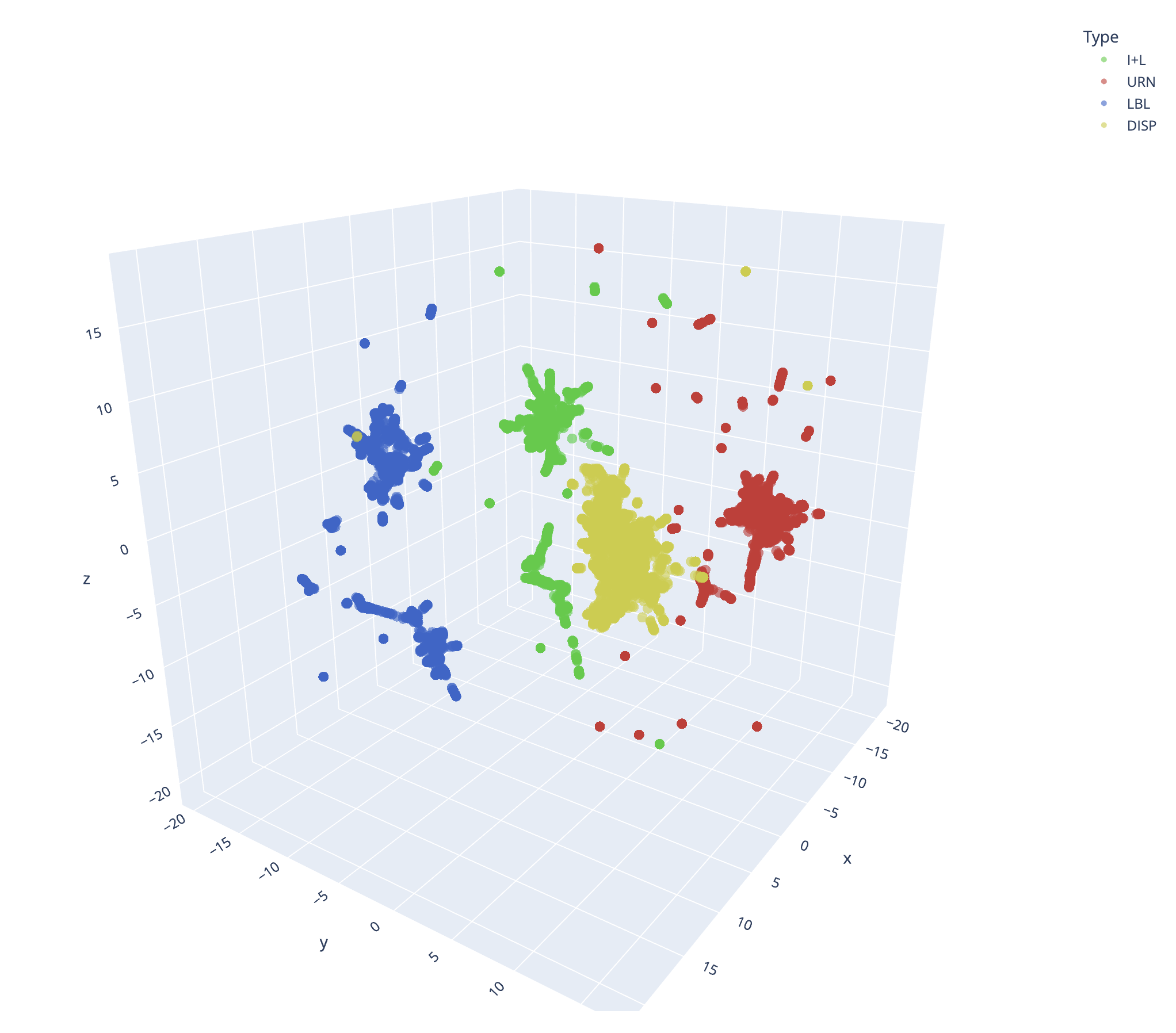}
    \caption{Three-dimensional PACMAP reduction of the embedding vectors. 
    The plot shows four distinct clusters corresponding to different vector types: 
    \texttt{LBL} (labels), \texttt{URN} (identifiers), \texttt{DISP} (content), 
    and \texttt{I+L} (combined identifier-plus-label). 
    }
    \label{fig:pacmap-polyvector}
\end{figure*}

In addition to the retrieval experiments, we performed a dimensionality reduction with PACMAP on the high-dimensional embeddings, then plotted them in three-dimensional space. As shown in Figure~\ref{fig:pacmap-polyvector}, the points neatly segregate into four distinct clusters: \texttt{DISP} for the full-text content embeddings, \texttt{LBL} for label-only embeddings, \texttt{URN} for identifier-only embeddings, and \texttt{I+L} for the combined identifier-plus-label vectors. This spatial separation indicates that Poly-Vector’s multiple representations (content, label, identifier) do not overlap or blur in the vector space, even when label and identifier are merged. Instead, each embedding type holds its own region, corroborating our approach of keeping reference-based embeddings (label and URN) distinct from content embeddings. 

\section{Experimental Setup}
\label{sec:experiments}

\subsection{Corpus and Preprocessing}
The experiment uses the text of the \textit{Constituição Federal do Brasil de 1988} (Brazil’s 1988 Federal Constitution) as the knowledge corpus. This document contains Titles, Chapters, Sections, Articles, Paragraphs (\textit{parágrafos}), Items (\textit{incisos}), and Sub-items (\textit{alíneas}) enumerating various legal provisions. Its structured nature makes it suitable for testing hierarchical retrieval methods and label-based lookup. The preprocessing steps for each method (a–h) involve segmenting the text and generating embeddings as described in Section~\ref{sec:methodology}. For the Poly-Vector methods (e–h), each structural unit is assigned \emph{three distinct label embeddings}: one for the textual label (e.g., \emph{“CFRB, Artigo 5º”}), another for its URN (e.g., \emph{“urn:lex:br:federal:constituicao:1988-10-05;1988!art5”}), and a combined label+URN string. These label-based vectors are embedded separately and linked to their corresponding content chunks, thereby supporting queries referencing a norm by name, by URN, or both.

\subsection{Retrieval Methods}
The eight retrieval methods detailed in Section \ref{sec:methodology} are implemented and compared: (a) Blind, (b) Flat, (c) Multi-layer \cite{lima2024unlocking}, (d) Multi-layer + Query Normalization \cite{lima2025improvingragretrievalpropositional}, (e) Poly-Vector + Blind, (f) Poly-Vector + Flat, (g) Poly-Vector + Multi-layer, and (h) Poly-Vector + Multi-layer + Query Normalization.

\subsection{Query Set}
We crafted a set of 8 representative questions in Portuguese to evaluate the retrieval performance of each method. These questions are designed to probe different scenarios: purely semantic, verbose semantic, and various types of label-based references. The queries are summarized in Table \ref{tab:queries}.

\begin{table*}
\footnotesize
\caption{Query Set for Evaluating Retrieval Methods on the Brazilian Constitution}
\begin{tabular}{p{0.02\textwidth} p{0.3\textwidth} p{0.3\textwidth} p{0.3\textwidth}}
\toprule
\textbf{\#} & \textbf{Query (Portuguese)} & \textbf{Query (English Translation)} & \textbf{Brief Comment/Purpose} \\
\midrule
Q1 & Quais são os objetivos fundamentais da República Federativa do Brasil? & What are the fundamental objectives of the Federative Republic of Brazil? & Control 1: Purely semantic query, no label mentioned. Tests baseline content retrieval. (Expected: Art. 3, caput) \\
Q2 & Por favor, você poderia me explicar quais direitos a Constituição garante aos povos indígenas? & Please, could you explain to me what rights the Constitution guarantees to indigenous peoples? & Control 2: Verbose semantic query, no label. Tests impact of query normalization. (Expected: Art. 231) \\
Q3 & Quais direitos são assegurados pelo art. 5º da Constituição? & What rights are guaranteed by Article 5 of the Constitution? & Label-based query (Article). Tests direct retrieval by article number. (Expected: Art. 5) \\
Q4 & Quais são os direitos previstos no art. 7º da Constituição? & What are the rights  provided for in Article 7 of the Constitution? & Label-based query (Article) + content keywords. Tests retrieval by article number for specific content. (Expected: Art. 7) \\
Q5 & Qual o tema do Capítulo VI do Título VIII da Constituição de 1988? & What is the theme of Chapter VI of Title VIII of the Constitution? & Label-based query (Title). Tests retrieval using higher-level structural labels. (Expected: Chapter I of Title VIII) \\
Q6 & Explique o art. 69 da Constituição. & Explain art. 69 of the Constitution. & Label-based query (Article) (Expected: Art. 69) \\
Q7 & Explique a norma urn:lex:br:federal:\allowbreak constituicao:1988-10-05;1988!art69 & Explain the norm urn:lex:br:federal:\allowbreak constitution:1988-10-05;1988!art69 & Identifier-based query. (Expected: Art. 69) \\
Q8 & Quais as diferenças entre o Art. 51 e o Art. 52 da Constituição? & What are the differences between Article 51 and Article 52 of the Constitution? & Tests retrieval and comparison involving multiple specific labels. (Expected: Arts. 51 \& 52) \\
\bottomrule
\label{tab:queries}
\end{tabular}
\end{table*}

\subsection{Evaluation Metrics}

For each query and each retrieval method, document chunks are retrieved based on cosine similarity between the query embedding and the document/label embeddings in the respective index. The selection of chunks to form the context follows an algorithm inspired by \cite{lima2025improvingragretrievalpropositional}, designed to gather relevant context dynamically:

\begin{enumerate}
    \item Embeddings are initially ranked by cosine similarity to the query.
    \item The algorithm iteratively selects the highest-ranked, non-duplicate text chunks.
    \item Selection continues until one of the following conditions is met:
        \begin{itemize}
            \item The cumulative token count of selected chunks reaches a target budget (here, 4000 tokens).
            \item The similarity score of the next best chunk drops by more than a specified threshold (here, 20\%) compared to the highest similarity score obtained in the initial ranking.
            \item A minimum number of chunks (here, 5) is selected, provided the similarity threshold hasn't been breached and candidates remain.
        \end{itemize}
    \item Crucially, for methods employing hierarchical structures (like Multi-Layer), an additional check prevents redundancy: if a smaller text segment (e.g., a paragraph) is retrieved but a larger segment containing it (e.g., the full article) has already been selected due to higher similarity, the smaller segment is discarded. This prioritizes the most encompassing relevant chunk identified.
\end{enumerate}

Once this variable set of context chunks is selected, we calculate the following metrics to evaluate the quality and relevance of the retrieved context, as presented in the Appendix tables (Tables \ref{tab:question_1} through \ref{tab:question_8}):

\begin{itemize}
    \item \textbf{Maximum Similarity (Max Sim.):} The highest cosine similarity score among the selected chunks. This indicates the relevance of the single best match found and is crucial for label-based queries where a direct match (e.g., to a label embedding) might yield a very high score.
    \item \textbf{Mean Similarity (Mean Sim.):} The average cosine similarity across all selected chunks. This reflects the overall relevance and consistency of the retrieved context set. Higher values suggest that, on average, the selected chunks are closely related to the query.
    \item \textbf{Minimum Similarity (Min Sim.):} The lowest cosine similarity score among the selected chunks. This shows the relevance floor of the retrieved context – how relevant even the least similar selected item is. A high minimum score indicates that all retrieved items maintain a strong connection to the query.
    \item \textbf{Standard Deviation (Std Dev.):} Measures the dispersion of similarity scores within the selected set. Lower values indicate more consistent relevance levels among the chunks.
    \item \textbf{Segments:} The total number of unique text chunks selected by the algorithm for the context.
    \item \textbf{Total Tokens:} The sum of tokens across all selected chunks, reflecting the size of the context provided.
\end{itemize}

The Appendix tables also show the retrieved items, indicating their embedding type (\texttt{ART}, \texttt{LBL}, \texttt{URN}, etc.) and their specific label or identifier, providing concrete examples of what the retrieval process identified as most relevant. No re-ranking beyond the initial cosine similarity scoring and the selection logic described above is applied. This set of metrics allows for a quantitative comparison of the effectiveness and characteristics of each retrieval method across different query types.

\section{Expected Results}
\label{sec:expected_results}

Before obtaining the experimental results, we outline our expectations for how each strategy will perform, based on their design:

\begin{itemize}
    \item \textbf{Baselines vs.\ Enhanced Methods:} We anticipate that the naive approaches (a: Blind Segmentation, b: Flat Per-Article) will generally be outperformed by methods leveraging structure or query processing. Blind segmentation (a) may struggle with relevance consistency (lower average/minimum similarity) due to arbitrary chunk boundaries. Flat per-article (b) is expected to handle larger units reasonably but might dilute relevance for queries targeting more specific or fine-grained provisions (e.g., Q7 with a direct URN reference). In contrast, Multi-layer indexing (c) should yield higher maximum similarities for these targeted queries by retrieving precisely defined smaller segments.

    \item \textbf{Effect of Query Normalization:} We expect Method (d) (Multi-layer + Query Norm) to outperform its counterpart (c) particularly on the verbose query (Q2), where removing extraneous phrasing and polite markers improves alignment with the text. As shown in prior work \cite{lima2025improvingragretrievalpropositional}, such preprocessing can yield higher similarity scores. For concise queries (Q1, Q3--Q8), the impact of normalization might be minimal or context-dependent.

    \item \textbf{Effect of Poly-Vector (Label Embeddings):} We hypothesize a clear improvement for methods (e)--(h) on label-oriented queries (Q3--Q8). Queries that explicitly reference articles or higher-level labels (e.g., ``Artigo 5º'') should match directly with the corresponding label embeddings and thus achieve high similarity scores. This advantage should be most evident where content-only similarity is limited. For instance, Q7 uses a URN identifier (``urn:lex:br:federal:constituicao:1988-10-05;1988!art69''), and Q8 compares multiple article numbers. Poly-Vector methods, especially those encoding fine-grained identifiers (g, h), are therefore expected to excel at resolving these references accurately.

    \item \textbf{Poly-Vector vs.\ Non-Poly Counterparts:} In comparing pairs of methods (e vs.\ a, f vs.\ b, g vs.\ c, h vs.\ d), we expect the Poly-Vector variant in each pair to demonstrate equal or better performance on the entire query set. Gains should be most pronounced for label-focused queries (Q3--Q8), where maximum and average similarity scores will likely be higher under Poly-Vector. 

    \item \textbf{Overall Best Approach:} Method (h) (\emph{Poly-Vector + Multi-layer + Query Norm}) combines all enhancements. It should handle verbose queries (Q2) via normalization, label-based queries (Q3--Q8) via Poly-Vector embeddings, and queries referencing structural subdivisions (Q5) or specific URNs (Q7) via the multi-layer index. Consequently, we predict it will attain both the highest and most consistent similarity scores across the entire query set.

    \item \textbf{Potential Trade-offs:} While Poly-Vector adds robust label-awareness, its efficacy depends on the quality and coverage of label extraction, as well as on how frequently users invoke such identifiers in their queries. Methods without label embeddings may occasionally yield high similarity on certain semantically rich queries, but their overall performance on label-driven or reference-heavy requests is expected to be significantly lower.
\end{itemize}

In summary, we anticipate that hierarchical segmentation (Multi-layer) and explicit label indexing (Poly-Vector) will yield notable benefits for retrieval accuracy in a legal context. Query normalization should further refine alignment for verbose or conversational queries. These hypotheses will be assessed against actual similarity metrics, followed by an in-depth analysis of the results.

\section{Results and Discussion}
\label{sec:results}

The experimental results, detailed in Tables \ref{tab:question_1} through \ref{tab:question_8} in the Appendix (using the updated data), provide quantitative insights into the performance of the eight retrieval strategies across different query types. We analyze these results query by query, comparing them against the expectations outlined in Section \ref{sec:expected_results}.  

\begin{figure*}[t]
    \centering
    \includegraphics[width=0.8\textwidth]{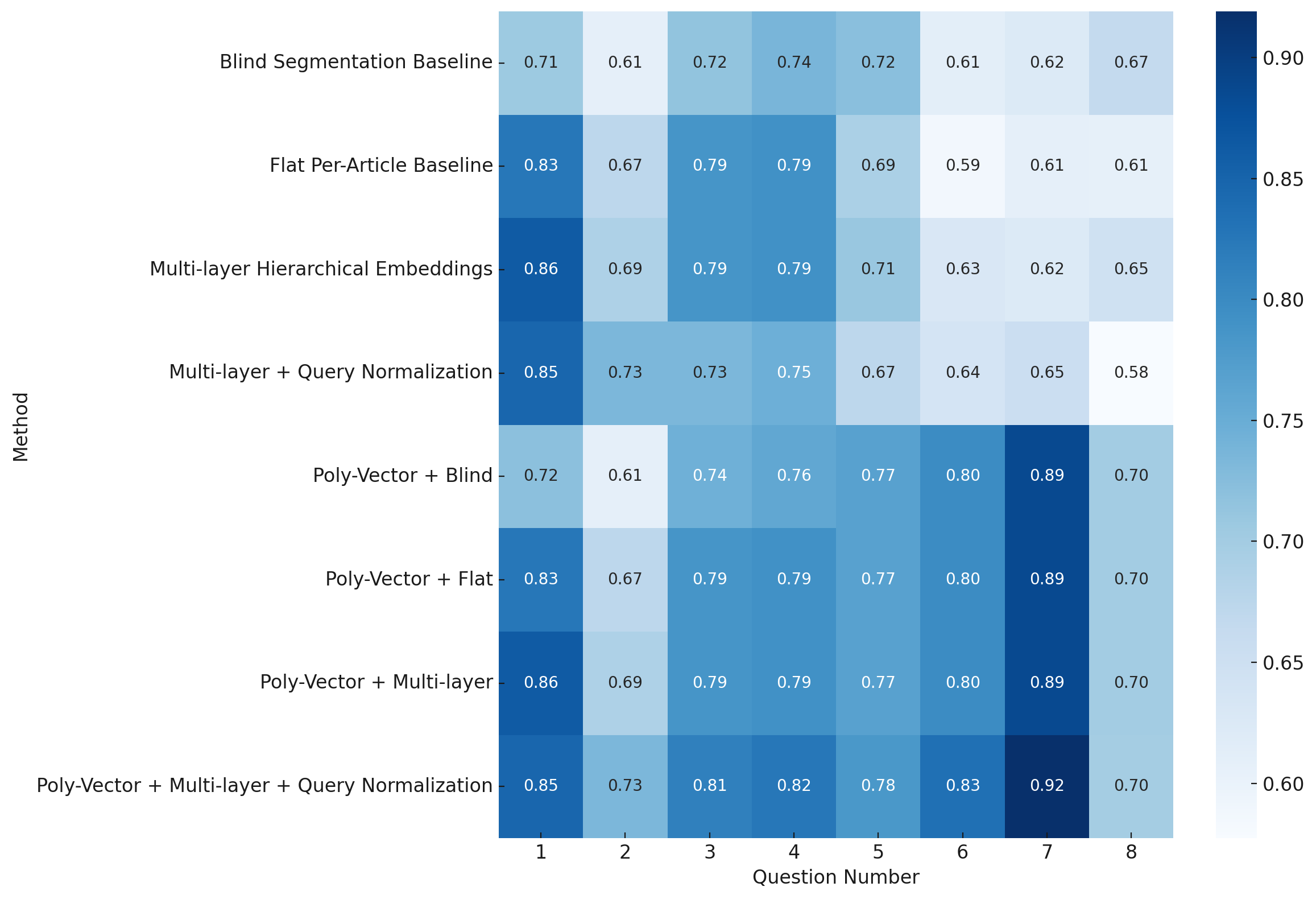}
    \caption{Heatmap of Maximum Similarity Scores Across All Queries and Methods. Each row corresponds to one retrieval method, and each column to one of the eight questions. Darker cells indicate higher maximum similarity scores.}
    \label{fig:polyvector-heatmap}
\end{figure*}

As a further illustration, Figure~\ref{fig:polyvector-heatmap} shows a \emph{heatmap} of the \textbf{maximum similarity scores} achieved by each method on each question. Along the vertical axis are the eight retrieval methods, and along the horizontal axis are the eight questions. Each cell in the matrix is colored according to the method's highest similarity for that particular question, with darker shades of blue indicating higher scores. This grid makes it easy to see, at a glance, which methods excel at particular question types. For example, we can confirm the advantage of Poly-Vector (especially the Poly-Vector + Multi-layer + Query Normalization variant) in queries that contain explicit article numbers (Questions~3, 4, 6, 7, and 8). Meanwhile, queries without direct references, such as Question~1, reveal that both the Multi-layer baseline and its Poly-Vector counterpart tie for the top-scoring approach, reinforcing the conclusion that Poly-Vector does not harm purely semantic queries while significantly improving label-centric retrieval.

\subsection{Query-Specific Analysis}

We now present a detailed analysis of how each retrieval method performed on the eight updated queries from Table~\ref{tab:queries}, referencing the revised results shown in Tables~\ref{tab:question_1}--\ref{tab:question_8}.

\subsubsection{Q1: Purely Semantic Query (``Fundamental objectives of Brazil?'')}
\textit{Expectation:} Methods leveraging the document structure (Multi-layer) should do well, and the addition of label embeddings (Poly-Vector) should not degrade performance. Query normalization was expected to have little impact on a concise query.

\textit{Results (Table~\ref{tab:question_1}):} 
Blind Segmentation (Method~a) attained a maximum similarity of only 0.7067, while both the Multi-layer baseline (c) and Poly-Vector + Multi-layer (g) reached the highest maximum similarity (0.8622). This reflects their success in pinpointing the \emph{caput} of Article~3 (the relevant text chunk) as top-1.  
Flat Per-Article methods (b, f) also performed decently (max.\ similarity 0.8256).  
Importantly, adding label embeddings (Methods~e--h) did not harm performance on this purely semantic query.  
Query normalization (d,~h) slightly lowered the \emph{peak} similarity (0.8479) but slightly \emph{increased} the average similarity of the retrieved set, suggesting that while the top-1 match remained strong, the overall retrieval consistency improved after removing conversational fluff from the query.

\subsubsection{Q2: Verbose Semantic Query (``Rights of indigenous peoples?'')}
\textit{Expectation:} Query normalization should help more conversational queries. Multi-layer indexing was also expected to outperform blind or purely flat segmentation on a strongly semantic task.

\textit{Results (Table~\ref{tab:question_2}):}
Methods using query normalization (d,~h) showed a clear benefit (max.\ similarities around 0.72--0.73) compared to their non-normalized counterparts (c,~g at 0.68).  
Although the Multi-layer approaches consistently retrieved the correct thematic section label for Indigenous rights (TÍTULO VIII, CAPÍTULO VIII), they occasionally included an \emph{inciso} from Article~20 in the top-2, which is less directly relevant than Article~231.  
In contrast, the flat baselines (b,~f) more directly retrieved Article~231 but did not achieve as high a top-1 similarity.  
Blind segmentation (a,~e) was the weakest, indicating that, for long queries with specialized terms, purely overlapping chunks can misalign with the sought-after passages.  

\subsubsection{Q3: Label-Based Query (``Rights guaranteed by Art.\ 5?'')}
\textit{Expectation:} Poly-Vector methods (e--h), which include a separate embedding for the label ``Art.\ 5\textdegree,'' should outperform purely content-based strategies.

\textit{Results (Table~\ref{tab:question_3}):}
All methods correctly retrieved portions of Article~5, but the Poly-Vector approaches had a clear edge in confidence.  
While Flat (b) and Multi-layer (c) both reached a respectable max.\ similarity of 0.7859 by matching the content itself, Poly-Vector + Multi-layer + Norm (h) achieved 0.8144 by directly matching the label embedding (\texttt{LBL}) for ``Art.\ 5\textdegree.''  
Thus, for direct article-number queries, adding label embeddings yields higher similarity scores and more precise top-1 retrieval.

\subsubsection{Q4: Label-Based Query + Keywords (``Rights in Art.\ 7?'')}
\textit{Expectation:} As with Q3, Poly-Vector methods should capitalize on the direct article label.

\textit{Results (Table~\ref{tab:question_4}):}
Poly-Vector + Multi-layer + Norm (h) again took the top spot (max.\ sim.\ 0.8248), retrieving \texttt{LBL} for ``Art.\ 7\textdegree'' first.  
Non-Poly multi-layer (c) and flat (b) baselines found the correct article content chunk (max.\ sim.\ 0.7912), but the label-based match in Poly-Vector methods gave a slight advantage.  
Hence, explicitly indexing article labels is beneficial even when a user appends semantic keywords.

\subsubsection{Q5: Higher-Level Label Query (``Theme of Chapter VI of Title VIII?'')}
\textit{Expectation:} Conventional content-only approaches struggle with references to broad structural units (``Título VIII,'' ``Capítulo VI''), whereas Poly-Vector should excel.

\textit{Results (Table~\ref{tab:question_5}):}
Non-Poly methods (a--d) fared poorly: their top similarities ranged from about 0.65 to 0.71 and did not retrieve the correct portion describing Chapter~VI of Title~VIII.  
By contrast, all Poly-Vector variants (e--h) promptly returned the exact label embeddings for \emph{``TÍTULO VIII, CAPÍTULO VI''} at high similarity (roughly 0.76--0.78).  
Among them, Poly-Vector + Multi-layer + Norm (h) was the best (max.\ sim.\ 0.7819).  
Thus, referencing large-scale headings is far more effective with a label embedding strategy.

\subsubsection{Q6: Purely Referential Query (``Explain Art.\ 69.'')}
\textit{Expectation:} Since the query is minimal and relies on a numeric label, Poly-Vector methods should dominate. Content-based approaches typically cannot match short numeric references effectively.

\textit{Results (Table~\ref{tab:question_6}):}
All Non-Poly methods (a--d) failed to retrieve the correct provision, with maximum similarities under 0.64.  
In contrast, each Poly-Vector method (e--h) found the exact label embedding (\texttt{LBL} = ``Art.\ 69''), returning maximum similarities from 0.7982 up to 0.8349, the latter achieved by Poly-Vector + Multi-layer + Norm (h).  
This starkly highlights how label embeddings resolve purely referential queries that do not share semantic overlap with the content.

\subsubsection{Q7: URN-Based Query (``Explain the norm \texttt{urn:lex:br:federal:constituicao:\allowbreak 1988-10-05;1988!art69}'')}
\textit{Expectation:} This query uses a URN identifier rather than the plain label. Poly-Vector storing URN embeddings (e--h) should outperform content-only baselines.

\textit{Results (Table~\ref{tab:question_7}):}
Non-Poly approaches (a--d) again exhibited low maximum similarity (under 0.65), failing to connect the URN to the actual text of Article~69.  
Poly-Vector indexing (e--g) soared to 0.8860 by matching the \texttt{URN} embedding directly.  
Further, combining label indexing with query normalization (method~h) reached 0.9191, the highest of all.  
Hence, modeling the URN as a distinct label embedding is highly effective for retrieving the correct chunk via purely referential identifiers.

\subsubsection{Q8: Multiple Label Query (``Differences between Art.\ 51 and Art.\ 52?'')}
\textit{Expectation:} Retrieving two distinct articles in one query is difficult for content-only methods, but Poly-Vector label embeddings can capture both references.

\textit{Results (Table~\ref{tab:question_8}):}
None of the Non-Poly methods returned both Article~51 and 52 in their top chunks; their maximum similarities mostly stayed under 0.65.  
By contrast, the four Poly-Vector methods successfully retrieved embeddings for both articles, with maximum similarities around 0.70.  
Method (e) (\emph{Poly-Vector + Blind}), for instance, identified \texttt{LBL} = ``Art.\ 51'' and \texttt{LBL} / \texttt{URN} for ``Art.\ 52,'' clarifying each article’s text.  
While Poly-Vector + Multi-layer + Norm (h) had a slightly lower max.\ similarity (0.6976), it still provided both references in the top results, confirming that Poly-Vector indexing effectively handles multi-label queries.

\subsection{Overall Discussion}

\begin{figure*}[p]
\centering
\includegraphics[width=\textwidth]{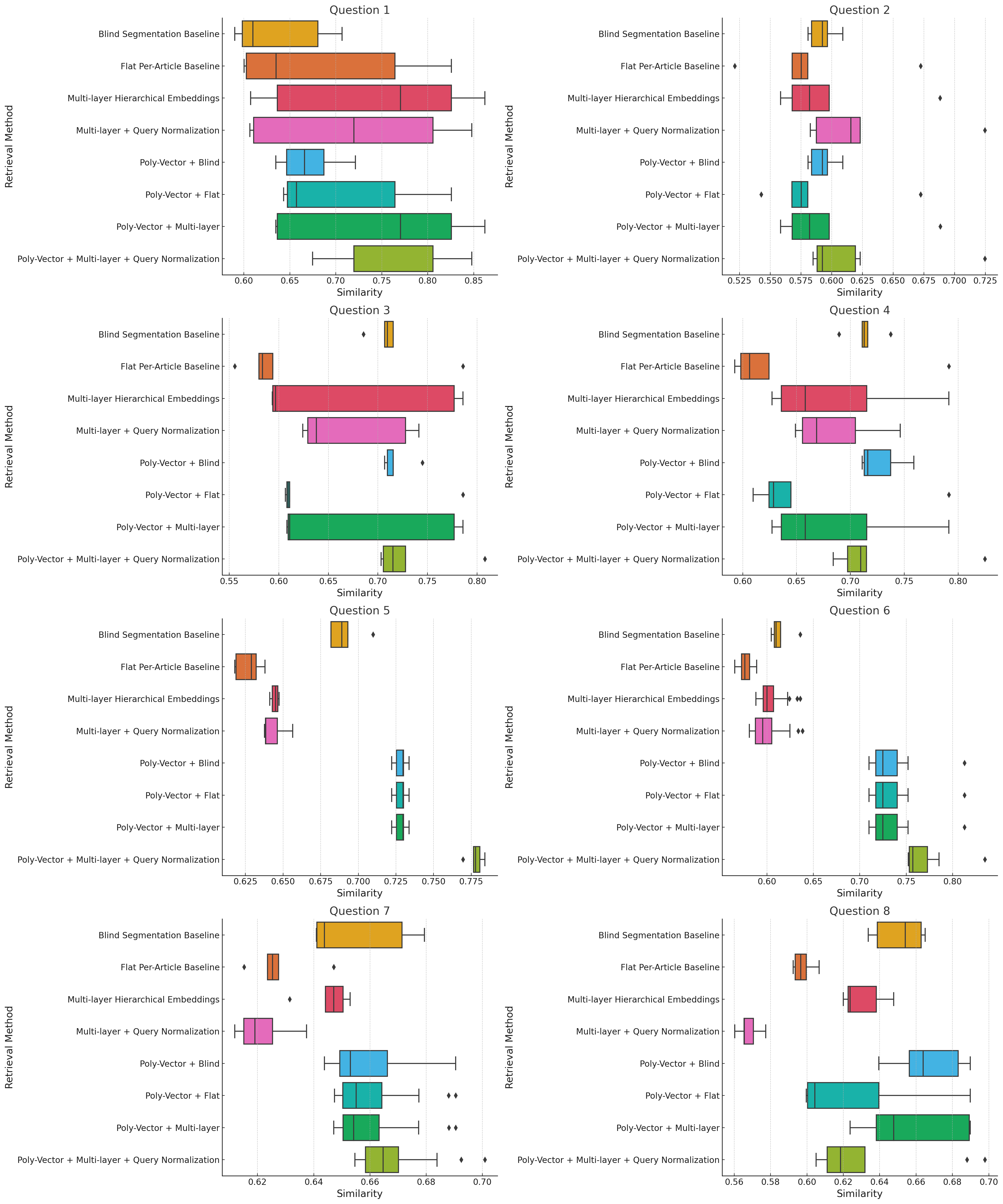}%
\caption{Distribution of Cosine Similarity Scores for Retrieved Chunks Across All Questions. Each box plot shows the median (center line), interquartile range (box), whiskers (extending to 1.5$\times$~IQR), and outliers (points) for the similarity scores achieved by each retrieval method for the corresponding question.}
\label{fig:boxplots}
\end{figure*}

Figure~\ref{fig:boxplots} visually summarizes the performance trends across the eight updated queries and the eight retrieval methods. Each box plot in the figure displays the distribution of cosine similarity scores for the retrieved chunks, grouped by query. The boxes represent the interquartile range (IQR), with the central horizontal line indicating the median. Whiskers typically span up to 1.5 times the IQR, and individual points outside this range are marked as outliers. Through this visualization, one can quickly gauge both the median retrieval quality (central tendency) and the spread (consistency) of results for each approach.

A principal conclusion from these experiments is that \textbf{Poly-Vector Retrieval} markedly enhances the handling of label-based or identifier-based queries in legal texts by assigning \emph{distinct} embeddings to labels (references) and their underlying content. In our evaluation on Brazil's 1988 Constitution, Poly-Vector approaches significantly outperformed content-only baselines (including Blind, Flat, and even hierarchical Multi-layer) on referential queries (e.g., Q3--Q8), all while preserving robust performance on purely semantic requests (Q1--Q2), especially when used in conjunction with multi-level chunking.

From a broader perspective, our experimental findings validate the following key points:

\begin{enumerate}
\item \textbf{Poly-Vector Excels at Referential Queries:}  
Whenever a query directly specifies an article number (Q3--Q4, Q6, Q8), a high-level section (Q5), or a URN identifier (Q7), the Poly-Vector variants provided more accurate and higher-confidence retrievals than standard content-only methods. This is most visible for the short, numeric queries in Q6 and the URN-based Q7, where non-Poly systems could not match the minimal textual overlap.

\item \textbf{No Harm to Semantic Queries:}  
Poly-Vector’s introduction of label embeddings did not degrade results for conceptually focused queries (Q1--Q2). For example, Poly-Vector + Multi-layer (Method~g) tied with its non-Poly counterpart (Method~c) on Q1 and consistently performed well on Q2, confirming that separating sense (content) and reference (label) can coexist with standard semantic retrieval mechanisms.

\item \textbf{Benefits of Multi-layer Hierarchical Embeddings:}  
The multi-level chunking approaches (Methods~c, d, g, h) generally surpassed their Blind or Flat counterparts on semantic tasks. For instance, on Q1 and Q2, the hierarchical indexing of articles, paragraphs, and higher-level groupings helped isolate the most relevant textual provisions. Even so, occasional mismatches highlight that the multi-layer approach may retrieve some adjacent but nonessential segments, underscoring the importance of robust selection heuristics.

\item \textbf{Query Normalization Helps Conversational Requests:}  
In line with previous work, normalizing verbose queries (Methods~d, h) yielded tangible improvements for Q2 (a more conversational prompt), as well as for certain label-centric queries. By stripping politeness and indirect speech from user questions, the system aligned better with the legal text’s phrasing and labels. For shorter or already direct queries (e.g., Q1, Q6), the effect was modest.

\item \textbf{Combining All Enhancements is Most Effective:}  
Method~(h) (\emph{Poly-Vector + Multi-layer + Query Normalization}) generally maintained top or near-top performance across the board. It captures label-based pointers, leverages hierarchical content chunking, and benefits from simplified queries, thus balancing both semantic breadth and referential precision.

\item \textbf{URN-based Queries Show the Power of Label Embeddings:}  
Q7, which referenced Article~69 purely by its URN (\texttt{urn:lex:br:federal:constituicao:\allowbreak 1988-10-05;1988!art69}), clearly demonstrated how non-Poly systems fail (similarities under 0.65), while Poly-Vector methods attained scores as high as 0.9191. This underscores the value of explicitly embedding structured identifiers in a domain where rigid, standardized references are common.
\end{enumerate}

These trends are visually apparent in Figure~\ref{fig:boxplots}. For queries heavily relying on explicit labels or identifiers (Q3--Q8), the teal and green box plots (Poly-Vector methods) show higher medians and tighter dispersions of similarity scores compared to orange and red (non-Poly) plots. By contrast, for the more purely semantic Q1 and the verbose Q2, the hierarchical approaches appear to dominate, reflecting the advantage of multi-layer indexing for concept-based questions.

From an applied standpoint, this shows the practical necessity of modeling explicit legal references or identifiers when designing Retrieval-Augmented Generation (RAG) systems in law. Traditional embedding-based retrieval often falters when faced with minimal textual cues and numeric or URN-like designators. Poly-Vector remedies this by assigning dedicated “label” embeddings that effectively function as rigid designators, bridging the gap between a user’s referential query and the actual text chunk. Overall, our experiments confirm that \emph{both} sense (semantic content) \emph{and} reference (article labels, URNs) are essential ingredients for robust legal document retrieval. 

When examining retrieval logs and similarity rankings (see Section~\ref{sec:results}), we found that the \texttt{I+L} embeddings did \emph{not} consistently outperform the simpler \texttt{LBL} or \texttt{URN} vectors in practice. In most label-based queries, either \texttt{LBL} or \texttt{URN} alone produced equally high (or higher) similarity scores, indicating that the embedding model was already effective in recognizing the referential link from isolated identifiers. Consequently, our empirical results did not confirm the initial hypothesis that the combined \texttt{I+L} vector would be prioritized during retrieval. Nonetheless, we retain \texttt{I+L} as an option for potential use cases involving less conventional referencing patterns.

\section{Conclusions}
\label{sec:conclusion}

This paper introduced the \textbf{Poly-Vector Retrieval} approach, designed to enhance Retrieval-Augmented Generation (RAG) systems in domains like law, where referencing documents or provisions by specific labels (e.g., “Article 5,” “Consumer Protection Code”) or standardized identifiers (e.g., URNs) is common. Inspired by the philosophical distinction between \emph{Sense}  and \emph{Reference}, Poly-Vector assigns separate embeddings to the substantive text (semantic sense) of each document chunk and to any associated labels (rigid references). This mechanism aims to significantly improve retrieval accuracy whenever queries explicitly cite those labels or identifiers.

Our experiments on the 1988 Brazilian Federal Constitution confirm the efficacy of Poly-Vector Retrieval under specific conditions:

\begin{itemize}
    \item \textbf{Label-Centric and URN-Based Queries:} Poly-Vector consistently delivered superior performance for questions mentioning articles by number, referencing larger structural headings (e.g., “Title VIII, Chapter VI”), or specifying URNs (e.g., ``urn:lex:br:federal:constituicao:1988-10-05;1988!art69''). On these purely referential queries (e.g., “Explain Art.~69”), content-only methods struggled, but Poly-Vector reliably retrieved the correct passage thanks to explicit label embeddings.
    
    \item \textbf{No Degradation on Semantic Queries:} Crucially, adding label embeddings did not impede retrieval on purely semantic queries lacking any label references. Poly-Vector methods performed on par with content-only approaches for these tasks, indicating compatibility with standard semantic search.

    \item \textbf{Synergy with Multi-Layer and Normalization:} Combining Poly-Vector with multi-layer hierarchical chunking and optional query normalization produced the most consistent performance across varied query styles. Hierarchical chunking aided in matching semantic scope, normalization boosted alignment on verbose prompts, and label indexing ensured direct resolution for referential queries.

\end{itemize}

By explicitly distinguishing between reference (label/URN) and sense (content), Poly-Vector Retrieval proposes a targeted upgrade for RAG systems, especially valuable in legal information retrieval where precise citations are paramount. Its demonstrated ability to address referential queries without reducing semantic performance makes it an attractive addition to structured, identifier-heavy corpora.

\paragraph{Future Directions} Several areas warrant further exploration:

\begin{itemize}
    \item \textbf{Advanced Label Management:} Investigate automated methods to extract, normalize, and manage a broader spectrum of identifiers, including hierarchies, abbreviations, and aliases.
    \item \textbf{Cross-Domain Application:} Validate Poly-Vector in other fields with rich structural labeling (e.g., financial regulations, medical guidelines, or engineering standards).
    \item \textbf{Scalability and Indexing Efficiency:} Explore robust indexing solutions that can handle large-scale Poly-Vector databases consisting of numerous content and label embeddings.
    \item \textbf{Automated Cross-Reference Resolution:} Deploy Poly-Vector as a mechanism to detect internal cross-references (e.g., “pursuant to Article~34”) within retrieved content, automatically expanding the context to reduce omissions.

\end{itemize}

It should be noted that our evaluation concentrated on cosine similarity scores as a measure of retrieval quality, which, while informative, does not equate to a full end-to-end assessment using classical IR metrics (precision, recall, F1) nor does it measure final LLM output quality. Furthermore, the reliance on a single (though complex) legal text---the Brazilian Constitution---and a handcrafted query set may limit the generality of our conclusions. These factors point to a need for broader validation in diverse settings.

In sum, Poly-Vector Retrieval stands as a promising enhancement to RAG systems operating on references-rich documents. Its advantage in handling label-based or URN-based queries, without compromising semantic retrieval, suggests a strong potential to boost the reliability and accuracy of information retrieval in legal and other specialized domains.

\bibliographystyle{unsrt}
\bibliography{Polyvector}

\section*{Acknowledgments}
The author is grateful to Ari Hershowitz for his comments on the draft of this paper, which have contributed to its improvement. 

\end{multicols}

\clearpage

\clearpage

\hypertarget{appendix-questions}{%
\section{Appendix}\label{appendix-questions}}

\subsection{Presentation and Interpretation of Query Metrics}
The data is systematically presented in a series of tables, each organized by question number. Within each table, rows represent different retrieval methods applied to the corpus, followed by two supplementary rows that provide detailed information about the top-performing chunks.

Each table presents a comprehensive set of metrics for comparative analysis:

\textbf{Method} describes the retrieval approach used, as defined by the filter description;

\textbf{Min Sim.} and \textbf{Max Sim.} represent the lowest and highest similarity scores obtained by each method, offering insight into the range of semantic relevance;

\textbf{Mean Sim.} provides the average similarity score, indicating overall retrieval quality;

\textbf{Std Dev.} quantifies the variability in similarity scores, with lower values suggesting more consistent performance;

\textbf{Total Tokens} represents the cumulative size of processed text in token units.

\textbf{Segments} counts the total number of text chunks processed during retrieval; and

For each retrieval method,  additional rows display the highest-ranked selected chunks. presents the embedding type of the chunk with the highest similarity score, accompanied by its label (where "CONSTITUIÇÃO DA REPÚBLICA FEDERATIVA DO BRASIL" is abbreviated as "CFRB").

This enhanced tabular organization facilitates not only statistical comparison between different retrieval methods for each question but also provides contextual examples of the most relevant content retrieved.

% Add these lines to your preamble if not already included
% \usepackage{longtable}
% \usepackage{pdflscape}
% \usepackage{geometry}

\begin{landscape}
\small
\textbf{Question 1} \\
\textbf{Original Question:} Quais são os objetivos fundamentais da República Federativa do Brasil? \\
\textbf{Normalized Question:} Objetivos fundamentais da República Federativa do Brasil \\
\textbf{Expected Top-1 Segment according to the chunking strategy:} 
\begin{itemize}
\item \textbf{Blind:} Chunk \#1  
\item \textbf{Flat:} Constituição da República Federativa do Brasil, Art. 3º  
\item \textbf{Multi-layer:} Constituição da República Federativa do Brasil, Art. 3º, caput 
\item \textbf{Poly-Vector:}  \begin{enumerate} \item (Label:) Constituição da República Federativa do Brasil, Art. 3º, caput   \item (URN+Label:) urn:lex:br:federal:constituicao:1988-10-05;1988!art3\_cpt, Constituição da República Federativa do Brasil, Art. 3º, caput   \item  (URN:) urn:lex:br:federal:constituicao:1988-10-05;1988!art3\_cpt 
\end{enumerate} \end{itemize}
\begin{longtable}{|l|r|r|r|r|r|r|}
\caption{Statistics for Question 1} \label{tab:question_1} \\
\hline
\multicolumn{7}{|c|}{\textbf{Statistics for Question 1}} \\
\hline
\textbf{Method} & \textbf{Min Sim.} & \textbf{Max Sim.} & \textbf{Mean Sim.} & \textbf{Std Dev.} & \textbf{Tokens} & \textbf{Segments} \\
\hline
\endfirsthead
\multicolumn{7}{c}%
{\tablename\ \thetable\ -- \textit{Continued from previous page}} \\
\hline
\textbf{Method} & \textbf{Min Sim.} & \textbf{Max Sim.} & \textbf{Mean Sim.} & \textbf{Std Dev.} & \textbf{Tokens} & \textbf{Segments} \\
\hline
\endhead
\hline \multicolumn{7}{|r|}{{Continued on next page}} \\ \hline
\endfoot
\hline
\endlastfoot
Blind Segmentation Baseline & 0.5899 & 0.7067 & 0.6370 & 0.0530 & 4000 & 5 \\
\multicolumn{1}{|r|}{\textit{\#1:} Blind} & \multicolumn{4}{l|}{\textbf{Chunk \#1} ["CONSTITUIÇÃO DA REPÚBLICA FEDERATIVA DO  ... rma da lei, a proteção aos loc"]} & 800 & 0.7067 \\
\multicolumn{1}{|r|}{\textit{\#2:} Blind} & \multicolumn{4}{l|}{Chunk \#2 ["ução pacífica dos conflitos; VIII – repú ... o à informação e resguardado o"]} & 800 & 0.6805 \\
\multicolumn{1}{|r|}{\textit{\#3:} Blind} & \multicolumn{4}{l|}{Chunk \#24 [" às respectivas candidatas, deverão ser  ... ias fluviais; IV – as ilhas fl"]} & 800 & 0.6096 \\
\multicolumn{1}{|r|}{\textit{\#4:} Blind} & \multicolumn{4}{l|}{Chunk \#23 [" elegido pelo menos quinze Deputados Fed ... nal, por lei complementar. § 4"]} & 800 & 0.5983 \\
\multicolumn{1}{|r|}{\textit{\#5:} Blind} & \multicolumn{4}{l|}{Chunk \#18 ["e da Câmara dos Deputados; III – de Pres ... tes do pleito. § 7º São ineleg"]} & 800 & 0.5899 \\
\hline
Flat Per-Article Baseline & 0.6001 & 0.8256 & 0.6855 & 0.1032 & 852 & 5 \\
\multicolumn{1}{|r|}{\textit{\#1:} ART} & \multicolumn{4}{l|}{\textbf{CRFB, Art. 3º}} & 123 & 0.8256 \\
\multicolumn{1}{|r|}{\textit{\#2:} ART} & \multicolumn{4}{l|}{CRFB, Art. 1º} & 148 & 0.7643 \\
\multicolumn{1}{|r|}{\textit{\#3:} ART} & \multicolumn{4}{l|}{CRFB, Art. 4º} & 194 & 0.6349 \\
\multicolumn{1}{|r|}{\textit{\#4:} ART} & \multicolumn{4}{l|}{CRFB, Art. 170.} & 279 & 0.6026 \\
\multicolumn{1}{|r|}{\textit{\#5:} ART} & \multicolumn{4}{l|}{CRFB, Art. 193.} & 108 & 0.6001 \\
\hline
Multi-layer Hierarchical Embeddings & 0.6072 & 0.8622 & 0.7403 & 0.1135 & 940 & 5 \\
\multicolumn{1}{|r|}{\textit{\#1:} CPT} & \multicolumn{4}{l|}{\textbf{CRFB, Art. 3º, caput}} & 97 & \textbf{0.8622} \\
\multicolumn{1}{|r|}{\textit{\#2:} ART} & \multicolumn{4}{l|}{CRFB, Art. 3º} & 123 & 0.8256 \\
\multicolumn{1}{|r|}{\textit{\#3:} TIT} & \multicolumn{4}{l|}{CRFB, TÍTULO I} & 490 & 0.7703 \\
\multicolumn{1}{|r|}{\textit{\#4:} CPT} & \multicolumn{4}{l|}{CRFB, Art. 170., caput} & 172 & 0.6362 \\
\multicolumn{1}{|r|}{\textit{\#5:} PAR} & \multicolumn{4}{l|}{CRFB, Art. 60., § 4º} & 58 & 0.6072 \\
\hline
Multi-layer + Query Normalization & 0.6064 & 0.8479 & 0.7180 & 0.1101 & 940 & 5 \\
\multicolumn{1}{|r|}{\textit{\#1:} CPT} & \multicolumn{4}{l|}{\textbf{CRFB, Art. 3º, caput}} & 97 & 0.8479 \\
\multicolumn{1}{|r|}{\textit{\#2:} ART} & \multicolumn{4}{l|}{CRFB, Art. 3º} & 123 & 0.8056 \\
\multicolumn{1}{|r|}{\textit{\#3:} TIT} & \multicolumn{4}{l|}{CRFB, TÍTULO I} & 490 & 0.7196 \\
\multicolumn{1}{|r|}{\textit{\#4:} PAR} & \multicolumn{4}{l|}{CRFB, Art. 60., § 4º} & 58 & 0.6107 \\
\multicolumn{1}{|r|}{\textit{\#5:} CPT} & \multicolumn{4}{l|}{CRFB, Art. 170., caput} & 172 & 0.6064 \\
\hline
Poly-Vector + Blind & 0.6347 & 0.7213 & 0.6625 & 0.0273 & 6702 & 13 \\
\multicolumn{1}{|r|}{\textit{\#1:} LBL} & \multicolumn{4}{l|}{CRFB, Art. 1º} & 148 & 0.7213 \\
\multicolumn{1}{|r|}{\textit{\#2:} Blind} & \multicolumn{4}{l|}{\textbf{Chunk \#1} ["CONSTITUIÇÃO DA REPÚBLICA FEDERATIVA DO  ... rma da lei, a proteção aos loc"]} & 800 & 0.7067 \\
\multicolumn{1}{|r|}{\textit{\#3:} Blind} & \multicolumn{4}{l|}{Chunk \#2 ["ução pacífica dos conflitos; VIII – repú ... o à informação e resguardado o"]} & 800 & 0.6805 \\
\multicolumn{1}{|r|}{\textit{\#4:} LBL} & \multicolumn{4}{l|}{CRFB, Art. 3º} & 123 & 0.6747 \\
\multicolumn{1}{|r|}{\textit{\#5:} I,L} & \multicolumn{4}{l|}{urn:lex:br:federal:constituicao:1988-10-05;1988!art1\_cpt\_inc4, CRFB, Art. 1º, caput, Inciso IV} & 68 & 0.6690 \\
\multicolumn{1}{|r|}{\textit{\#6:} I,L} & \multicolumn{4}{l|}{urn:lex:br:federal:constituicao:1988-10-05;1988!art2, CRFB, Art. 2º} & 51 & 0.6597 \\
\multicolumn{1}{|r|}{\textit{\#7:} I,L} & \multicolumn{4}{l|}{urn:lex:br:federal:constituicao:1988-10-05;1988!art3\_cpt\_inc4, CRFB, Art. 3º, caput, Inciso IV} & 58 & 0.6521 \\
\multicolumn{1}{|r|}{\textit{\#8:} I,L} & \multicolumn{4}{l|}{urn:lex:br:federal:constituicao:1988-10-05;1988!art1\_cpt\_inc2, CRFB, Art. 1º, caput, Inciso II} & 61 & 0.6509 \\
\multicolumn{1}{|r|}{\textit{\#9:} LBL} & \multicolumn{4}{l|}{CRFB, Art. 4º} & 194 & 0.6474 \\
\multicolumn{1}{|r|}{\textit{\#10:} LBL} & \multicolumn{4}{l|}{CRFB, TÍTULO I} & 490 & 0.6433 \\
\multicolumn{1}{|r|}{\textit{\#11:} I,L} & \multicolumn{4}{l|}{urn:lex:br:federal:constituicao:1988-10-05;1988!art1\_cpt\_inc3, CRFB, Art. 1º, caput, Inciso III} & 64 & 0.6367 \\
\multicolumn{1}{|r|}{\textit{\#12:} I,L} & \multicolumn{4}{l|}{urn:lex:br:federal:constituicao:1988-10-05;1988!art3\_cpt\_inc2, CRFB, Art. 3º, caput, Inciso II} & 30 & 0.6356 \\
\multicolumn{1}{|r|}{\textit{\#13:} LBL} & \multicolumn{4}{l|}{CRFB, TÍTULO II, CAPÍTULO I} & 3815 & 0.6347 \\
\hline
Poly-Vector + Flat & 0.6521 & 0.8256 & 0.7142 & 0.0771 & 448 & 5 \\
\multicolumn{1}{|r|}{\textit{\#1:} ART} & \multicolumn{4}{l|}{\textbf{CRFB, Art. 3º}} & 123 & 0.8256 \\
\multicolumn{1}{|r|}{\textit{\#2:} ART} & \multicolumn{4}{l|}{CRFB, Art. 1º} & 148 & 0.7643 \\
\multicolumn{1}{|r|}{\textit{\#3:} I,L} & \multicolumn{4}{l|}{urn:lex:br:federal:constituicao:1988-10-05;1988!art1\_cpt\_inc4, CRFB, Art. 1º, caput, Inciso IV} & 68 & 0.6690 \\
\multicolumn{1}{|r|}{\textit{\#4:} I,L} & \multicolumn{4}{l|}{urn:lex:br:federal:constituicao:1988-10-05;1988!art2, CRFB, Art. 2º} & 51 & 0.6597 \\
\multicolumn{1}{|r|}{\textit{\#5:} I,L} & \multicolumn{4}{l|}{urn:lex:br:federal:constituicao:1988-10-05;1988!art3\_cpt\_inc4, CRFB, Art. 3º, caput, Inciso IV} & 58 & 0.6521 \\
\hline
Poly-Vector + Multi-layer & 0.6362 & 0.8622 & 0.7508 & 0.0998 & 933 & 5 \\
\multicolumn{1}{|r|}{\textit{\#1:} CPT} & \multicolumn{4}{l|}{\textbf{CRFB, Art. 3º, caput}} & 97 & \textbf{0.8622} \\
\multicolumn{1}{|r|}{\textit{\#2:} ART} & \multicolumn{4}{l|}{CRFB, Art. 3º} & 123 & 0.8256 \\
\multicolumn{1}{|r|}{\textit{\#3:} TIT} & \multicolumn{4}{l|}{CRFB, TÍTULO I} & 490 & 0.7703 \\
\multicolumn{1}{|r|}{\textit{\#4:} I,L} & \multicolumn{4}{l|}{urn:lex:br:federal:constituicao:1988-10-05;1988!art2, CRFB, Art. 2º} & 51 & 0.6597 \\
\multicolumn{1}{|r|}{\textit{\#5:} CPT} & \multicolumn{4}{l|}{CRFB, Art. 170., caput} & 172 & 0.6362 \\
\hline
Poly-Vector + Multi-layer + Q. Norm. & 0.6747 & 0.8479 & 0.7535 & 0.0709 & 4463 & 5 \\
\multicolumn{1}{|r|}{\textit{\#1:} CPT} & \multicolumn{4}{l|}{\textbf{CRFB, Art. 3º, caput}} & 97 & 0.8479 \\
\multicolumn{1}{|r|}{\textit{\#2:} ART} & \multicolumn{4}{l|}{CRFB, Art. 3º} & 123 & 0.8056 \\
\multicolumn{1}{|r|}{\textit{\#3:} LBL} & \multicolumn{4}{l|}{CRFB, Art. 1º} & 148 & 0.7197 \\
\multicolumn{1}{|r|}{\textit{\#4:} TIT} & \multicolumn{4}{l|}{CRFB, TÍTULO I} & 490 & 0.7196 \\
\multicolumn{1}{|r|}{\textit{\#5:} LBL} & \multicolumn{4}{l|}{CRFB, TÍTULO V} & 3605 & 0.6747 \\
\hline
\end{longtable}
\clearpage

\textbf{Question 2} \\
\textbf{Original Question:} Por favor, você poderia me explicar quais direitos a Constituição garante aos povos indígenas? \\
\textbf{Normalized Question:} Direitos garantidos aos povos indígenas pela Constituição" \\
\textbf{Expected Top-1 Segment according to the chunking strategy:} 
\begin{itemize}
\item \textbf{Blind:} Chunk \#276  
\item \textbf{Flat:} Constituição da República Federativa do Brasil, Art. 231.  
\item \textbf{Multi-layer:} Constituição da República Federativa do Brasil, TÍTULO VIII, CAPÍTULO VIII 
\item \textbf{Poly-Vector:}  \begin{enumerate} \item (Label:) Constituição da República Federativa do Brasil, TÍTULO VIII, CAPÍTULO VIII   \item (URN+Label:) urn:lex:br:federal:constituicao:1988-10-05;1988!tit8\_cap8, Constituição da República Federativa do Brasil, TÍTULO VIII, CAPÍTULO VIII   \item  (URN:) urn:lex:br:federal:constituicao:1988-10-05;1988!tit8\_cap8 
\end{enumerate} \end{itemize}
\begin{longtable}{|l|r|r|r|r|r|r|}
\caption{Statistics for Question 2} \label{tab:question_2} \\
\hline
\multicolumn{7}{|c|}{\textbf{Statistics for Question 2}} \\
\hline
\textbf{Method} & \textbf{Min Sim.} & \textbf{Max Sim.} & \textbf{Mean Sim.} & \textbf{Std Dev.} & \textbf{Tokens} & \textbf{Segments} \\
\hline
\endfirsthead
\multicolumn{7}{c}%
{\tablename\ \thetable\ -- \textit{Continued from previous page}} \\
\hline
\textbf{Method} & \textbf{Min Sim.} & \textbf{Max Sim.} & \textbf{Mean Sim.} & \textbf{Std Dev.} & \textbf{Tokens} & \textbf{Segments} \\
\hline
\endhead
\hline \multicolumn{7}{|r|}{{Continued on next page}} \\ \hline
\endfoot
\hline
\endlastfoot
Blind Segmentation Baseline & 0.5807 & 0.6090 & 0.5924 & 0.0113 & 4000 & 5 \\
\multicolumn{1}{|r|}{\textit{\#1:} Blind} & \multicolumn{4}{l|}{Chunk \#3 ["ais de culto e a suas liturgias; VII – é ... e perigo público, a autoridade"]} & 800 & 0.6090 \\
\multicolumn{1}{|r|}{\textit{\#2:} Blind} & \multicolumn{4}{l|}{Chunk \#4 [" sigilo da fonte, quando necessário ao e ... do promoverá, na forma da lei,"]} & 800 & 0.5965 \\
\multicolumn{1}{|r|}{\textit{\#3:} Blind} & \multicolumn{4}{l|}{\textbf{Chunk \#276} [" têm o dever de assistir, criar e educar ...  4º. Art. 232. Os índios, suas"]} & 800 & 0.5923 \\
\multicolumn{1}{|r|}{\textit{\#4:} Blind} & \multicolumn{4}{l|}{Chunk \#10 [" público; LXX – o mandado de segurança c ... ncia, a assistência aos desamp"]} & 800 & 0.5836 \\
\multicolumn{1}{|r|}{\textit{\#5:} Blind} & \multicolumn{4}{l|}{Chunk \#2 ["ução pacífica dos conflitos; VIII – repú ... o à informação e resguardado o"]} & 800 & 0.5807 \\
\hline
Flat Per-Article Baseline & 0.5210 & 0.6724 & 0.5834 & 0.0551 & 6759 & 5 \\
\multicolumn{1}{|r|}{\textit{\#1:} ART} & \multicolumn{4}{l|}{\textbf{CRFB, Art. 231.}} & 631 & 0.6724 \\
\multicolumn{1}{|r|}{\textit{\#2:} ART} & \multicolumn{4}{l|}{CRFB, Art. 232.} & 70 & 0.5805 \\
\multicolumn{1}{|r|}{\textit{\#3:} ART} & \multicolumn{4}{l|}{CRFB, Art. 5º} & 3791 & 0.5752 \\
\multicolumn{1}{|r|}{\textit{\#4:} ART} & \multicolumn{4}{l|}{CRFB, Art. 7º} & 1297 & 0.5678 \\
\multicolumn{1}{|r|}{\textit{\#5:} ART} & \multicolumn{4}{l|}{CRFB, Art. 225.} & 970 & 0.5210 \\
\hline
Multi-layer Hierarchical Embeddings & 0.5584 & 0.6882 & 0.5989 & 0.0521 & 7934 & 5 \\
\multicolumn{1}{|r|}{\textit{\#1:} CAP} & \multicolumn{4}{l|}{\textbf{CRFB, TÍTULO VIII, CAPÍTULO VIII}} & 698 & 0.6882 \\
\multicolumn{1}{|r|}{\textit{\#2:} INC} & \multicolumn{4}{l|}{CRFB, Art. 20., caput, Inciso XI} & 29 & 0.5980 \\
\multicolumn{1}{|r|}{\textit{\#3:} CAP} & \multicolumn{4}{l|}{CRFB, TÍTULO II, CAPÍTULO I} & 3815 & 0.5819 \\
\multicolumn{1}{|r|}{\textit{\#4:} ART} & \multicolumn{4}{l|}{CRFB, Art. 7º} & 1297 & 0.5678 \\
\multicolumn{1}{|r|}{\textit{\#5:} CAP} & \multicolumn{4}{l|}{CRFB, TÍTULO II, CAPÍTULO II} & 2095 & 0.5584 \\
\hline
Multi-layer + Query Normalization & 0.5864 & 0.7343 & 0.6275 & 0.0612 & 865 & 5 \\
\multicolumn{1}{|r|}{\textit{\#1:} CAP} & \multicolumn{4}{l|}{\textbf{CRFB, TÍTULO VIII, CAPÍTULO VIII}} & 698 & \textbf{0.7343} \\
\multicolumn{1}{|r|}{\textit{\#2:} INC} & \multicolumn{4}{l|}{CRFB, Art. 20., caput, Inciso XI} & 29 & 0.6226 \\
\multicolumn{1}{|r|}{\textit{\#3:} PAR} & \multicolumn{4}{l|}{CRFB, Art. 5º, § 2º} & 50 & 0.5974 \\
\multicolumn{1}{|r|}{\textit{\#4:} INC} & \multicolumn{4}{l|}{CRFB, Art. 22., caput, Inciso XIV} & 29 & 0.5969 \\
\multicolumn{1}{|r|}{\textit{\#5:} INC} & \multicolumn{4}{l|}{CRFB, Art. 49., caput, Inciso XVI} & 59 & 0.5864 \\
\hline
Poly-Vector + Blind & 0.5808 & 0.6092 & 0.5925 & 0.0113 & 4000 & 5 \\
\multicolumn{1}{|r|}{\textit{\#1:} Blind} & \multicolumn{4}{l|}{Chunk \#3 ["ais de culto e a suas liturgias; VII – é ... e perigo público, a autoridade"]} & 800 & 0.6092 \\
\multicolumn{1}{|r|}{\textit{\#2:} Blind} & \multicolumn{4}{l|}{Chunk \#4 [" sigilo da fonte, quando necessário ao e ... do promoverá, na forma da lei,"]} & 800 & 0.5967 \\
\multicolumn{1}{|r|}{\textit{\#3:} Blind} & \multicolumn{4}{l|}{\textbf{Chunk \#276} [" têm o dever de assistir, criar e educar ...  4º. Art. 232. Os índios, suas"]} & 800 & 0.5923 \\
\multicolumn{1}{|r|}{\textit{\#4:} Blind} & \multicolumn{4}{l|}{Chunk \#10 [" público; LXX – o mandado de segurança c ... ncia, a assistência aos desamp"]} & 800 & 0.5836 \\
\multicolumn{1}{|r|}{\textit{\#5:} Blind} & \multicolumn{4}{l|}{Chunk \#2 ["ução pacífica dos conflitos; VIII – repú ... o à informação e resguardado o"]} & 800 & 0.5808 \\
\hline
Poly-Vector + Flat & 0.5429 & 0.6723 & 0.5878 & 0.0494 & 5933 & 5 \\
\multicolumn{1}{|r|}{\textit{\#1:} ART} & \multicolumn{4}{l|}{\textbf{CRFB, Art. 231.}} & 631 & 0.6723 \\
\multicolumn{1}{|r|}{\textit{\#2:} ART} & \multicolumn{4}{l|}{CRFB, Art. 232.} & 70 & 0.5805 \\
\multicolumn{1}{|r|}{\textit{\#3:} ART} & \multicolumn{4}{l|}{CRFB, Art. 5º} & 3791 & 0.5753 \\
\multicolumn{1}{|r|}{\textit{\#4:} ART} & \multicolumn{4}{l|}{CRFB, Art. 7º} & 1297 & 0.5680 \\
\multicolumn{1}{|r|}{\textit{\#5:} LBL} & \multicolumn{4}{l|}{CRFB, TÍTULO IV, CAPÍTULO I, Seção VIII, Subseção I} & 144 & 0.5429 \\
\hline
Poly-Vector + Multi-layer & 0.5584 & 0.6882 & 0.5989 & 0.0521 & 7934 & 5 \\
\multicolumn{1}{|r|}{\textit{\#1:} CAP} & \multicolumn{4}{l|}{\textbf{CRFB, TÍTULO VIII, CAPÍTULO VIII}} & 698 & 0.6882 \\
\multicolumn{1}{|r|}{\textit{\#2:} INC} & \multicolumn{4}{l|}{CRFB, Art. 20., caput, Inciso XI} & 29 & 0.5980 \\
\multicolumn{1}{|r|}{\textit{\#3:} CAP} & \multicolumn{4}{l|}{CRFB, TÍTULO II, CAPÍTULO I} & 3815 & 0.5819 \\
\multicolumn{1}{|r|}{\textit{\#4:} ART} & \multicolumn{4}{l|}{CRFB, Art. 7º} & 1297 & 0.5678 \\
\multicolumn{1}{|r|}{\textit{\#5:} CAP} & \multicolumn{4}{l|}{CRFB, TÍTULO II, CAPÍTULO II} & 2095 & 0.5584 \\
\hline
Poly-Vector + Multi-layer + Q. Norm. & 0.5964 & 0.7343 & 0.6218 & 0.0505 & 5548 & 7 \\
\multicolumn{1}{|r|}{\textit{\#1:} CAP} & \multicolumn{4}{l|}{\textbf{CRFB, TÍTULO VIII, CAPÍTULO VIII}} & 698 & \textbf{0.7343} \\
\multicolumn{1}{|r|}{\textit{\#2:} INC} & \multicolumn{4}{l|}{CRFB, Art. 20., caput, Inciso XI} & 29 & 0.6226 \\
\multicolumn{1}{|r|}{\textit{\#3:} LBL} & \multicolumn{4}{l|}{CRFB, Art. 51.} & 239 & 0.6070 \\
\multicolumn{1}{|r|}{\textit{\#4:} LBL} & \multicolumn{4}{l|}{CRFB, TÍTULO IV, CAPÍTULO I, Seção VIII, Subseção I} & 144 & 0.5983 \\
\multicolumn{1}{|r|}{\textit{\#5:} PAR} & \multicolumn{4}{l|}{CRFB, Art. 5º, § 2º} & 50 & 0.5974 \\
\multicolumn{1}{|r|}{\textit{\#6:} INC} & \multicolumn{4}{l|}{CRFB, Art. 22., caput, Inciso XIV} & 29 & 0.5969 \\
\multicolumn{1}{|r|}{\textit{\#7:} I,L} & \multicolumn{4}{l|}{urn:lex:br:federal:constituicao:1988-10-05;1988!tit7, CRFB, TÍTULO VII} & 4359 & 0.5964 \\
\hline
\end{longtable}
\clearpage

\textbf{Question 3} \\
\textbf{Original Question:} Quais direitos são assegurados pelo art. 5º da Constituição? \\
\textbf{Normalized Question:} Direitos assegurados pelo art. 5º da Constituição \\
\textbf{Expected Top-1 Segment according to the chunking strategy:} 
\begin{itemize}
\item \textbf{Blind:} Chunk \#2  
\item \textbf{Flat:} Constituição da República Federativa do Brasil, Art. 5º  
\item \textbf{Multi-layer:} Constituição da República Federativa do Brasil, Art. 5º 
\item \textbf{Poly-Vector:}  \begin{enumerate} \item (Label:) Constituição da República Federativa do Brasil, Art. 5º   \item (URN+Label:) urn:lex:br:federal:constituicao:1988-10-05;1988!art5, Constituição da República Federativa do Brasil, Art. 5º   \item  (URN:) urn:lex:br:federal:constituicao:1988-10-05;1988!art5 
\end{enumerate} \end{itemize}
\begin{longtable}{|l|r|r|r|r|r|r|}
\caption{Statistics for Question 3} \label{tab:question_3} \\
\hline
\multicolumn{7}{|c|}{\textbf{Statistics for Question 3}} \\
\hline
\textbf{Method} & \textbf{Min Sim.} & \textbf{Max Sim.} & \textbf{Mean Sim.} & \textbf{Std Dev.} & \textbf{Tokens} & \textbf{Segments} \\
\hline
\endfirsthead
\multicolumn{7}{c}%
{\tablename\ \thetable\ -- \textit{Continued from previous page}} \\
\hline
\textbf{Method} & \textbf{Min Sim.} & \textbf{Max Sim.} & \textbf{Mean Sim.} & \textbf{Std Dev.} & \textbf{Tokens} & \textbf{Segments} \\
\hline
\endhead
\hline \multicolumn{7}{|r|}{{Continued on next page}} \\ \hline
\endfoot
\hline
\endlastfoot
Blind Segmentation Baseline & 0.6853 & 0.7155 & 0.7065 & 0.0124 & 4000 & 5 \\
\multicolumn{1}{|r|}{\textit{\#1:} Blind} & \multicolumn{4}{l|}{\textbf{Chunk \#2} ["ução pacífica dos conflitos; VIII – repú ... o à informação e resguardado o"]} & 800 & 0.7155 \\
\multicolumn{1}{|r|}{\textit{\#2:} Blind} & \multicolumn{4}{l|}{Chunk \#3 ["ais de culto e a suas liturgias; VII – é ... e perigo público, a autoridade"]} & 800 & 0.7155 \\
\multicolumn{1}{|r|}{\textit{\#3:} Blind} & \multicolumn{4}{l|}{Chunk \#1 ["CONSTITUIÇÃO DA REPÚBLICA FEDERATIVA DO  ... rma da lei, a proteção aos loc"]} & 800 & 0.7095 \\
\multicolumn{1}{|r|}{\textit{\#4:} Blind} & \multicolumn{4}{l|}{Chunk \#5 [" competente poderá usar de propriedade p ... a prática do racismo constitui"]} & 800 & 0.7067 \\
\multicolumn{1}{|r|}{\textit{\#5:} Blind} & \multicolumn{4}{l|}{Chunk \#4 [" sigilo da fonte, quando necessário ao e ... do promoverá, na forma da lei,"]} & 800 & 0.6853 \\
\hline
Flat Per-Article Baseline & 0.5555 & 0.7859 & 0.6198 & 0.0939 & 9305 & 5 \\
\multicolumn{1}{|r|}{\textit{\#1:} ART} & \multicolumn{4}{l|}{\textbf{CRFB, Art. 5º}} & 3791 & 0.7859 \\
\multicolumn{1}{|r|}{\textit{\#2:} ART} & \multicolumn{4}{l|}{CRFB, Art. 37.} & 3036 & 0.5939 \\
\multicolumn{1}{|r|}{\textit{\#3:} ART} & \multicolumn{4}{l|}{CRFB, Art. 7º} & 1297 & 0.5836 \\
\multicolumn{1}{|r|}{\textit{\#4:} ART} & \multicolumn{4}{l|}{CRFB, Art. 9º} & 129 & 0.5800 \\
\multicolumn{1}{|r|}{\textit{\#5:} ART} & \multicolumn{4}{l|}{CRFB, Art. 14.} & 1052 & 0.5555 \\
\hline
Multi-layer Hierarchical Embeddings & 0.5935 & 0.7859 & 0.6694 & 0.1023 & 14312 & 5 \\
\multicolumn{1}{|r|}{\textit{\#1:} ART} & \multicolumn{4}{l|}{\textbf{CRFB, Art. 5º}} & 3791 & 0.7859 \\
\multicolumn{1}{|r|}{\textit{\#2:} CAP} & \multicolumn{4}{l|}{CRFB, TÍTULO II, CAPÍTULO I} & 3815 & 0.7770 \\
\multicolumn{1}{|r|}{\textit{\#3:} CPT} & \multicolumn{4}{l|}{CRFB, Art. 37., caput} & 1575 & 0.5967 \\
\multicolumn{1}{|r|}{\textit{\#4:} ART} & \multicolumn{4}{l|}{CRFB, Art. 37.} & 3036 & 0.5939 \\
\multicolumn{1}{|r|}{\textit{\#5:} CAP} & \multicolumn{4}{l|}{CRFB, TÍTULO II, CAPÍTULO II} & 2095 & 0.5935 \\
\hline
Multi-layer + Query Normalization & 0.6140 & 0.7338 & 0.6615 & 0.0613 & 9264 & 5 \\
\multicolumn{1}{|r|}{\textit{\#1:} ART} & \multicolumn{4}{l|}{\textbf{CRFB, Art. 5º}} & 3791 & 0.7338 \\
\multicolumn{1}{|r|}{\textit{\#2:} CAP} & \multicolumn{4}{l|}{CRFB, TÍTULO II, CAPÍTULO I} & 3815 & 0.7231 \\
\multicolumn{1}{|r|}{\textit{\#3:} INC} & \multicolumn{4}{l|}{CRFB, Art. 60., § 4º, Inciso IV} & 41 & 0.6212 \\
\multicolumn{1}{|r|}{\textit{\#4:} CPT} & \multicolumn{4}{l|}{CRFB, Art. 37., caput} & 1575 & 0.6155 \\
\multicolumn{1}{|r|}{\textit{\#5:} PAR} & \multicolumn{4}{l|}{CRFB, Art. 170., Parágrafo único.} & 42 & 0.6140 \\
\hline
Poly-Vector + Blind & 0.7137 & 0.7450 & 0.7215 & 0.0132 & 8788 & 5 \\
\multicolumn{1}{|r|}{\textit{\#1:} LBL} & \multicolumn{4}{l|}{\textbf{CRFB, Art. 5º}} & 3791 & 0.7450 \\
\multicolumn{1}{|r|}{\textit{\#2:} I,L} & \multicolumn{4}{l|}{urn:lex:br:federal:constituicao:1988-10-05;1988!art5\_cpt\_inc1, CRFB, Art. 5º, caput, Inciso I} & 100 & 0.7176 \\
\multicolumn{1}{|r|}{\textit{\#3:} Blind} & \multicolumn{4}{l|}{\textbf{Chunk \#2} ["ução pacífica dos conflitos; VIII – repú ... o à informação e resguardado o"]} & 800 & 0.7155 \\
\multicolumn{1}{|r|}{\textit{\#4:} Blind} & \multicolumn{4}{l|}{Chunk \#3 ["ais de culto e a suas liturgias; VII – é ... e perigo público, a autoridade"]} & 800 & 0.7155 \\
\multicolumn{1}{|r|}{\textit{\#5:} I,L} & \multicolumn{4}{l|}{urn:lex:br:federal:constituicao:1988-10-05;1988!art5\_cpt, CRFB, Art. 5º, caput} & 3297 & 0.7137 \\
\hline
Poly-Vector + Flat & 0.6638 & 0.7859 & 0.7129 & 0.0465 & 7236 & 5 \\
\multicolumn{1}{|r|}{\textit{\#1:} ART} & \multicolumn{4}{l|}{\textbf{CRFB, Art. 5º}} & 3791 & 0.7859 \\
\multicolumn{1}{|r|}{\textit{\#2:} I,L} & \multicolumn{4}{l|}{urn:lex:br:federal:constituicao:1988-10-05;1988!art5\_cpt\_inc1, CRFB, Art. 5º, caput, Inciso I} & 100 & 0.7176 \\
\multicolumn{1}{|r|}{\textit{\#3:} I,L} & \multicolumn{4}{l|}{urn:lex:br:federal:constituicao:1988-10-05;1988!art5\_cpt, CRFB, Art. 5º, caput} & 3297 & 0.7137 \\
\multicolumn{1}{|r|}{\textit{\#4:} I,L} & \multicolumn{4}{l|}{urn:lex:br:federal:constituicao:1988-10-05;1988!art5\_par1, CRFB, Art. 5º, § 1º} & 21 & 0.6833 \\
\multicolumn{1}{|r|}{\textit{\#5:} URN} & \multicolumn{4}{l|}{urn:lex:br:federal:constituicao:1988-10-05;1988!art5\_par4} & 27 & 0.6638 \\
\hline
Poly-Vector + Multi-layer & 0.6833 & 0.7859 & 0.7317 & 0.0471 & 7824 & 5 \\
\multicolumn{1}{|r|}{\textit{\#1:} ART} & \multicolumn{4}{l|}{\textbf{CRFB, Art. 5º}} & 3791 & 0.7859 \\
\multicolumn{1}{|r|}{\textit{\#2:} CAP} & \multicolumn{4}{l|}{CRFB, TÍTULO II, CAPÍTULO I} & 3815 & 0.7770 \\
\multicolumn{1}{|r|}{\textit{\#3:} I,L} & \multicolumn{4}{l|}{urn:lex:br:federal:constituicao:1988-10-05;1988!art5\_cpt\_inc1, CRFB, Art. 5º, caput, Inciso I} & 100 & 0.7176 \\
\multicolumn{1}{|r|}{\textit{\#4:} I,L} & \multicolumn{4}{l|}{urn:lex:br:federal:constituicao:1988-10-05;1988!art5\_cpt\_inc20, CRFB, Art. 5º, caput, Inciso XX} & 97 & 0.6949 \\
\multicolumn{1}{|r|}{\textit{\#5:} I,L} & \multicolumn{4}{l|}{urn:lex:br:federal:constituicao:1988-10-05;1988!art5\_par1, CRFB, Art. 5º, § 1º} & 21 & 0.6833 \\
\hline
Poly-Vector + Multi-layer + Q. Norm. & 0.7102 & 0.8144 & 0.7349 & 0.0447 & 8289 & 5 \\
\multicolumn{1}{|r|}{\textit{\#1:} LBL} & \multicolumn{4}{l|}{\textbf{CRFB, Art. 5º}} & 3791 & \textbf{0.8144} \\
\multicolumn{1}{|r|}{\textit{\#2:} CAP} & \multicolumn{4}{l|}{CRFB, TÍTULO II, CAPÍTULO I} & 3815 & 0.7231 \\
\multicolumn{1}{|r|}{\textit{\#3:} LBL} & \multicolumn{4}{l|}{CRFB, Art. 6º} & 170 & 0.7137 \\
\multicolumn{1}{|r|}{\textit{\#4:} LBL} & \multicolumn{4}{l|}{CRFB, Art. 51.} & 239 & 0.7130 \\
\multicolumn{1}{|r|}{\textit{\#5:} LBL} & \multicolumn{4}{l|}{CRFB, Art. 31.} & 274 & 0.7102 \\
\hline
\end{longtable}
\clearpage

\textbf{Question 4} \\
\textbf{Original Question:} Quais são os direitos previstos no art. 7º da Constituição? \\
\textbf{Normalized Question:} Direitos previstos no art. 7º da Constituição \\
\textbf{Expected Top-1 Segment according to the chunking strategy:} 
\begin{itemize}
\item \textbf{Blind:} Chunk \#12  
\item \textbf{Flat:} Constituição da República Federativa do Brasil, Art. 7º  
\item \textbf{Multi-layer:} Constituição da República Federativa do Brasil, Art. 7º 
\item \textbf{Poly-Vector:}  \begin{enumerate} \item (Label:) Constituição da República Federativa do Brasil, Art. 7º   \item (URN+Label:) urn:lex:br:federal:constituicao:1988-10-05;1988!art7, Constituição da República Federativa do Brasil, Art. 7º   \item  (URN:) urn:lex:br:federal:constituicao:1988-10-05;1988!art7 
\end{enumerate} \end{itemize}
\begin{longtable}{|l|r|r|r|r|r|r|}
\caption{Statistics for Question 4} \label{tab:question_4} \\
\hline
\multicolumn{7}{|c|}{\textbf{Statistics for Question 4}} \\
\hline
\textbf{Method} & \textbf{Min Sim.} & \textbf{Max Sim.} & \textbf{Mean Sim.} & \textbf{Std Dev.} & \textbf{Tokens} & \textbf{Segments} \\
\hline
\endfirsthead
\multicolumn{7}{c}%
{\tablename\ \thetable\ -- \textit{Continued from previous page}} \\
\hline
\textbf{Method} & \textbf{Min Sim.} & \textbf{Max Sim.} & \textbf{Mean Sim.} & \textbf{Std Dev.} & \textbf{Tokens} & \textbf{Segments} \\
\hline
\endhead
\hline \multicolumn{7}{|r|}{{Continued on next page}} \\ \hline
\endfoot
\hline
\endlastfoot
Blind Segmentation Baseline & 0.6893 & 0.7374 & 0.7133 & 0.0171 & 4000 & 5 \\
\multicolumn{1}{|r|}{\textit{\#1:} Blind} & \multicolumn{4}{l|}{Chunk \#11 [" pobres, na forma da lei: a) o registro  ... ho noturno superior à do diurn"]} & 800 & 0.7374 \\
\multicolumn{1}{|r|}{\textit{\#2:} Blind} & \multicolumn{4}{l|}{Chunk \#13 ["o; X – proteção do salário na forma da l ... te e o trabalhador avulso. Par"]} & 800 & 0.7159 \\
\multicolumn{1}{|r|}{\textit{\#3:} Blind} & \multicolumn{4}{l|}{Chunk \#15 ["ágrafo único. São assegurados à categori ... s o entendimento direto com os"]} & 800 & 0.7128 \\
\multicolumn{1}{|r|}{\textit{\#4:} Blind} & \multicolumn{4}{l|}{\textbf{Chunk \#12} ["arados, na forma desta Constituição. Par ... , higiene e segurança; XXIII –"]} & 800 & 0.7110 \\
\multicolumn{1}{|r|}{\textit{\#5:} Blind} & \multicolumn{4}{l|}{Chunk \#10 [" público; LXX – o mandado de segurança c ... ncia, a assistência aos desamp"]} & 800 & 0.6893 \\
\hline
Flat Per-Article Baseline & 0.5925 & 0.7912 & 0.6426 & 0.0840 & 5842 & 5 \\
\multicolumn{1}{|r|}{\textit{\#1:} ART} & \multicolumn{4}{l|}{\textbf{CRFB, Art. 7º}} & 1297 & 0.7912 \\
\multicolumn{1}{|r|}{\textit{\#2:} ART} & \multicolumn{4}{l|}{CRFB, Art. 5º} & 3791 & 0.6244 \\
\multicolumn{1}{|r|}{\textit{\#3:} ART} & \multicolumn{4}{l|}{CRFB, Art. 8º} & 455 & 0.6064 \\
\multicolumn{1}{|r|}{\textit{\#4:} ART} & \multicolumn{4}{l|}{CRFB, Art. 6º} & 170 & 0.5983 \\
\multicolumn{1}{|r|}{\textit{\#5:} ART} & \multicolumn{4}{l|}{CRFB, Art. 9º} & 129 & 0.5925 \\
\hline
Multi-layer Hierarchical Embeddings & 0.6272 & 0.7912 & 0.6855 & 0.0683 & 7424 & 5 \\
\multicolumn{1}{|r|}{\textit{\#1:} ART} & \multicolumn{4}{l|}{\textbf{CRFB, Art. 7º}} & 1297 & 0.7912 \\
\multicolumn{1}{|r|}{\textit{\#2:} CAP} & \multicolumn{4}{l|}{CRFB, TÍTULO II, CAPÍTULO II} & 2095 & 0.7152 \\
\multicolumn{1}{|r|}{\textit{\#3:} PAR} & \multicolumn{4}{l|}{CRFB, Art. 39., § 3º} & 81 & 0.6581 \\
\multicolumn{1}{|r|}{\textit{\#4:} CAP} & \multicolumn{4}{l|}{CRFB, TÍTULO II, CAPÍTULO I} & 3815 & 0.6360 \\
\multicolumn{1}{|r|}{\textit{\#5:} INC} & \multicolumn{4}{l|}{CRFB, Art. 142., § 3º, Inciso VIII} & 136 & 0.6272 \\
\hline
Multi-layer + Query Normalization & 0.6489 & 0.7463 & 0.6848 & 0.0405 & 5147 & 5 \\
\multicolumn{1}{|r|}{\textit{\#1:} ART} & \multicolumn{4}{l|}{\textbf{CRFB, Art. 7º}} & 1297 & 0.7463 \\
\multicolumn{1}{|r|}{\textit{\#2:} CAP} & \multicolumn{4}{l|}{CRFB, TÍTULO II, CAPÍTULO II} & 2095 & 0.7045 \\
\multicolumn{1}{|r|}{\textit{\#3:} PAR} & \multicolumn{4}{l|}{CRFB, Art. 39., § 3º} & 81 & 0.6685 \\
\multicolumn{1}{|r|}{\textit{\#4:} INC} & \multicolumn{4}{l|}{CRFB, Art. 37., caput, Inciso VII} & 99 & 0.6556 \\
\multicolumn{1}{|r|}{\textit{\#5:} CPT} & \multicolumn{4}{l|}{CRFB, Art. 37., caput} & 1575 & 0.6489 \\
\hline
Poly-Vector + Blind & 0.7159 & 0.7588 & 0.7356 & 0.0130 & 4217 & 7 \\
\multicolumn{1}{|r|}{\textit{\#1:} LBL} & \multicolumn{4}{l|}{\textbf{CRFB, Art. 7º}} & 1297 & 0.7588 \\
\multicolumn{1}{|r|}{\textit{\#2:} I,L} & \multicolumn{4}{l|}{urn:lex:br:federal:constituicao:1988-10-05;1988!art7\_cpt\_inc7, CRFB, Art. 7º, caput, Inciso VII} & 61 & 0.7419 \\
\multicolumn{1}{|r|}{\textit{\#3:} Blind} & \multicolumn{4}{l|}{Chunk \#11 [" pobres, na forma da lei: a) o registro  ... ho noturno superior à do diurn"]} & 800 & 0.7374 \\
\multicolumn{1}{|r|}{\textit{\#4:} I,L} & \multicolumn{4}{l|}{urn:lex:br:federal:constituicao:1988-10-05;1988!art7\_cpt\_inc1, CRFB, Art. 7º, caput, Inciso I} & 83 & 0.7336 \\
\multicolumn{1}{|r|}{\textit{\#5:} I,L} & \multicolumn{4}{l|}{urn:lex:br:federal:constituicao:1988-10-05;1988!art7\_cpt, CRFB, Art. 7º, caput} & 1019 & 0.7319 \\
\multicolumn{1}{|r|}{\textit{\#6:} I,L} & \multicolumn{4}{l|}{urn:lex:br:federal:constituicao:1988-10-05;1988!art7\_par1u, CRFB, Art. 7º, Parágrafo único.} & 157 & 0.7298 \\
\multicolumn{1}{|r|}{\textit{\#7:} Blind} & \multicolumn{4}{l|}{Chunk \#13 ["o; X – proteção do salário na forma da l ... te e o trabalhador avulso. Par"]} & 800 & 0.7159 \\
\hline
Poly-Vector + Flat & 0.6395 & 0.7912 & 0.7181 & 0.0520 & 2710 & 7 \\
\multicolumn{1}{|r|}{\textit{\#1:} ART} & \multicolumn{4}{l|}{\textbf{CRFB, Art. 7º}} & 1297 & 0.7912 \\
\multicolumn{1}{|r|}{\textit{\#2:} I,L} & \multicolumn{4}{l|}{urn:lex:br:federal:constituicao:1988-10-05;1988!art7\_cpt\_inc7, CRFB, Art. 7º, caput, Inciso VII} & 61 & 0.7419 \\
\multicolumn{1}{|r|}{\textit{\#3:} I,L} & \multicolumn{4}{l|}{urn:lex:br:federal:constituicao:1988-10-05;1988!art7\_cpt\_inc1, CRFB, Art. 7º, caput, Inciso I} & 83 & 0.7336 \\
\multicolumn{1}{|r|}{\textit{\#4:} I,L} & \multicolumn{4}{l|}{urn:lex:br:federal:constituicao:1988-10-05;1988!art7\_cpt, CRFB, Art. 7º, caput} & 1019 & 0.7319 \\
\multicolumn{1}{|r|}{\textit{\#5:} I,L} & \multicolumn{4}{l|}{urn:lex:br:federal:constituicao:1988-10-05;1988!art7\_par1u, CRFB, Art. 7º, Parágrafo único.} & 157 & 0.7298 \\
\multicolumn{1}{|r|}{\textit{\#6:} I,L} & \multicolumn{4}{l|}{urn:lex:br:federal:constituicao:1988-10-05;1988!art8\_cpt\_inc7, CRFB, Art. 8º, caput, Inciso VII} & 51 & 0.6585 \\
\multicolumn{1}{|r|}{\textit{\#7:} I,L} & \multicolumn{4}{l|}{urn:lex:br:federal:constituicao:1988-10-05;1988!art4\_cpt\_inc7, CRFB, Art. 4º, caput, Inciso VII} & 42 & 0.6395 \\
\hline
Poly-Vector + Multi-layer & 0.7152 & 0.7912 & 0.7406 & 0.0263 & 4712 & 6 \\
\multicolumn{1}{|r|}{\textit{\#1:} ART} & \multicolumn{4}{l|}{\textbf{CRFB, Art. 7º}} & 1297 & 0.7912 \\
\multicolumn{1}{|r|}{\textit{\#2:} I,L} & \multicolumn{4}{l|}{urn:lex:br:federal:constituicao:1988-10-05;1988!art7\_cpt\_inc7, CRFB, Art. 7º, caput, Inciso VII} & 61 & 0.7419 \\
\multicolumn{1}{|r|}{\textit{\#3:} I,L} & \multicolumn{4}{l|}{urn:lex:br:federal:constituicao:1988-10-05;1988!art7\_cpt\_inc1, CRFB, Art. 7º, caput, Inciso I} & 83 & 0.7336 \\
\multicolumn{1}{|r|}{\textit{\#4:} I,L} & \multicolumn{4}{l|}{urn:lex:br:federal:constituicao:1988-10-05;1988!art7\_cpt, CRFB, Art. 7º, caput} & 1019 & 0.7319 \\
\multicolumn{1}{|r|}{\textit{\#5:} I,L} & \multicolumn{4}{l|}{urn:lex:br:federal:constituicao:1988-10-05;1988!art7\_par1u, CRFB, Art. 7º, Parágrafo único.} & 157 & 0.7298 \\
\multicolumn{1}{|r|}{\textit{\#6:} CAP} & \multicolumn{4}{l|}{CRFB, TÍTULO II, CAPÍTULO II} & 2095 & 0.7152 \\
\hline
Poly-Vector + Multi-layer + Q. Norm. & 0.6948 & 0.8248 & 0.7303 & 0.0476 & 4011 & 6 \\
\multicolumn{1}{|r|}{\textit{\#1:} LBL} & \multicolumn{4}{l|}{\textbf{CRFB, Art. 7º}} & 1297 & \textbf{0.8248} \\
\multicolumn{1}{|r|}{\textit{\#2:} I,L} & \multicolumn{4}{l|}{urn:lex:br:federal:constituicao:1988-10-05;1988!art7\_cpt\_inc18, CRFB, Art. 7º, caput, Inciso XVIII} & 71 & 0.7280 \\
\multicolumn{1}{|r|}{\textit{\#3:} LBL} & \multicolumn{4}{l|}{CRFB, Art. 8º, caput, Inciso VII} & 51 & 0.7148 \\
\multicolumn{1}{|r|}{\textit{\#4:} LBL} & \multicolumn{4}{l|}{CRFB, Art. 8º} & 455 & 0.7146 \\
\multicolumn{1}{|r|}{\textit{\#5:} CAP} & \multicolumn{4}{l|}{CRFB, TÍTULO II, CAPÍTULO II} & 2095 & 0.7045 \\
\multicolumn{1}{|r|}{\textit{\#6:} LBL} & \multicolumn{4}{l|}{CRFB, Art. 4º, caput, Inciso VII} & 42 & 0.6948 \\
\hline
\end{longtable}
\clearpage

\textbf{Question 5} \\
\textbf{Original Question:} Qual o tema do Capítulo VI do Título VIII da Constituição? \\
\textbf{Normalized Question:} Tema do Capítulo VI do Título VIII da Constituição \\
\textbf{Expected Top-1 Segment according to the chunking strategy:} 
\begin{itemize}
\item \textbf{Blind:} Chunk \#272  
\item \textbf{Flat:} Constituição da República Federativa do Brasil, Art. 272.  
\item \textbf{Multi-layer:} Constituição da República Federativa do Brasil, TÍTULO VIII, CAPÍTULO VI 
\item \textbf{Poly-Vector:}  \begin{enumerate} \item (Label:) Constituição da República Federativa do Brasil, TÍTULO VIII, CAPÍTULO VI   \item (URN+Label:) urn:lex:br:federal:constituicao:1988-10-05;1988!tit8\_cap6, Constituição da República Federativa do Brasil, TÍTULO VIII, CAPÍTULO VI   \item  (URN:) urn:lex:br:federal:constituicao:1988-10-05;1988!tit8\_cap6 
\end{enumerate} \end{itemize}
\begin{longtable}{|l|r|r|r|r|r|r|}
\caption{Statistics for Question 5} \label{tab:question_5} \\
\hline
\multicolumn{7}{|c|}{\textbf{Statistics for Question 5}} \\
\hline
\textbf{Method} & \textbf{Min Sim.} & \textbf{Max Sim.} & \textbf{Mean Sim.} & \textbf{Std Dev.} & \textbf{Tokens} & \textbf{Segments} \\
\hline
\endfirsthead
\multicolumn{7}{c}%
{\tablename\ \thetable\ -- \textit{Continued from previous page}} \\
\hline
\textbf{Method} & \textbf{Min Sim.} & \textbf{Max Sim.} & \textbf{Mean Sim.} & \textbf{Std Dev.} & \textbf{Tokens} & \textbf{Segments} \\
\hline
\endhead
\hline \multicolumn{7}{|r|}{{Continued on next page}} \\ \hline
\endfoot
\hline
\endlastfoot
Blind Segmentation Baseline & 0.7120 & 0.7226 & 0.7165 & 0.0044 & 4000 & 5 \\
\multicolumn{1}{|r|}{\textit{\#1:} Blind} & \multicolumn{4}{l|}{Chunk \#47 ["ções e serviços públicos de saúde; IV –  ...  cargo ou emprego, na carreira"]} & 800 & 0.7226 \\
\multicolumn{1}{|r|}{\textit{\#2:} Blind} & \multicolumn{4}{l|}{Chunk \#2 ["ução pacífica dos conflitos; VIII – repú ... o à informação e resguardado o"]} & 800 & 0.7191 \\
\multicolumn{1}{|r|}{\textit{\#3:} Blind} & \multicolumn{4}{l|}{Chunk \#279 [" idoneidade e saber jurídico, com dez an ... e 3 de dezembro de 1970, passa"]} & 800 & 0.7155 \\
\multicolumn{1}{|r|}{\textit{\#4:} Blind} & \multicolumn{4}{l|}{Chunk \#99 [" delegar as atribuições mencionadas nos  ... sa Nacional Subseção I Do Cons"]} & 800 & 0.7131 \\
\multicolumn{1}{|r|}{\textit{\#5:} Blind} & \multicolumn{4}{l|}{Chunk \#48 ["ia Legislativa, o decreto limitar-se-á a ... ou não, incluídas as vantagens"]} & 800 & 0.7120 \\
\hline
Flat Per-Article Baseline & 0.6418 & 0.6914 & 0.6672 & 0.0185 & 6905 & 5 \\
\multicolumn{1}{|r|}{\textit{\#1:} ART} & \multicolumn{4}{l|}{CRFB, Art. 37.} & 3036 & 0.6914 \\
\multicolumn{1}{|r|}{\textit{\#2:} ART} & \multicolumn{4}{l|}{CRFB, Art. 24.} & 481 & 0.6735 \\
\multicolumn{1}{|r|}{\textit{\#3:} ART} & \multicolumn{4}{l|}{CRFB, Art. 21.} & 1101 & 0.6712 \\
\multicolumn{1}{|r|}{\textit{\#4:} ART} & \multicolumn{4}{l|}{CRFB, Art. 167.} & 1566 & 0.6583 \\
\multicolumn{1}{|r|}{\textit{\#5:} ART} & \multicolumn{4}{l|}{CRFB, Art. 22.} & 721 & 0.6418 \\
\hline
Multi-layer Hierarchical Embeddings & 0.6794 & 0.7107 & 0.6956 & 0.0125 & 16182 & 5 \\
\multicolumn{1}{|r|}{\textit{\#1:} CAP} & \multicolumn{4}{l|}{CRFB, TÍTULO III, CAPÍTULO II} & 3271 & 0.7107 \\
\multicolumn{1}{|r|}{\textit{\#2:} TIT} & \multicolumn{4}{l|}{CRFB, TÍTULO V} & 3605 & 0.7054 \\
\multicolumn{1}{|r|}{\textit{\#3:} ART} & \multicolumn{4}{l|}{CRFB, Art. 37.} & 3036 & 0.6914 \\
\multicolumn{1}{|r|}{\textit{\#4:} TIT} & \multicolumn{4}{l|}{CRFB, TÍTULO IX} & 2957 & 0.6908 \\
\multicolumn{1}{|r|}{\textit{\#5:} CAP} & \multicolumn{4}{l|}{CRFB, TÍTULO IV, CAPÍTULO II} & 3313 & 0.6794 \\
\hline
Multi-layer + Query Normalization & 0.6364 & 0.6730 & 0.6531 & 0.0159 & 9554 & 5 \\
\multicolumn{1}{|r|}{\textit{\#1:} CAP} & \multicolumn{4}{l|}{CRFB, TÍTULO III, CAPÍTULO II} & 3271 & 0.6730 \\
\multicolumn{1}{|r|}{\textit{\#2:} TIT} & \multicolumn{4}{l|}{CRFB, TÍTULO V} & 3605 & 0.6669 \\
\multicolumn{1}{|r|}{\textit{\#3:} ART} & \multicolumn{4}{l|}{CRFB, Art. 237.} & 67 & 0.6450 \\
\multicolumn{1}{|r|}{\textit{\#4:} CAP} & \multicolumn{4}{l|}{CRFB, TÍTULO VII, CAPÍTULO I} & 2519 & 0.6440 \\
\multicolumn{1}{|r|}{\textit{\#5:} INC} & \multicolumn{4}{l|}{CRFB, Art. 48., caput, Inciso V} & 92 & 0.6364 \\
\hline
Poly-Vector + Blind & 0.7563 & 0.7675 & 0.7597 & 0.0046 & 4620 & 5 \\
\multicolumn{1}{|r|}{\textit{\#1:} LBL} & \multicolumn{4}{l|}{\textbf{CRFB, TÍTULO VIII, CAPÍTULO VI}} & 981 & 0.7675 \\
\multicolumn{1}{|r|}{\textit{\#2:} LBL} & \multicolumn{4}{l|}{CRFB, TÍTULO III, CAPÍTULO VI} & 931 & 0.7597 \\
\multicolumn{1}{|r|}{\textit{\#3:} LBL} & \multicolumn{4}{l|}{CRFB, TÍTULO VII, CAPÍTULO II} & 552 & 0.7577 \\
\multicolumn{1}{|r|}{\textit{\#4:} LBL} & \multicolumn{4}{l|}{CRFB, TÍTULO VIII, CAPÍTULO VIII} & 698 & 0.7571 \\
\multicolumn{1}{|r|}{\textit{\#5:} LBL} & \multicolumn{4}{l|}{CRFB, TÍTULO VIII, CAPÍTULO VII} & 1458 & 0.7563 \\
\hline
Poly-Vector + Flat & 0.7563 & 0.7675 & 0.7597 & 0.0046 & 4620 & 5 \\
\multicolumn{1}{|r|}{\textit{\#1:} LBL} & \multicolumn{4}{l|}{\textbf{CRFB, TÍTULO VIII, CAPÍTULO VI}} & 981 & 0.7675 \\
\multicolumn{1}{|r|}{\textit{\#2:} LBL} & \multicolumn{4}{l|}{CRFB, TÍTULO III, CAPÍTULO VI} & 931 & 0.7597 \\
\multicolumn{1}{|r|}{\textit{\#3:} LBL} & \multicolumn{4}{l|}{CRFB, TÍTULO VII, CAPÍTULO II} & 552 & 0.7577 \\
\multicolumn{1}{|r|}{\textit{\#4:} LBL} & \multicolumn{4}{l|}{CRFB, TÍTULO VIII, CAPÍTULO VIII} & 698 & 0.7571 \\
\multicolumn{1}{|r|}{\textit{\#5:} LBL} & \multicolumn{4}{l|}{CRFB, TÍTULO VIII, CAPÍTULO VII} & 1458 & 0.7563 \\
\hline
Poly-Vector + Multi-layer & 0.7563 & 0.7675 & 0.7597 & 0.0046 & 4620 & 5 \\
\multicolumn{1}{|r|}{\textit{\#1:} LBL} & \multicolumn{4}{l|}{\textbf{CRFB, TÍTULO VIII, CAPÍTULO VI}} & 981 & 0.7675 \\
\multicolumn{1}{|r|}{\textit{\#2:} LBL} & \multicolumn{4}{l|}{CRFB, TÍTULO III, CAPÍTULO VI} & 931 & 0.7597 \\
\multicolumn{1}{|r|}{\textit{\#3:} LBL} & \multicolumn{4}{l|}{CRFB, TÍTULO VII, CAPÍTULO II} & 552 & 0.7577 \\
\multicolumn{1}{|r|}{\textit{\#4:} LBL} & \multicolumn{4}{l|}{CRFB, TÍTULO VIII, CAPÍTULO VIII} & 698 & 0.7571 \\
\multicolumn{1}{|r|}{\textit{\#5:} LBL} & \multicolumn{4}{l|}{CRFB, TÍTULO VIII, CAPÍTULO VII} & 1458 & 0.7563 \\
\hline
Poly-Vector + Multi-layer + Q. Norm. & 0.7696 & 0.7819 & 0.7734 & 0.0052 & 5298 & 5 \\
\multicolumn{1}{|r|}{\textit{\#1:} LBL} & \multicolumn{4}{l|}{\textbf{CRFB, TÍTULO VIII, CAPÍTULO VI}} & 981 & \textbf{0.7819} \\
\multicolumn{1}{|r|}{\textit{\#2:} LBL} & \multicolumn{4}{l|}{CRFB, TÍTULO III, CAPÍTULO VI} & 931 & 0.7749 \\
\multicolumn{1}{|r|}{\textit{\#3:} LBL} & \multicolumn{4}{l|}{CRFB, TÍTULO VIII, CAPÍTULO V} & 1230 & 0.7710 \\
\multicolumn{1}{|r|}{\textit{\#4:} LBL} & \multicolumn{4}{l|}{CRFB, TÍTULO VIII, CAPÍTULO VIII} & 698 & 0.7696 \\
\multicolumn{1}{|r|}{\textit{\#5:} LBL} & \multicolumn{4}{l|}{CRFB, TÍTULO VIII, CAPÍTULO VII} & 1458 & 0.7696 \\
\hline
\end{longtable}
\clearpage

\textbf{Question 6} \\
\textbf{Original Question:} Explique o art. 69 da Constituição \\
\textbf{Normalized Question:} Art. 69 da Constituição \\
\textbf{Expected Top-1 Segment according to the chunking strategy:} 
\begin{itemize}
\item \textbf{Blind:} Chunk \#89  
\item \textbf{Flat:} Constituição da República Federativa do Brasil, Art. 69.  
\item \textbf{Multi-layer:} Constituição da República Federativa do Brasil, Art. 69. 
\item \textbf{Poly-Vector:}  \begin{enumerate} \item (Label:) Constituição da República Federativa do Brasil, Art. 69.   \item (URN+Label:) urn:lex:br:federal:constituicao:1988-10-05;1988!art69, Constituição da República Federativa do Brasil, Art. 69.   \item  (URN:) urn:lex:br:federal:constituicao:1988-10-05;1988!art69 
\end{enumerate} \end{itemize}
\begin{longtable}{|l|r|r|r|r|r|r|}
\caption{Statistics for Question 6} \label{tab:question_6} \\
\hline
\multicolumn{7}{|c|}{\textbf{Statistics for Question 6}} \\
\hline
\textbf{Method} & \textbf{Min Sim.} & \textbf{Max Sim.} & \textbf{Mean Sim.} & \textbf{Std Dev.} & \textbf{Tokens} & \textbf{Segments} \\
\hline
\endfirsthead
\multicolumn{7}{c}%
{\tablename\ \thetable\ -- \textit{Continued from previous page}} \\
\hline
\textbf{Method} & \textbf{Min Sim.} & \textbf{Max Sim.} & \textbf{Mean Sim.} & \textbf{Std Dev.} & \textbf{Tokens} & \textbf{Segments} \\
\hline
\endhead
\hline \multicolumn{7}{|r|}{{Continued on next page}} \\ \hline
\endfoot
\hline
\endlastfoot
Blind Segmentation Baseline & 0.5906 & 0.6125 & 0.5981 & 0.0086 & 4000 & 5 \\
\multicolumn{1}{|r|}{\textit{\#1:} Blind} & \multicolumn{4}{l|}{Chunk \#88 ["gação, ao Presidente da República. § 6º  ...  e mantidas pelo poder público"]} & 800 & 0.6125 \\
\multicolumn{1}{|r|}{\textit{\#2:} Blind} & \multicolumn{4}{l|}{Chunk \#82 [". § 1º A Constituição não poderá ser eme ... nacionalidade, cidadania, dire"]} & 800 & 0.5993 \\
\multicolumn{1}{|r|}{\textit{\#3:} Blind} & \multicolumn{4}{l|}{Chunk \#20 ["íveis, no território de jurisdição do ti ... undamentais da pessoa humana e"]} & 800 & 0.5942 \\
\multicolumn{1}{|r|}{\textit{\#4:} Blind} & \multicolumn{4}{l|}{Chunk \#71 [" renovação de concessão de emissoras de  ... ional de Justiça e do Conselho"]} & 800 & 0.5939 \\
\multicolumn{1}{|r|}{\textit{\#5:} Blind} & \multicolumn{4}{l|}{\textbf{Chunk \#89} [" seu conteúdo e os termos de seu exercíc ...  e sobre resultados de auditor"]} & 800 & 0.5906 \\
\hline
Flat Per-Article Baseline & 0.5529 & 0.5873 & 0.5697 & 0.0140 & 4123 & 7 \\
\multicolumn{1}{|r|}{\textit{\#1:} ART} & \multicolumn{4}{l|}{CRFB, Art. 167-C.} & 207 & 0.5873 \\
\multicolumn{1}{|r|}{\textit{\#2:} ART} & \multicolumn{4}{l|}{CRFB, Art. 171.} & 46 & 0.5873 \\
\multicolumn{1}{|r|}{\textit{\#3:} ART} & \multicolumn{4}{l|}{CRFB, Art. 50.} & 291 & 0.5755 \\
\multicolumn{1}{|r|}{\textit{\#4:} ART} & \multicolumn{4}{l|}{CRFB, Art. 37.} & 3036 & 0.5659 \\
\multicolumn{1}{|r|}{\textit{\#5:} ART} & \multicolumn{4}{l|}{CRFB, Art. 233.} & 31 & 0.5623 \\
\multicolumn{1}{|r|}{\textit{\#6:} ART} & \multicolumn{4}{l|}{CRFB, Art. 72.} & 235 & 0.5567 \\
\multicolumn{1}{|r|}{\textit{\#7:} ART} & \multicolumn{4}{l|}{CRFB, Art. 68.} & 277 & 0.5529 \\
\hline
Multi-layer Hierarchical Embeddings & 0.5851 & 0.6307 & 0.5968 & 0.0115 & 6741 & 27 \\
\multicolumn{1}{|r|}{\textit{\#1:} INC} & \multicolumn{4}{l|}{CRFB, Art. 49., caput, Inciso XI} & 47 & 0.6307 \\
\multicolumn{1}{|r|}{\textit{\#2:} INC} & \multicolumn{4}{l|}{CRFB, Art. 49., caput, Inciso XIII} & 37 & 0.6183 \\
\multicolumn{1}{|r|}{\textit{\#3:} INC} & \multicolumn{4}{l|}{CRFB, Art. 49., caput, Inciso V} & 52 & 0.6155 \\
\multicolumn{1}{|r|}{\textit{\#4:} INC} & \multicolumn{4}{l|}{CRFB, Art. 49., caput, Inciso XV} & 33 & 0.6108 \\
\multicolumn{1}{|r|}{\textit{\#5:} ALI} & \multicolumn{4}{l|}{CRFB, Art. 52., caput, Inciso III, Alínea a} & 58 & 0.6070 \\
\multicolumn{1}{|r|}{\textit{\#6:} ALI} & \multicolumn{4}{l|}{CRFB, Art. 96., caput, Inciso I, Alínea c} & 49 & 0.6036 \\
\multicolumn{1}{|r|}{\textit{\#7:} INC} & \multicolumn{4}{l|}{CRFB, Art. 23., caput, Inciso I} & 61 & 0.6019 \\
\multicolumn{1}{|r|}{\textit{\#8:} INC} & \multicolumn{4}{l|}{CRFB, Art. 84., caput, Inciso XV} & 48 & 0.6019 \\
\multicolumn{1}{|r|}{\textit{\#9:} INC} & \multicolumn{4}{l|}{CRFB, Art. 49., caput, Inciso X} & 58 & 0.6016 \\
\multicolumn{1}{|r|}{\textit{\#10:} INC} & \multicolumn{4}{l|}{CRFB, Art. 71., caput, Inciso IX} & 80 & 0.5978 \\
\multicolumn{1}{|r|}{\textit{\#11:} ALI} & \multicolumn{4}{l|}{CRFB, Art. 52., caput, Inciso III, Alínea b} & 63 & 0.5966 \\
\multicolumn{1}{|r|}{\textit{\#12:} INC} & \multicolumn{4}{l|}{CRFB, Art. 49., caput, Inciso VIII} & 88 & 0.5950 \\
\multicolumn{1}{|r|}{\textit{\#13:} INC} & \multicolumn{4}{l|}{CRFB, Art. 84., caput, Inciso XXII} & 59 & 0.5928 \\
\multicolumn{1}{|r|}{\textit{\#14:} INC} & \multicolumn{4}{l|}{CRFB, Art. 49., caput, Inciso II} & 80 & 0.5924 \\
\multicolumn{1}{|r|}{\textit{\#15:} INC} & \multicolumn{4}{l|}{CRFB, Art. 5º, caput, Inciso LXIX} & 149 & 0.5921 \\
\multicolumn{1}{|r|}{\textit{\#16:} ALI} & \multicolumn{4}{l|}{CRFB, Art. 102., caput, Inciso I, Alínea h} & 48 & 0.5920 \\
\multicolumn{1}{|r|}{\textit{\#17:} INC} & \multicolumn{4}{l|}{CRFB, Art. 49., caput, Inciso III} & 55 & 0.5907 \\
\multicolumn{1}{|r|}{\textit{\#18:} INC} & \multicolumn{4}{l|}{CRFB, Art. 146., caput, Inciso II} & 29 & 0.5891 \\
\multicolumn{1}{|r|}{\textit{\#19:} INC} & \multicolumn{4}{l|}{CRFB, Art. 51., caput, Inciso V} & 45 & 0.5891 \\
\multicolumn{1}{|r|}{\textit{\#20:} SEC} & \multicolumn{4}{l|}{CRFB, TÍTULO IV, CAPÍTULO I, Seção II} & 1462 & 0.5890 \\
\multicolumn{1}{|r|}{\textit{\#21:} INC} & \multicolumn{4}{l|}{CRFB, Art. 51., caput, Inciso I} & 60 & 0.5876 \\
\multicolumn{1}{|r|}{\textit{\#22:} ART} & \multicolumn{4}{l|}{CRFB, Art. 167-C.} & 207 & 0.5873 \\
\multicolumn{1}{|r|}{\textit{\#23:} ART} & \multicolumn{4}{l|}{CRFB, Art. 171.} & 46 & 0.5873 \\
\multicolumn{1}{|r|}{\textit{\#24:} PAR} & \multicolumn{4}{l|}{CRFB, Art. 163., Parágrafo único.} & 39 & 0.5868 \\
\multicolumn{1}{|r|}{\textit{\#25:} ALI} & \multicolumn{4}{l|}{CRFB, Art. 102., caput, Inciso I, Alínea q} & 131 & 0.5864 \\
\multicolumn{1}{|r|}{\textit{\#26:} INC} & \multicolumn{4}{l|}{CRFB, Art. 52., caput, Inciso X} & 52 & 0.5854 \\
\multicolumn{1}{|r|}{\textit{\#27:} TIT} & \multicolumn{4}{l|}{CRFB, TÍTULO V} & 3605 & 0.5851 \\
\hline
Multi-layer + Query Normalization & 0.5809 & 0.6384 & 0.5984 & 0.0135 & 4068 & 64 \\
\multicolumn{1}{|r|}{\textit{\#1:} INC} & \multicolumn{4}{l|}{CRFB, Art. 49., caput, Inciso V} & 52 & 0.6384 \\
\multicolumn{1}{|r|}{\textit{\#2:} ART} & \multicolumn{4}{l|}{CRFB, Art. 171.} & 46 & 0.6338 \\
\multicolumn{1}{|r|}{\textit{\#3:} ALI} & \multicolumn{4}{l|}{CRFB, Art. 102., caput, Inciso I, Alínea h} & 48 & 0.6246 \\
\multicolumn{1}{|r|}{\textit{\#4:} INC} & \multicolumn{4}{l|}{CRFB, Art. 71., caput, Inciso IX} & 80 & 0.6233 \\
\multicolumn{1}{|r|}{\textit{\#5:} INC} & \multicolumn{4}{l|}{CRFB, Art. 146., caput, Inciso II} & 29 & 0.6209 \\
\multicolumn{1}{|r|}{\textit{\#6:} INC} & \multicolumn{4}{l|}{CRFB, Art. 167-A., § 5º, Inciso II} & 68 & 0.6198 \\
\multicolumn{1}{|r|}{\textit{\#7:} INC} & \multicolumn{4}{l|}{CRFB, Art. 49., caput, Inciso XI} & 47 & 0.6171 \\
\multicolumn{1}{|r|}{\textit{\#8:} INC} & \multicolumn{4}{l|}{CRFB, Art. 49., caput, Inciso IV} & 51 & 0.6133 \\
\multicolumn{1}{|r|}{\textit{\#9:} INC} & \multicolumn{4}{l|}{CRFB, Art. 60., § 4º, Inciso I} & 39 & 0.6125 \\
\multicolumn{1}{|r|}{\textit{\#10:} CPT} & \multicolumn{4}{l|}{CRFB, Art. 117., caput} & 5 & 0.6122 \\
\multicolumn{1}{|r|}{\textit{\#11:} INC} & \multicolumn{4}{l|}{CRFB, Art. 150., caput, Inciso I} & 64 & 0.6119 \\
\multicolumn{1}{|r|}{\textit{\#12:} INC} & \multicolumn{4}{l|}{CRFB, Art. 160., § 1º, Inciso II} & 64 & 0.6113 \\
\multicolumn{1}{|r|}{\textit{\#13:} INC} & \multicolumn{4}{l|}{CRFB, Art. 167-A., § 5º, Inciso I} & 50 & 0.6100 \\
\multicolumn{1}{|r|}{\textit{\#14:} ART} & \multicolumn{4}{l|}{CRFB, Art. 167-C.} & 207 & 0.6077 \\
\multicolumn{1}{|r|}{\textit{\#15:} INC} & \multicolumn{4}{l|}{CRFB, Art. 60., § 4º, Inciso IV} & 41 & 0.6077 \\
\multicolumn{1}{|r|}{\textit{\#16:} ART} & \multicolumn{4}{l|}{CRFB, Art. 233.} & 31 & 0.6075 \\
\multicolumn{1}{|r|}{\textit{\#17:} ALI} & \multicolumn{4}{l|}{CRFB, Art. 166., § 3º, Inciso III, Alínea a} & 65 & 0.6061 \\
\multicolumn{1}{|r|}{\textit{\#18:} INC} & \multicolumn{4}{l|}{CRFB, Art. 49., caput, Inciso XIII} & 37 & 0.6046 \\
\multicolumn{1}{|r|}{\textit{\#19:} INC} & \multicolumn{4}{l|}{CRFB, Art. 160., § 1º, Inciso I} & 54 & 0.6039 \\
\multicolumn{1}{|r|}{\textit{\#20:} INC} & \multicolumn{4}{l|}{CRFB, Art. 5º, caput, Inciso LXIX} & 149 & 0.6035 \\
\multicolumn{1}{|r|}{\textit{\#21:} INC} & \multicolumn{4}{l|}{CRFB, Art. 49., caput, Inciso II} & 80 & 0.6025 \\
\multicolumn{1}{|r|}{\textit{\#22:} PAR} & \multicolumn{4}{l|}{CRFB, Art. 163., Parágrafo único.} & 39 & 0.6022 \\
\multicolumn{1}{|r|}{\textit{\#23:} INC} & \multicolumn{4}{l|}{CRFB, Art. 23., caput, Inciso I} & 61 & 0.6019 \\
\multicolumn{1}{|r|}{\textit{\#24:} ALI} & \multicolumn{4}{l|}{CRFB, Art. 102., caput, Inciso I, Alínea p} & 63 & 0.6016 \\
\multicolumn{1}{|r|}{\textit{\#25:} INC} & \multicolumn{4}{l|}{CRFB, Art. 59., caput, Inciso I} & 28 & 0.6014 \\
\multicolumn{1}{|r|}{\textit{\#26:} INC} & \multicolumn{4}{l|}{CRFB, Art. 68., § 1º, Inciso II} & 88 & 0.6003 \\
\multicolumn{1}{|r|}{\textit{\#27:} INC} & \multicolumn{4}{l|}{CRFB, Art. 48., caput, Inciso I} & 81 & 0.6001 \\
\multicolumn{1}{|r|}{\textit{\#28:} PAR} & \multicolumn{4}{l|}{CRFB, Art. 36., § 3º} & 78 & 0.5996 \\
\multicolumn{1}{|r|}{\textit{\#29:} PAR} & \multicolumn{4}{l|}{CRFB, Art. 160., § 1º} & 66 & 0.5988 \\
\multicolumn{1}{|r|}{\textit{\#30:} ALI} & \multicolumn{4}{l|}{CRFB, Art. 102., caput, Inciso I, Alínea q} & 131 & 0.5983 \\
\multicolumn{1}{|r|}{\textit{\#31:} INC} & \multicolumn{4}{l|}{CRFB, Art. 48., caput, Inciso XIV} & 87 & 0.5981 \\
\multicolumn{1}{|r|}{\textit{\#32:} INC} & \multicolumn{4}{l|}{CRFB, Art. 156-B., § 2º, Inciso VII} & 57 & 0.5965 \\
\multicolumn{1}{|r|}{\textit{\#33:} INC} & \multicolumn{4}{l|}{CRFB, Art. 166., § 3º, Inciso III} & 78 & 0.5956 \\
\multicolumn{1}{|r|}{\textit{\#34:} INC} & \multicolumn{4}{l|}{CRFB, Art. 49., caput, Inciso XII} & 46 & 0.5950 \\
\multicolumn{1}{|r|}{\textit{\#35:} PAR} & \multicolumn{4}{l|}{CRFB, Art. 167-A., § 2º} & 38 & 0.5945 \\
\multicolumn{1}{|r|}{\textit{\#36:} ALI} & \multicolumn{4}{l|}{CRFB, Art. 62., § 1º, Inciso I, Alínea a} & 58 & 0.5943 \\
\multicolumn{1}{|r|}{\textit{\#37:} PAR} & \multicolumn{4}{l|}{CRFB, Art. 149-C., § 3º} & 82 & 0.5942 \\
\multicolumn{1}{|r|}{\textit{\#38:} INC} & \multicolumn{4}{l|}{CRFB, Art. 49., caput, Inciso X} & 58 & 0.5940 \\
\multicolumn{1}{|r|}{\textit{\#39:} INC} & \multicolumn{4}{l|}{CRFB, Art. 165., § 11., Inciso I} & 92 & 0.5934 \\
\multicolumn{1}{|r|}{\textit{\#40:} INC} & \multicolumn{4}{l|}{CRFB, Art. 156-B., § 2º, Inciso VI} & 81 & 0.5930 \\
\multicolumn{1}{|r|}{\textit{\#41:} INC} & \multicolumn{4}{l|}{CRFB, Art. 49., caput, Inciso XV} & 33 & 0.5928 \\
\multicolumn{1}{|r|}{\textit{\#42:} PAR} & \multicolumn{4}{l|}{CRFB, Art. 165., § 16.} & 49 & 0.5921 \\
\multicolumn{1}{|r|}{\textit{\#43:} INC} & \multicolumn{4}{l|}{CRFB, Art. 60., § 4º, Inciso III} & 40 & 0.5918 \\
\multicolumn{1}{|r|}{\textit{\#44:} INC} & \multicolumn{4}{l|}{CRFB, Art. 48., caput, Inciso VII} & 76 & 0.5899 \\
\multicolumn{1}{|r|}{\textit{\#45:} INC} & \multicolumn{4}{l|}{CRFB, Art. 68., § 1º, Inciso III} & 88 & 0.5885 \\
\multicolumn{1}{|r|}{\textit{\#46:} INC} & \multicolumn{4}{l|}{CRFB, Art. 21., caput, Inciso VII} & 20 & 0.5883 \\
\multicolumn{1}{|r|}{\textit{\#47:} INC} & \multicolumn{4}{l|}{CRFB, Art. 5º, caput, Inciso LXXIII} & 174 & 0.5883 \\
\multicolumn{1}{|r|}{\textit{\#48:} INC} & \multicolumn{4}{l|}{CRFB, Art. 167., caput, Inciso VIII} & 74 & 0.5875 \\
\multicolumn{1}{|r|}{\textit{\#49:} INC} & \multicolumn{4}{l|}{CRFB, Art. 24., caput, Inciso V} & 34 & 0.5872 \\
\multicolumn{1}{|r|}{\textit{\#50:} INC} & \multicolumn{4}{l|}{CRFB, Art. 21., caput, Inciso VI} & 33 & 0.5865 \\
\multicolumn{1}{|r|}{\textit{\#51:} INC} & \multicolumn{4}{l|}{CRFB, Art. 48., caput, Inciso VIII} & 72 & 0.5863 \\
\multicolumn{1}{|r|}{\textit{\#52:} INC} & \multicolumn{4}{l|}{CRFB, Art. 59., caput, Inciso V} & 28 & 0.5857 \\
\multicolumn{1}{|r|}{\textit{\#53:} INC} & \multicolumn{4}{l|}{CRFB, Art. 161., caput, Inciso I} & 40 & 0.5851 \\
\multicolumn{1}{|r|}{\textit{\#54:} INC} & \multicolumn{4}{l|}{CRFB, Art. 103., caput, Inciso VI} & 47 & 0.5840 \\
\multicolumn{1}{|r|}{\textit{\#55:} ALI} & \multicolumn{4}{l|}{CRFB, Art. 102., caput, Inciso II, Alínea b} & 47 & 0.5834 \\
\multicolumn{1}{|r|}{\textit{\#56:} INC} & \multicolumn{4}{l|}{CRFB, Art. 167., caput, Inciso VI} & 55 & 0.5833 \\
\multicolumn{1}{|r|}{\textit{\#57:} CPT} & \multicolumn{4}{l|}{CRFB, Art. 175., caput} & 40 & 0.5830 \\
\multicolumn{1}{|r|}{\textit{\#58:} INC} & \multicolumn{4}{l|}{CRFB, Art. 48., caput, Inciso V} & 92 & 0.5820 \\
\multicolumn{1}{|r|}{\textit{\#59:} INC} & \multicolumn{4}{l|}{CRFB, Art. 48., caput, Inciso XII} & 76 & 0.5819 \\
\multicolumn{1}{|r|}{\textit{\#60:} INC} & \multicolumn{4}{l|}{CRFB, Art. 102., caput, Inciso II} & 95 & 0.5819 \\
\multicolumn{1}{|r|}{\textit{\#61:} INC} & \multicolumn{4}{l|}{CRFB, Art. 163., caput, Inciso VII} & 58 & 0.5817 \\
\multicolumn{1}{|r|}{\textit{\#62:} INC} & \multicolumn{4}{l|}{CRFB, Art. 62., § 1º, Inciso II} & 56 & 0.5814 \\
\multicolumn{1}{|r|}{\textit{\#63:} INC} & \multicolumn{4}{l|}{CRFB, Art. 177., § 2º, Inciso I} & 51 & 0.5809 \\
\multicolumn{1}{|r|}{\textit{\#64:} ALI} & \multicolumn{4}{l|}{CRFB, Art. 102., caput, Inciso III, Alínea a} & 71 & 0.5809 \\
\hline
Poly-Vector + Blind & 0.7019 & 0.7982 & 0.7188 & 0.0235 & 4128 & 17 \\
\multicolumn{1}{|r|}{\textit{\#1:} LBL} & \multicolumn{4}{l|}{\textbf{CRFB, Art. 69.}} & 53 & 0.7982 \\
\multicolumn{1}{|r|}{\textit{\#2:} LBL} & \multicolumn{4}{l|}{CRFB, Art. 71.} & 898 & 0.7413 \\
\multicolumn{1}{|r|}{\textit{\#3:} LBL} & \multicolumn{4}{l|}{CRFB, Art. 68.} & 277 & 0.7335 \\
\multicolumn{1}{|r|}{\textit{\#4:} LBL} & \multicolumn{4}{l|}{CRFB, Art. 67.} & 91 & 0.7291 \\
\multicolumn{1}{|r|}{\textit{\#5:} I,L} & \multicolumn{4}{l|}{urn:lex:br:federal:constituicao:1988-10-05;1988!art5\_cpt\_inc69, CRFB, Art. 5º, caput, Inciso LXIX} & 149 & 0.7213 \\
\multicolumn{1}{|r|}{\textit{\#6:} LBL} & \multicolumn{4}{l|}{CRFB, Art. 70.} & 211 & 0.7191 \\
\multicolumn{1}{|r|}{\textit{\#7:} LBL} & \multicolumn{4}{l|}{CRFB, Art. 5º, caput, Inciso LXVIII} & 133 & 0.7141 \\
\multicolumn{1}{|r|}{\textit{\#8:} LBL} & \multicolumn{4}{l|}{CRFB, Art. 75.} & 141 & 0.7134 \\
\multicolumn{1}{|r|}{\textit{\#9:} LBL} & \multicolumn{4}{l|}{CRFB, Art. 66.} & 456 & 0.7097 \\
\multicolumn{1}{|r|}{\textit{\#10:} LBL} & \multicolumn{4}{l|}{CRFB, Art. 74.} & 334 & 0.7079 \\
\multicolumn{1}{|r|}{\textit{\#11:} LBL} & \multicolumn{4}{l|}{CRFB, Art. 78.} & 168 & 0.7071 \\
\multicolumn{1}{|r|}{\textit{\#12:} LBL} & \multicolumn{4}{l|}{CRFB, Art. 79.} & 121 & 0.7049 \\
\multicolumn{1}{|r|}{\textit{\#13:} LBL} & \multicolumn{4}{l|}{CRFB, Art. 47.} & 77 & 0.7048 \\
\multicolumn{1}{|r|}{\textit{\#14:} LBL} & \multicolumn{4}{l|}{CRFB, Art. 72.} & 235 & 0.7045 \\
\multicolumn{1}{|r|}{\textit{\#15:} I,L} & \multicolumn{4}{l|}{urn:lex:br:federal:constituicao:1988-10-05;1988!art65\_cpt, CRFB, Art. 65., caput} & 58 & 0.7045 \\
\multicolumn{1}{|r|}{\textit{\#16:} LBL} & \multicolumn{4}{l|}{CRFB, Art. 48.} & 512 & 0.7037 \\
\multicolumn{1}{|r|}{\textit{\#17:} LBL} & \multicolumn{4}{l|}{CRFB, Art. 119.} & 214 & 0.7019 \\
\hline
Poly-Vector + Flat & 0.7019 & 0.7982 & 0.7188 & 0.0235 & 4128 & 17 \\
\multicolumn{1}{|r|}{\textit{\#1:} LBL} & \multicolumn{4}{l|}{\textbf{CRFB, Art. 69.}} & 53 & 0.7982 \\
\multicolumn{1}{|r|}{\textit{\#2:} LBL} & \multicolumn{4}{l|}{CRFB, Art. 71.} & 898 & 0.7413 \\
\multicolumn{1}{|r|}{\textit{\#3:} LBL} & \multicolumn{4}{l|}{CRFB, Art. 68.} & 277 & 0.7335 \\
\multicolumn{1}{|r|}{\textit{\#4:} LBL} & \multicolumn{4}{l|}{CRFB, Art. 67.} & 91 & 0.7291 \\
\multicolumn{1}{|r|}{\textit{\#5:} I,L} & \multicolumn{4}{l|}{urn:lex:br:federal:constituicao:1988-10-05;1988!art5\_cpt\_inc69, CRFB, Art. 5º, caput, Inciso LXIX} & 149 & 0.7213 \\
\multicolumn{1}{|r|}{\textit{\#6:} LBL} & \multicolumn{4}{l|}{CRFB, Art. 70.} & 211 & 0.7191 \\
\multicolumn{1}{|r|}{\textit{\#7:} LBL} & \multicolumn{4}{l|}{CRFB, Art. 5º, caput, Inciso LXVIII} & 133 & 0.7141 \\
\multicolumn{1}{|r|}{\textit{\#8:} LBL} & \multicolumn{4}{l|}{CRFB, Art. 75.} & 141 & 0.7134 \\
\multicolumn{1}{|r|}{\textit{\#9:} LBL} & \multicolumn{4}{l|}{CRFB, Art. 66.} & 456 & 0.7097 \\
\multicolumn{1}{|r|}{\textit{\#10:} LBL} & \multicolumn{4}{l|}{CRFB, Art. 74.} & 334 & 0.7079 \\
\multicolumn{1}{|r|}{\textit{\#11:} LBL} & \multicolumn{4}{l|}{CRFB, Art. 78.} & 168 & 0.7071 \\
\multicolumn{1}{|r|}{\textit{\#12:} LBL} & \multicolumn{4}{l|}{CRFB, Art. 79.} & 121 & 0.7049 \\
\multicolumn{1}{|r|}{\textit{\#13:} LBL} & \multicolumn{4}{l|}{CRFB, Art. 47.} & 77 & 0.7048 \\
\multicolumn{1}{|r|}{\textit{\#14:} LBL} & \multicolumn{4}{l|}{CRFB, Art. 72.} & 235 & 0.7045 \\
\multicolumn{1}{|r|}{\textit{\#15:} I,L} & \multicolumn{4}{l|}{urn:lex:br:federal:constituicao:1988-10-05;1988!art65\_cpt, CRFB, Art. 65., caput} & 58 & 0.7045 \\
\multicolumn{1}{|r|}{\textit{\#16:} LBL} & \multicolumn{4}{l|}{CRFB, Art. 48.} & 512 & 0.7037 \\
\multicolumn{1}{|r|}{\textit{\#17:} LBL} & \multicolumn{4}{l|}{CRFB, Art. 119.} & 214 & 0.7019 \\
\hline
Poly-Vector + Multi-layer & 0.7019 & 0.7982 & 0.7188 & 0.0235 & 4128 & 17 \\
\multicolumn{1}{|r|}{\textit{\#1:} LBL} & \multicolumn{4}{l|}{\textbf{CRFB, Art. 69.}} & 53 & 0.7982 \\
\multicolumn{1}{|r|}{\textit{\#2:} LBL} & \multicolumn{4}{l|}{CRFB, Art. 71.} & 898 & 0.7413 \\
\multicolumn{1}{|r|}{\textit{\#3:} LBL} & \multicolumn{4}{l|}{CRFB, Art. 68.} & 277 & 0.7335 \\
\multicolumn{1}{|r|}{\textit{\#4:} LBL} & \multicolumn{4}{l|}{CRFB, Art. 67.} & 91 & 0.7291 \\
\multicolumn{1}{|r|}{\textit{\#5:} I,L} & \multicolumn{4}{l|}{urn:lex:br:federal:constituicao:1988-10-05;1988!art5\_cpt\_inc69, CRFB, Art. 5º, caput, Inciso LXIX} & 149 & 0.7213 \\
\multicolumn{1}{|r|}{\textit{\#6:} LBL} & \multicolumn{4}{l|}{CRFB, Art. 70.} & 211 & 0.7191 \\
\multicolumn{1}{|r|}{\textit{\#7:} LBL} & \multicolumn{4}{l|}{CRFB, Art. 5º, caput, Inciso LXVIII} & 133 & 0.7141 \\
\multicolumn{1}{|r|}{\textit{\#8:} LBL} & \multicolumn{4}{l|}{CRFB, Art. 75.} & 141 & 0.7134 \\
\multicolumn{1}{|r|}{\textit{\#9:} LBL} & \multicolumn{4}{l|}{CRFB, Art. 66.} & 456 & 0.7097 \\
\multicolumn{1}{|r|}{\textit{\#10:} LBL} & \multicolumn{4}{l|}{CRFB, Art. 74.} & 334 & 0.7079 \\
\multicolumn{1}{|r|}{\textit{\#11:} LBL} & \multicolumn{4}{l|}{CRFB, Art. 78.} & 168 & 0.7071 \\
\multicolumn{1}{|r|}{\textit{\#12:} LBL} & \multicolumn{4}{l|}{CRFB, Art. 79.} & 121 & 0.7049 \\
\multicolumn{1}{|r|}{\textit{\#13:} LBL} & \multicolumn{4}{l|}{CRFB, Art. 47.} & 77 & 0.7048 \\
\multicolumn{1}{|r|}{\textit{\#14:} LBL} & \multicolumn{4}{l|}{CRFB, Art. 72.} & 235 & 0.7045 \\
\multicolumn{1}{|r|}{\textit{\#15:} I,L} & \multicolumn{4}{l|}{urn:lex:br:federal:constituicao:1988-10-05;1988!art65\_cpt, CRFB, Art. 65., caput} & 58 & 0.7045 \\
\multicolumn{1}{|r|}{\textit{\#16:} LBL} & \multicolumn{4}{l|}{CRFB, Art. 48.} & 512 & 0.7037 \\
\multicolumn{1}{|r|}{\textit{\#17:} LBL} & \multicolumn{4}{l|}{CRFB, Art. 119.} & 214 & 0.7019 \\
\hline
Poly-Vector + Multi-layer + Q. Norm. & 0.7523 & 0.8349 & 0.7670 & 0.0220 & 4047 & 15 \\
\multicolumn{1}{|r|}{\textit{\#1:} LBL} & \multicolumn{4}{l|}{\textbf{CRFB, Art. 69.}} & 53 & \textbf{0.8349} \\
\multicolumn{1}{|r|}{\textit{\#2:} LBL} & \multicolumn{4}{l|}{CRFB, Art. 70.} & 211 & 0.7855 \\
\multicolumn{1}{|r|}{\textit{\#3:} LBL} & \multicolumn{4}{l|}{CRFB, Art. 68.} & 277 & 0.7825 \\
\multicolumn{1}{|r|}{\textit{\#4:} LBL} & \multicolumn{4}{l|}{CRFB, Art. 67.} & 91 & 0.7808 \\
\multicolumn{1}{|r|}{\textit{\#5:} LBL} & \multicolumn{4}{l|}{CRFB, Art. 90.} & 167 & 0.7650 \\
\multicolumn{1}{|r|}{\textit{\#6:} LBL} & \multicolumn{4}{l|}{CRFB, Art. 65., caput} & 58 & 0.7605 \\
\multicolumn{1}{|r|}{\textit{\#7:} LBL} & \multicolumn{4}{l|}{CRFB, Art. 91.} & 388 & 0.7603 \\
\multicolumn{1}{|r|}{\textit{\#8:} LBL} & \multicolumn{4}{l|}{CRFB, Art. 71.} & 898 & 0.7572 \\
\multicolumn{1}{|r|}{\textit{\#9:} LBL} & \multicolumn{4}{l|}{CRFB, Art. 59., caput, Inciso VI} & 29 & 0.7569 \\
\multicolumn{1}{|r|}{\textit{\#10:} LBL} & \multicolumn{4}{l|}{CRFB, Art. 5º, caput, Inciso LXIX} & 149 & 0.7554 \\
\multicolumn{1}{|r|}{\textit{\#11:} LBL} & \multicolumn{4}{l|}{CRFB, Art. 75.} & 141 & 0.7534 \\
\multicolumn{1}{|r|}{\textit{\#12:} LBL} & \multicolumn{4}{l|}{CRFB, Art. 105.} & 1306 & 0.7532 \\
\multicolumn{1}{|r|}{\textit{\#13:} LBL} & \multicolumn{4}{l|}{CRFB, Art. 59., caput, Inciso VII} & 26 & 0.7532 \\
\multicolumn{1}{|r|}{\textit{\#14:} LBL} & \multicolumn{4}{l|}{CRFB, Art. 59.} & 136 & 0.7532 \\
\multicolumn{1}{|r|}{\textit{\#15:} LBL} & \multicolumn{4}{l|}{CRFB, Art. 65.} & 117 & 0.7523 \\
\hline
\end{longtable}
\clearpage

\textbf{Question 7} \\
\textbf{Original Question:} Explique a norma urn:lex:br:federal:constituicao:1988-10-05;1988!art69 \\
\textbf{Normalized Question:} Norma urn:lex:br:federal:constituicao:1988-10-05;1988!art69 \\
\textbf{Expected Top-1 Segment according to the chunking strategy:} 
\begin{itemize}
\item \textbf{Blind:} Chunk \#89  
\item \textbf{Flat:} Constituição da República Federativa do Brasil, Art. 69.  
\item \textbf{Multi-layer:} Constituição da República Federativa do Brasil, Art. 69. 
\item \textbf{Poly-Vector:}  \begin{enumerate} \item (Label:) Constituição da República Federativa do Brasil, Art. 69.   \item (URN+Label:) urn:lex:br:federal:constituicao:1988-10-05;1988!art69, Constituição da República Federativa do Brasil, Art. 69.   \item  (URN:) urn:lex:br:federal:constituicao:1988-10-05;1988!art69 
\end{enumerate} \end{itemize}
\begin{longtable}{|l|r|r|r|r|r|r|}
\caption{Statistics for Question 7} \label{tab:question_7} \\
\hline
\multicolumn{7}{|c|}{\textbf{Statistics for Question 7}} \\
\hline
\textbf{Method} & \textbf{Min Sim.} & \textbf{Max Sim.} & \textbf{Mean Sim.} & \textbf{Std Dev.} & \textbf{Tokens} & \textbf{Segments} \\
\hline
\endfirsthead
\multicolumn{7}{c}%
{\tablename\ \thetable\ -- \textit{Continued from previous page}} \\
\hline
\textbf{Method} & \textbf{Min Sim.} & \textbf{Max Sim.} & \textbf{Mean Sim.} & \textbf{Std Dev.} & \textbf{Tokens} & \textbf{Segments} \\
\hline
\endhead
\hline \multicolumn{7}{|r|}{{Continued on next page}} \\ \hline
\endfoot
\hline
\endlastfoot
Blind Segmentation Baseline & 0.6026 & 0.6232 & 0.6104 & 0.0087 & 4000 & 5 \\
\multicolumn{1}{|r|}{\textit{\#1:} Blind} & \multicolumn{4}{l|}{Chunk \#82 [". § 1º A Constituição não poderá ser eme ... nacionalidade, cidadania, dire"]} & 800 & 0.6232 \\
\multicolumn{1}{|r|}{\textit{\#2:} Blind} & \multicolumn{4}{l|}{Chunk \#48 ["ia Legislativa, o decreto limitar-se-á a ... ou não, incluídas as vantagens"]} & 800 & 0.6152 \\
\multicolumn{1}{|r|}{\textit{\#3:} Blind} & \multicolumn{4}{l|}{Chunk \#47 ["ções e serviços públicos de saúde; IV –  ...  cargo ou emprego, na carreira"]} & 800 & 0.6075 \\
\multicolumn{1}{|r|}{\textit{\#4:} Blind} & \multicolumn{4}{l|}{Chunk \#18 ["e da Câmara dos Deputados; III – de Pres ... tes do pleito. § 7º São ineleg"]} & 800 & 0.6037 \\
\multicolumn{1}{|r|}{\textit{\#5:} Blind} & \multicolumn{4}{l|}{Chunk \#68 [" para que nenhuma daquelas unidades da F ... as estrangeiras transitem pelo"]} & 800 & 0.6026 \\
\hline
Flat Per-Article Baseline & 0.5695 & 0.6087 & 0.5799 & 0.0113 & 4280 & 13 \\
\multicolumn{1}{|r|}{\textit{\#1:} ART} & \multicolumn{4}{l|}{CRFB, Art. 167-C.} & 207 & 0.6087 \\
\multicolumn{1}{|r|}{\textit{\#2:} ART} & \multicolumn{4}{l|}{CRFB, Art. 68.} & 277 & 0.5925 \\
\multicolumn{1}{|r|}{\textit{\#3:} ART} & \multicolumn{4}{l|}{CRFB, Art. 121.} & 395 & 0.5883 \\
\multicolumn{1}{|r|}{\textit{\#4:} ART} & \multicolumn{4}{l|}{CRFB, Art. 247.} & 164 & 0.5838 \\
\multicolumn{1}{|r|}{\textit{\#5:} ART} & \multicolumn{4}{l|}{CRFB, Art. 75.} & 141 & 0.5827 \\
\multicolumn{1}{|r|}{\textit{\#6:} ART} & \multicolumn{4}{l|}{CRFB, Art. 223.} & 274 & 0.5780 \\
\multicolumn{1}{|r|}{\textit{\#7:} ART} & \multicolumn{4}{l|}{CRFB, Art. 14.} & 1052 & 0.5759 \\
\multicolumn{1}{|r|}{\textit{\#8:} ART} & \multicolumn{4}{l|}{CRFB, Art. 167-B.} & 153 & 0.5740 \\
\multicolumn{1}{|r|}{\textit{\#9:} ART} & \multicolumn{4}{l|}{CRFB, Art. 70.} & 211 & 0.5720 \\
\multicolumn{1}{|r|}{\textit{\#10:} ART} & \multicolumn{4}{l|}{CRFB, Art. 1º} & 148 & 0.5719 \\
\multicolumn{1}{|r|}{\textit{\#11:} ART} & \multicolumn{4}{l|}{CRFB, Art. 171.} & 46 & 0.5710 \\
\multicolumn{1}{|r|}{\textit{\#12:} ART} & \multicolumn{4}{l|}{CRFB, Art. 85.} & 211 & 0.5703 \\
\multicolumn{1}{|r|}{\textit{\#13:} ART} & \multicolumn{4}{l|}{CRFB, Art. 62.} & 1001 & 0.5695 \\
\hline
Multi-layer Hierarchical Embeddings & 0.5868 & 0.6231 & 0.5985 & 0.0092 & 5453 & 42 \\
\multicolumn{1}{|r|}{\textit{\#1:} INC} & \multicolumn{4}{l|}{CRFB, Art. 68., § 1º, Inciso II} & 88 & 0.6231 \\
\multicolumn{1}{|r|}{\textit{\#2:} INC} & \multicolumn{4}{l|}{CRFB, Art. 23., caput, Inciso I} & 61 & 0.6157 \\
\multicolumn{1}{|r|}{\textit{\#3:} INC} & \multicolumn{4}{l|}{CRFB, Art. 5º, caput, Inciso LXXIII} & 174 & 0.6127 \\
\multicolumn{1}{|r|}{\textit{\#4:} ALI} & \multicolumn{4}{l|}{CRFB, Art. 52., caput, Inciso III, Alínea a} & 58 & 0.6119 \\
\multicolumn{1}{|r|}{\textit{\#5:} INC} & \multicolumn{4}{l|}{CRFB, Art. 68., § 1º, Inciso I} & 98 & 0.6115 \\
\multicolumn{1}{|r|}{\textit{\#6:} INC} & \multicolumn{4}{l|}{CRFB, Art. 1º, caput, Inciso I} & 60 & 0.6115 \\
\multicolumn{1}{|r|}{\textit{\#7:} INC} & \multicolumn{4}{l|}{CRFB, Art. 51., caput, Inciso V} & 45 & 0.6114 \\
\multicolumn{1}{|r|}{\textit{\#8:} ALI} & \multicolumn{4}{l|}{CRFB, Art. 128., § 5º, Inciso II, Alínea e} & 96 & 0.6102 \\
\multicolumn{1}{|r|}{\textit{\#9:} ART} & \multicolumn{4}{l|}{CRFB, Art. 167-C.} & 207 & 0.6087 \\
\multicolumn{1}{|r|}{\textit{\#10:} INC} & \multicolumn{4}{l|}{CRFB, Art. 5º, caput, Inciso LXIX} & 149 & 0.6081 \\
\multicolumn{1}{|r|}{\textit{\#11:} INC} & \multicolumn{4}{l|}{CRFB, Art. 49., caput, Inciso XI} & 47 & 0.6071 \\
\multicolumn{1}{|r|}{\textit{\#12:} INC} & \multicolumn{4}{l|}{CRFB, Art. 49., caput, Inciso XIII} & 37 & 0.6048 \\
\multicolumn{1}{|r|}{\textit{\#13:} INC} & \multicolumn{4}{l|}{CRFB, Art. 84., caput, Inciso XV} & 48 & 0.6006 \\
\multicolumn{1}{|r|}{\textit{\#14:} INC} & \multicolumn{4}{l|}{CRFB, Art. 1º, caput, Inciso V} & 62 & 0.5998 \\
\multicolumn{1}{|r|}{\textit{\#15:} INC} & \multicolumn{4}{l|}{CRFB, Art. 71., caput, Inciso VIII} & 100 & 0.5989 \\
\multicolumn{1}{|r|}{\textit{\#16:} INC} & \multicolumn{4}{l|}{CRFB, Art. 71., caput, Inciso IX} & 80 & 0.5988 \\
\multicolumn{1}{|r|}{\textit{\#17:} ALI} & \multicolumn{4}{l|}{CRFB, Art. 96., caput, Inciso I, Alínea c} & 49 & 0.5974 \\
\multicolumn{1}{|r|}{\textit{\#18:} INC} & \multicolumn{4}{l|}{CRFB, Art. 52., caput, Inciso XIV} & 41 & 0.5974 \\
\multicolumn{1}{|r|}{\textit{\#19:} INC} & \multicolumn{4}{l|}{CRFB, Art. 1º, caput, Inciso II} & 61 & 0.5974 \\
\multicolumn{1}{|r|}{\textit{\#20:} ALI} & \multicolumn{4}{l|}{CRFB, Art. 52., caput, Inciso III, Alínea b} & 63 & 0.5971 \\
\multicolumn{1}{|r|}{\textit{\#21:} ALI} & \multicolumn{4}{l|}{CRFB, Art. 5º, caput, Inciso LXX, Alínea a} & 108 & 0.5963 \\
\multicolumn{1}{|r|}{\textit{\#22:} ALI} & \multicolumn{4}{l|}{CRFB, Art. 34., caput, Inciso VII, Alínea d} & 64 & 0.5957 \\
\multicolumn{1}{|r|}{\textit{\#23:} INC} & \multicolumn{4}{l|}{CRFB, Art. 84., caput, Inciso XXVII} & 35 & 0.5955 \\
\multicolumn{1}{|r|}{\textit{\#24:} INC} & \multicolumn{4}{l|}{CRFB, Art. 14., caput, Inciso I} & 53 & 0.5947 \\
\multicolumn{1}{|r|}{\textit{\#25:} INC} & \multicolumn{4}{l|}{CRFB, Art. 29., caput, Inciso XIV} & 117 & 0.5940 \\
\multicolumn{1}{|r|}{\textit{\#26:} INC} & \multicolumn{4}{l|}{CRFB, Art. 71., caput, Inciso XI} & 60 & 0.5935 \\
\multicolumn{1}{|r|}{\textit{\#27:} INC} & \multicolumn{4}{l|}{CRFB, Art. 48., caput, Inciso X} & 105 & 0.5934 \\
\multicolumn{1}{|r|}{\textit{\#28:} ALI} & \multicolumn{4}{l|}{CRFB, Art. 52., caput, Inciso III, Alínea f} & 56 & 0.5926 \\
\multicolumn{1}{|r|}{\textit{\#29:} ART} & \multicolumn{4}{l|}{CRFB, Art. 68.} & 277 & 0.5925 \\
\multicolumn{1}{|r|}{\textit{\#30:} INC} & \multicolumn{4}{l|}{CRFB, Art. 5º, caput, Inciso LXXI} & 140 & 0.5925 \\
\multicolumn{1}{|r|}{\textit{\#31:} INC} & \multicolumn{4}{l|}{CRFB, Art. 51., caput, Inciso I} & 60 & 0.5924 \\
\multicolumn{1}{|r|}{\textit{\#32:} ALI} & \multicolumn{4}{l|}{CRFB, Art. 128., § 5º, Inciso II, Alínea b} & 92 & 0.5921 \\
\multicolumn{1}{|r|}{\textit{\#33:} INC} & \multicolumn{4}{l|}{CRFB, Art. 5º, caput, Inciso LX} & 110 & 0.5906 \\
\multicolumn{1}{|r|}{\textit{\#34:} INC} & \multicolumn{4}{l|}{CRFB, Art. 14., caput, Inciso III} & 52 & 0.5901 \\
\multicolumn{1}{|r|}{\textit{\#35:} ALI} & \multicolumn{4}{l|}{CRFB, Art. 128., § 5º, Inciso II, Alínea d} & 109 & 0.5899 \\
\multicolumn{1}{|r|}{\textit{\#36:} INC} & \multicolumn{4}{l|}{CRFB, Art. 85., caput, Inciso II} & 75 & 0.5898 \\
\multicolumn{1}{|r|}{\textit{\#37:} ALI} & \multicolumn{4}{l|}{CRFB, Art. 34., caput, Inciso VII, Alínea a} & 58 & 0.5887 \\
\multicolumn{1}{|r|}{\textit{\#38:} INC} & \multicolumn{4}{l|}{CRFB, Art. 71., caput, Inciso II} & 131 & 0.5886 \\
\multicolumn{1}{|r|}{\textit{\#39:} INC} & \multicolumn{4}{l|}{CRFB, Art. 37., caput, Inciso VII} & 99 & 0.5886 \\
\multicolumn{1}{|r|}{\textit{\#40:} ART} & \multicolumn{4}{l|}{CRFB, Art. 121.} & 395 & 0.5883 \\
\multicolumn{1}{|r|}{\textit{\#41:} INC} & \multicolumn{4}{l|}{CRFB, Art. 49., caput, Inciso X} & 58 & 0.5869 \\
\multicolumn{1}{|r|}{\textit{\#42:} CPT} & \multicolumn{4}{l|}{CRFB, Art. 37., caput} & 1575 & 0.5868 \\
\hline
Multi-layer + Query Normalization & 0.6076 & 0.6546 & 0.6195 & 0.0119 & 4228 & 20 \\
\multicolumn{1}{|r|}{\textit{\#1:} INC} & \multicolumn{4}{l|}{CRFB, Art. 68., § 1º, Inciso II} & 88 & 0.6546 \\
\multicolumn{1}{|r|}{\textit{\#2:} ART} & \multicolumn{4}{l|}{CRFB, Art. 167-C.} & 207 & 0.6345 \\
\multicolumn{1}{|r|}{\textit{\#3:} INC} & \multicolumn{4}{l|}{CRFB, Art. 5º, caput, Inciso LXXIII} & 174 & 0.6340 \\
\multicolumn{1}{|r|}{\textit{\#4:} INC} & \multicolumn{4}{l|}{CRFB, Art. 167-A., caput, Inciso V} & 166 & 0.6299 \\
\multicolumn{1}{|r|}{\textit{\#5:} CPT} & \multicolumn{4}{l|}{CRFB, Art. 37., caput} & 1575 & 0.6280 \\
\multicolumn{1}{|r|}{\textit{\#6:} INC} & \multicolumn{4}{l|}{CRFB, Art. 68., § 1º, Inciso I} & 98 & 0.6246 \\
\multicolumn{1}{|r|}{\textit{\#7:} INC} & \multicolumn{4}{l|}{CRFB, Art. 167-A., caput, Inciso II} & 156 & 0.6234 \\
\multicolumn{1}{|r|}{\textit{\#8:} INC} & \multicolumn{4}{l|}{CRFB, Art. 167-A., caput, Inciso X} & 156 & 0.6189 \\
\multicolumn{1}{|r|}{\textit{\#9:} ALI} & \multicolumn{4}{l|}{CRFB, Art. 128., § 5º, Inciso II, Alínea f} & 128 & 0.6182 \\
\multicolumn{1}{|r|}{\textit{\#10:} INC} & \multicolumn{4}{l|}{CRFB, Art. 68., § 1º, Inciso III} & 88 & 0.6167 \\
\multicolumn{1}{|r|}{\textit{\#11:} INC} & \multicolumn{4}{l|}{CRFB, Art. 23., caput, Inciso I} & 61 & 0.6144 \\
\multicolumn{1}{|r|}{\textit{\#12:} ALI} & \multicolumn{4}{l|}{CRFB, Art. 128., § 5º, Inciso II, Alínea a} & 108 & 0.6135 \\
\multicolumn{1}{|r|}{\textit{\#13:} ART} & \multicolumn{4}{l|}{CRFB, Art. 175.} & 180 & 0.6123 \\
\multicolumn{1}{|r|}{\textit{\#14:} INC} & \multicolumn{4}{l|}{CRFB, Art. 49., caput, Inciso XI} & 47 & 0.6119 \\
\multicolumn{1}{|r|}{\textit{\#15:} ALI} & \multicolumn{4}{l|}{CRFB, Art. 128., § 5º, Inciso II, Alínea d} & 109 & 0.6108 \\
\multicolumn{1}{|r|}{\textit{\#16:} INC} & \multicolumn{4}{l|}{CRFB, Art. 5º, caput, Inciso LX} & 110 & 0.6101 \\
\multicolumn{1}{|r|}{\textit{\#17:} INC} & \multicolumn{4}{l|}{CRFB, Art. 5º, caput, Inciso XXIV} & 133 & 0.6096 \\
\multicolumn{1}{|r|}{\textit{\#18:} INC} & \multicolumn{4}{l|}{CRFB, Art. 48., caput, Inciso X} & 105 & 0.6093 \\
\multicolumn{1}{|r|}{\textit{\#19:} INC} & \multicolumn{4}{l|}{CRFB, Art. 167-A., caput, Inciso VI} & 265 & 0.6085 \\
\multicolumn{1}{|r|}{\textit{\#20:} ART} & \multicolumn{4}{l|}{CRFB, Art. 223.} & 274 & 0.6076 \\
\hline
Poly-Vector + Blind & 0.7971 & 0.8860 & 0.8122 & 0.0199 & 4363 & 22 \\
\multicolumn{1}{|r|}{\textit{\#1:} URN} & \multicolumn{4}{l|}{\textbf{urn:lex:br:federal:constituicao:1988-10-05;1988!art69}} & 53 & 0.8860 \\
\multicolumn{1}{|r|}{\textit{\#2:} URN} & \multicolumn{4}{l|}{urn:lex:br:federal:constituicao:1988-10-05;1988!art68} & 277 & 0.8350 \\
\multicolumn{1}{|r|}{\textit{\#3:} URN} & \multicolumn{4}{l|}{urn:lex:br:federal:constituicao:1988-10-05;1988!art71} & 898 & 0.8326 \\
\multicolumn{1}{|r|}{\textit{\#4:} URN} & \multicolumn{4}{l|}{urn:lex:br:federal:constituicao:1988-10-05;1988!art70} & 211 & 0.8267 \\
\multicolumn{1}{|r|}{\textit{\#5:} URN} & \multicolumn{4}{l|}{urn:lex:br:federal:constituicao:1988-10-05;1988!art65} & 117 & 0.8219 \\
\multicolumn{1}{|r|}{\textit{\#6:} URN} & \multicolumn{4}{l|}{urn:lex:br:federal:constituicao:1988-10-05;1988!art66\_par1} & 89 & 0.8150 \\
\multicolumn{1}{|r|}{\textit{\#7:} URN} & \multicolumn{4}{l|}{urn:lex:br:federal:constituicao:1988-10-05;1988!art5\_cpt\_inc69} & 149 & 0.8109 \\
\multicolumn{1}{|r|}{\textit{\#8:} URN} & \multicolumn{4}{l|}{urn:lex:br:federal:constituicao:1988-10-05;1988!art67} & 91 & 0.8109 \\
\multicolumn{1}{|r|}{\textit{\#9:} URN} & \multicolumn{4}{l|}{urn:lex:br:federal:constituicao:1988-10-05;1988!art66} & 456 & 0.8096 \\
\multicolumn{1}{|r|}{\textit{\#10:} URN} & \multicolumn{4}{l|}{urn:lex:br:federal:constituicao:1988-10-05;1988!art73\_par1} & 115 & 0.8078 \\
\multicolumn{1}{|r|}{\textit{\#11:} URN} & \multicolumn{4}{l|}{urn:lex:br:federal:constituicao:1988-10-05;1988!art78} & 168 & 0.8066 \\
\multicolumn{1}{|r|}{\textit{\#12:} URN} & \multicolumn{4}{l|}{urn:lex:br:federal:constituicao:1988-10-05;1988!art76} & 61 & 0.8049 \\
\multicolumn{1}{|r|}{\textit{\#13:} URN} & \multicolumn{4}{l|}{urn:lex:br:federal:constituicao:1988-10-05;1988!art74} & 334 & 0.8040 \\
\multicolumn{1}{|r|}{\textit{\#14:} URN} & \multicolumn{4}{l|}{urn:lex:br:federal:constituicao:1988-10-05;1988!art72\_par1} & 46 & 0.8034 \\
\multicolumn{1}{|r|}{\textit{\#15:} URN} & \multicolumn{4}{l|}{urn:lex:br:federal:constituicao:1988-10-05;1988!art165\_par9} & 176 & 0.8024 \\
\multicolumn{1}{|r|}{\textit{\#16:} URN} & \multicolumn{4}{l|}{urn:lex:br:federal:constituicao:1988-10-05;1988!art79\_par1u} & 45 & 0.8016 \\
\multicolumn{1}{|r|}{\textit{\#17:} URN} & \multicolumn{4}{l|}{urn:lex:br:federal:constituicao:1988-10-05;1988!art79} & 121 & 0.7995 \\
\multicolumn{1}{|r|}{\textit{\#18:} URN} & \multicolumn{4}{l|}{urn:lex:br:federal:constituicao:1988-10-05;1988!art86\_par1} & 59 & 0.7990 \\
\multicolumn{1}{|r|}{\textit{\#19:} URN} & \multicolumn{4}{l|}{urn:lex:br:federal:constituicao:1988-10-05;1988!art88} & 59 & 0.7984 \\
\multicolumn{1}{|r|}{\textit{\#20:} I,L} & \multicolumn{4}{l|}{urn:lex:br:federal:constituicao:1988-10-05;1988!art91\_par1, CRFB, Art. 91., § 1º} & 158 & 0.7976 \\
\multicolumn{1}{|r|}{\textit{\#21:} URN} & \multicolumn{4}{l|}{urn:lex:br:federal:constituicao:1988-10-05;1988!art89} & 223 & 0.7972 \\
\multicolumn{1}{|r|}{\textit{\#22:} URN} & \multicolumn{4}{l|}{urn:lex:br:federal:constituicao:1988-10-05;1988!art73} & 457 & 0.7971 \\
\hline
Poly-Vector + Flat & 0.7971 & 0.8860 & 0.8122 & 0.0199 & 4363 & 22 \\
\multicolumn{1}{|r|}{\textit{\#1:} URN} & \multicolumn{4}{l|}{\textbf{urn:lex:br:federal:constituicao:1988-10-05;1988!art69}} & 53 & 0.8860 \\
\multicolumn{1}{|r|}{\textit{\#2:} URN} & \multicolumn{4}{l|}{urn:lex:br:federal:constituicao:1988-10-05;1988!art68} & 277 & 0.8350 \\
\multicolumn{1}{|r|}{\textit{\#3:} URN} & \multicolumn{4}{l|}{urn:lex:br:federal:constituicao:1988-10-05;1988!art71} & 898 & 0.8326 \\
\multicolumn{1}{|r|}{\textit{\#4:} URN} & \multicolumn{4}{l|}{urn:lex:br:federal:constituicao:1988-10-05;1988!art70} & 211 & 0.8267 \\
\multicolumn{1}{|r|}{\textit{\#5:} URN} & \multicolumn{4}{l|}{urn:lex:br:federal:constituicao:1988-10-05;1988!art65} & 117 & 0.8219 \\
\multicolumn{1}{|r|}{\textit{\#6:} URN} & \multicolumn{4}{l|}{urn:lex:br:federal:constituicao:1988-10-05;1988!art66\_par1} & 89 & 0.8150 \\
\multicolumn{1}{|r|}{\textit{\#7:} URN} & \multicolumn{4}{l|}{urn:lex:br:federal:constituicao:1988-10-05;1988!art5\_cpt\_inc69} & 149 & 0.8109 \\
\multicolumn{1}{|r|}{\textit{\#8:} URN} & \multicolumn{4}{l|}{urn:lex:br:federal:constituicao:1988-10-05;1988!art67} & 91 & 0.8109 \\
\multicolumn{1}{|r|}{\textit{\#9:} URN} & \multicolumn{4}{l|}{urn:lex:br:federal:constituicao:1988-10-05;1988!art66} & 456 & 0.8096 \\
\multicolumn{1}{|r|}{\textit{\#10:} URN} & \multicolumn{4}{l|}{urn:lex:br:federal:constituicao:1988-10-05;1988!art73\_par1} & 115 & 0.8078 \\
\multicolumn{1}{|r|}{\textit{\#11:} URN} & \multicolumn{4}{l|}{urn:lex:br:federal:constituicao:1988-10-05;1988!art78} & 168 & 0.8066 \\
\multicolumn{1}{|r|}{\textit{\#12:} URN} & \multicolumn{4}{l|}{urn:lex:br:federal:constituicao:1988-10-05;1988!art76} & 61 & 0.8049 \\
\multicolumn{1}{|r|}{\textit{\#13:} URN} & \multicolumn{4}{l|}{urn:lex:br:federal:constituicao:1988-10-05;1988!art74} & 334 & 0.8040 \\
\multicolumn{1}{|r|}{\textit{\#14:} URN} & \multicolumn{4}{l|}{urn:lex:br:federal:constituicao:1988-10-05;1988!art72\_par1} & 46 & 0.8034 \\
\multicolumn{1}{|r|}{\textit{\#15:} URN} & \multicolumn{4}{l|}{urn:lex:br:federal:constituicao:1988-10-05;1988!art165\_par9} & 176 & 0.8024 \\
\multicolumn{1}{|r|}{\textit{\#16:} URN} & \multicolumn{4}{l|}{urn:lex:br:federal:constituicao:1988-10-05;1988!art79\_par1u} & 45 & 0.8016 \\
\multicolumn{1}{|r|}{\textit{\#17:} URN} & \multicolumn{4}{l|}{urn:lex:br:federal:constituicao:1988-10-05;1988!art79} & 121 & 0.7995 \\
\multicolumn{1}{|r|}{\textit{\#18:} URN} & \multicolumn{4}{l|}{urn:lex:br:federal:constituicao:1988-10-05;1988!art86\_par1} & 59 & 0.7990 \\
\multicolumn{1}{|r|}{\textit{\#19:} URN} & \multicolumn{4}{l|}{urn:lex:br:federal:constituicao:1988-10-05;1988!art88} & 59 & 0.7984 \\
\multicolumn{1}{|r|}{\textit{\#20:} I,L} & \multicolumn{4}{l|}{urn:lex:br:federal:constituicao:1988-10-05;1988!art91\_par1, CRFB, Art. 91., § 1º} & 158 & 0.7976 \\
\multicolumn{1}{|r|}{\textit{\#21:} URN} & \multicolumn{4}{l|}{urn:lex:br:federal:constituicao:1988-10-05;1988!art89} & 223 & 0.7972 \\
\multicolumn{1}{|r|}{\textit{\#22:} URN} & \multicolumn{4}{l|}{urn:lex:br:federal:constituicao:1988-10-05;1988!art73} & 457 & 0.7971 \\
\hline
Poly-Vector + Multi-layer & 0.7971 & 0.8860 & 0.8122 & 0.0199 & 4363 & 22 \\
\multicolumn{1}{|r|}{\textit{\#1:} URN} & \multicolumn{4}{l|}{\textbf{urn:lex:br:federal:constituicao:1988-10-05;1988!art69}} & 53 & 0.8860 \\
\multicolumn{1}{|r|}{\textit{\#2:} URN} & \multicolumn{4}{l|}{urn:lex:br:federal:constituicao:1988-10-05;1988!art68} & 277 & 0.8350 \\
\multicolumn{1}{|r|}{\textit{\#3:} URN} & \multicolumn{4}{l|}{urn:lex:br:federal:constituicao:1988-10-05;1988!art71} & 898 & 0.8326 \\
\multicolumn{1}{|r|}{\textit{\#4:} URN} & \multicolumn{4}{l|}{urn:lex:br:federal:constituicao:1988-10-05;1988!art70} & 211 & 0.8267 \\
\multicolumn{1}{|r|}{\textit{\#5:} URN} & \multicolumn{4}{l|}{urn:lex:br:federal:constituicao:1988-10-05;1988!art65} & 117 & 0.8219 \\
\multicolumn{1}{|r|}{\textit{\#6:} URN} & \multicolumn{4}{l|}{urn:lex:br:federal:constituicao:1988-10-05;1988!art66\_par1} & 89 & 0.8150 \\
\multicolumn{1}{|r|}{\textit{\#7:} URN} & \multicolumn{4}{l|}{urn:lex:br:federal:constituicao:1988-10-05;1988!art5\_cpt\_inc69} & 149 & 0.8109 \\
\multicolumn{1}{|r|}{\textit{\#8:} URN} & \multicolumn{4}{l|}{urn:lex:br:federal:constituicao:1988-10-05;1988!art67} & 91 & 0.8109 \\
\multicolumn{1}{|r|}{\textit{\#9:} URN} & \multicolumn{4}{l|}{urn:lex:br:federal:constituicao:1988-10-05;1988!art66} & 456 & 0.8096 \\
\multicolumn{1}{|r|}{\textit{\#10:} URN} & \multicolumn{4}{l|}{urn:lex:br:federal:constituicao:1988-10-05;1988!art73\_par1} & 115 & 0.8078 \\
\multicolumn{1}{|r|}{\textit{\#11:} URN} & \multicolumn{4}{l|}{urn:lex:br:federal:constituicao:1988-10-05;1988!art78} & 168 & 0.8066 \\
\multicolumn{1}{|r|}{\textit{\#12:} URN} & \multicolumn{4}{l|}{urn:lex:br:federal:constituicao:1988-10-05;1988!art76} & 61 & 0.8049 \\
\multicolumn{1}{|r|}{\textit{\#13:} URN} & \multicolumn{4}{l|}{urn:lex:br:federal:constituicao:1988-10-05;1988!art74} & 334 & 0.8040 \\
\multicolumn{1}{|r|}{\textit{\#14:} URN} & \multicolumn{4}{l|}{urn:lex:br:federal:constituicao:1988-10-05;1988!art72\_par1} & 46 & 0.8034 \\
\multicolumn{1}{|r|}{\textit{\#15:} URN} & \multicolumn{4}{l|}{urn:lex:br:federal:constituicao:1988-10-05;1988!art165\_par9} & 176 & 0.8024 \\
\multicolumn{1}{|r|}{\textit{\#16:} URN} & \multicolumn{4}{l|}{urn:lex:br:federal:constituicao:1988-10-05;1988!art79\_par1u} & 45 & 0.8016 \\
\multicolumn{1}{|r|}{\textit{\#17:} URN} & \multicolumn{4}{l|}{urn:lex:br:federal:constituicao:1988-10-05;1988!art79} & 121 & 0.7995 \\
\multicolumn{1}{|r|}{\textit{\#18:} URN} & \multicolumn{4}{l|}{urn:lex:br:federal:constituicao:1988-10-05;1988!art86\_par1} & 59 & 0.7990 \\
\multicolumn{1}{|r|}{\textit{\#19:} URN} & \multicolumn{4}{l|}{urn:lex:br:federal:constituicao:1988-10-05;1988!art88} & 59 & 0.7984 \\
\multicolumn{1}{|r|}{\textit{\#20:} I,L} & \multicolumn{4}{l|}{urn:lex:br:federal:constituicao:1988-10-05;1988!art91\_par1, CRFB, Art. 91., § 1º} & 158 & 0.7976 \\
\multicolumn{1}{|r|}{\textit{\#21:} URN} & \multicolumn{4}{l|}{urn:lex:br:federal:constituicao:1988-10-05;1988!art89} & 223 & 0.7972 \\
\multicolumn{1}{|r|}{\textit{\#22:} URN} & \multicolumn{4}{l|}{urn:lex:br:federal:constituicao:1988-10-05;1988!art73} & 457 & 0.7971 \\
\hline
Poly-Vector + Multi-layer + Q. Norm. & 0.8445 & 0.9191 & 0.8593 & 0.0168 & 4232 & 19 \\
\multicolumn{1}{|r|}{\textit{\#1:} URN} & \multicolumn{4}{l|}{\textbf{urn:lex:br:federal:constituicao:1988-10-05;1988!art69}} & 53 & \textbf{0.9191} \\
\multicolumn{1}{|r|}{\textit{\#2:} URN} & \multicolumn{4}{l|}{urn:lex:br:federal:constituicao:1988-10-05;1988!art68} & 277 & 0.8754 \\
\multicolumn{1}{|r|}{\textit{\#3:} URN} & \multicolumn{4}{l|}{urn:lex:br:federal:constituicao:1988-10-05;1988!art71} & 898 & 0.8737 \\
\multicolumn{1}{|r|}{\textit{\#4:} URN} & \multicolumn{4}{l|}{urn:lex:br:federal:constituicao:1988-10-05;1988!art70} & 211 & 0.8683 \\
\multicolumn{1}{|r|}{\textit{\#5:} URN} & \multicolumn{4}{l|}{urn:lex:br:federal:constituicao:1988-10-05;1988!art67} & 91 & 0.8605 \\
\multicolumn{1}{|r|}{\textit{\#6:} URN} & \multicolumn{4}{l|}{urn:lex:br:federal:constituicao:1988-10-05;1988!art66\_par1} & 89 & 0.8593 \\
\multicolumn{1}{|r|}{\textit{\#7:} URN} & \multicolumn{4}{l|}{urn:lex:br:federal:constituicao:1988-10-05;1988!art79\_par1u} & 45 & 0.8566 \\
\multicolumn{1}{|r|}{\textit{\#8:} URN} & \multicolumn{4}{l|}{urn:lex:br:federal:constituicao:1988-10-05;1988!art78} & 168 & 0.8553 \\
\multicolumn{1}{|r|}{\textit{\#9:} URN} & \multicolumn{4}{l|}{urn:lex:br:federal:constituicao:1988-10-05;1988!art66} & 456 & 0.8543 \\
\multicolumn{1}{|r|}{\textit{\#10:} URN} & \multicolumn{4}{l|}{urn:lex:br:federal:constituicao:1988-10-05;1988!art79} & 121 & 0.8539 \\
\multicolumn{1}{|r|}{\textit{\#11:} URN} & \multicolumn{4}{l|}{urn:lex:br:federal:constituicao:1988-10-05;1988!art65\_par1u} & 15 & 0.8537 \\
\multicolumn{1}{|r|}{\textit{\#12:} URN} & \multicolumn{4}{l|}{urn:lex:br:federal:constituicao:1988-10-05;1988!art96} & 660 & 0.8526 \\
\multicolumn{1}{|r|}{\textit{\#13:} URN} & \multicolumn{4}{l|}{urn:lex:br:federal:constituicao:1988-10-05;1988!art97} & 86 & 0.8523 \\
\multicolumn{1}{|r|}{\textit{\#14:} URN} & \multicolumn{4}{l|}{urn:lex:br:federal:constituicao:1988-10-05;1988!art73\_par1} & 115 & 0.8517 \\
\multicolumn{1}{|r|}{\textit{\#15:} URN} & \multicolumn{4}{l|}{urn:lex:br:federal:constituicao:1988-10-05;1988!art166\_par9-1} & 71 & 0.8513 \\
\multicolumn{1}{|r|}{\textit{\#16:} URN} & \multicolumn{4}{l|}{urn:lex:br:federal:constituicao:1988-10-05;1988!art65} & 117 & 0.8507 \\
\multicolumn{1}{|r|}{\textit{\#17:} URN} & \multicolumn{4}{l|}{urn:lex:br:federal:constituicao:1988-10-05;1988!art74} & 334 & 0.8484 \\
\multicolumn{1}{|r|}{\textit{\#18:} URN} & \multicolumn{4}{l|}{urn:lex:br:federal:constituicao:1988-10-05;1988!art166\_par9} & 74 & 0.8453 \\
\multicolumn{1}{|r|}{\textit{\#19:} URN} & \multicolumn{4}{l|}{urn:lex:br:federal:constituicao:1988-10-05;1988!art107} & 351 & 0.8445 \\
\hline
\end{longtable}
\clearpage

\textbf{Question 8} \\
\textbf{Original Question:} Quais as diferenças entre o art. 51 e o art. 52 da Constituição? \\
\textbf{Normalized Question:} Diferenças entre o art. 51 e o art. 52 da Constituição \\
\textbf{Expected Top-1 Segment according to the chunking strategy:} 
\begin{itemize}
\item \textbf{Blind:} Chunk \#72  
\item \textbf{Flat:} Constituição da República Federativa do Brasil, Art. 51.  
\item \textbf{Multi-layer:} Constituição da República Federativa do Brasil, Art. 51. 
\item \textbf{Poly-Vector:}  \begin{enumerate} \item (Label:) Constituição da República Federativa do Brasil, Art. 51.   \item (URN+Label:) urn:lex:br:federal:constituicao:1988-10-05;1988!art51, Constituição da República Federativa do Brasil, Art. 51.   \item  (URN:) urn:lex:br:federal:constituicao:1988-10-05;1988!art51 
\end{enumerate} \end{itemize}
\begin{longtable}{|l|r|r|r|r|r|r|}
\caption{Statistics for Question 8} \label{tab:question_8} \\
\hline
\multicolumn{7}{|c|}{\textbf{Statistics for Question 8}} \\
\hline
\textbf{Method} & \textbf{Min Sim.} & \textbf{Max Sim.} & \textbf{Mean Sim.} & \textbf{Std Dev.} & \textbf{Tokens} & \textbf{Segments} \\
\hline
\endfirsthead
\multicolumn{7}{c}%
{\tablename\ \thetable\ -- \textit{Continued from previous page}} \\
\hline
\textbf{Method} & \textbf{Min Sim.} & \textbf{Max Sim.} & \textbf{Mean Sim.} & \textbf{Std Dev.} & \textbf{Tokens} & \textbf{Segments} \\
\hline
\endhead
\hline \multicolumn{7}{|r|}{{Continued on next page}} \\ \hline
\endfoot
\hline
\endlastfoot
Blind Segmentation Baseline & 0.6337 & 0.6650 & 0.6508 & 0.0141 & 4000 & 5 \\
\multicolumn{1}{|r|}{\textit{\#1:} Blind} & \multicolumn{4}{l|}{\textbf{Chunk \#72} ["idos escritos de informações a Ministros ...  Distrito Federal e dos Municí"]} & 800 & 0.6650 \\
\multicolumn{1}{|r|}{\textit{\#2:} Blind} & \multicolumn{4}{l|}{Chunk \#71 [" renovação de concessão de emissoras de  ... ional de Justiça e do Conselho"]} & 800 & 0.6628 \\
\multicolumn{1}{|r|}{\textit{\#3:} Blind} & \multicolumn{4}{l|}{Chunk \#80 [" da convocação. § 8º Havendo medidas pro ... ioria relativa de seus membros"]} & 800 & 0.6541 \\
\multicolumn{1}{|r|}{\textit{\#4:} Blind} & \multicolumn{4}{l|}{Chunk \#81 ["ivas Casas, serão criadas pela Câmara do ...  a) criação de cargos, funções"]} & 800 & 0.6386 \\
\multicolumn{1}{|r|}{\textit{\#5:} Blind} & \multicolumn{4}{l|}{Chunk \#74 ["pios; X – suspender a execução, no todo  ... irão durante o estado de sítio"]} & 800 & 0.6337 \\
\hline
Flat Per-Article Baseline & 0.5925 & 0.6067 & 0.5978 & 0.0057 & 5846 & 5 \\
\multicolumn{1}{|r|}{\textit{\#1:} ART} & \multicolumn{4}{l|}{CRFB, Art. 47.} & 77 & 0.6067 \\
\multicolumn{1}{|r|}{\textit{\#2:} ART} & \multicolumn{4}{l|}{CRFB, Art. 37.} & 3036 & 0.5996 \\
\multicolumn{1}{|r|}{\textit{\#3:} ART} & \multicolumn{4}{l|}{CRFB, Art. 62.} & 1001 & 0.5966 \\
\multicolumn{1}{|r|}{\textit{\#4:} ART} & \multicolumn{4}{l|}{CRFB, Art. 102.} & 1220 & 0.5935 \\
\multicolumn{1}{|r|}{\textit{\#5:} ART} & \multicolumn{4}{l|}{CRFB, Art. 48.} & 512 & 0.5925 \\
\hline
Multi-layer Hierarchical Embeddings & 0.6199 & 0.6476 & 0.6304 & 0.0120 & 12007 & 5 \\
\multicolumn{1}{|r|}{\textit{\#1:} SEC} & \multicolumn{4}{l|}{CRFB, TÍTULO IV, CAPÍTULO I, Seção V} & 1450 & 0.6476 \\
\multicolumn{1}{|r|}{\textit{\#2:} SEC} & \multicolumn{4}{l|}{CRFB, TÍTULO III, CAPÍTULO VII, Seção I} & 3316 & 0.6382 \\
\multicolumn{1}{|r|}{\textit{\#3:} SSC} & \multicolumn{4}{l|}{CRFB, TÍTULO IV, CAPÍTULO I, Seção VIII, Subseção III} & 2659 & 0.6238 \\
\multicolumn{1}{|r|}{\textit{\#4:} SEC} & \multicolumn{4}{l|}{CRFB, TÍTULO IV, CAPÍTULO I, Seção II} & 1462 & 0.6225 \\
\multicolumn{1}{|r|}{\textit{\#5:} SEC} & \multicolumn{4}{l|}{CRFB, TÍTULO IV, CAPÍTULO I, Seção VIII} & 3120 & 0.6199 \\
\hline
Multi-layer + Query Normalization & 0.5602 & 0.5773 & 0.5678 & 0.0064 & 8975 & 5 \\
\multicolumn{1}{|r|}{\textit{\#1:} SEC} & \multicolumn{4}{l|}{CRFB, TÍTULO III, CAPÍTULO VII, Seção I} & 3316 & 0.5773 \\
\multicolumn{1}{|r|}{\textit{\#2:} INC} & \multicolumn{4}{l|}{CRFB, Art. 68., § 1º, Inciso II} & 88 & 0.5706 \\
\multicolumn{1}{|r|}{\textit{\#3:} SEC} & \multicolumn{4}{l|}{CRFB, TÍTULO IV, CAPÍTULO I, Seção II} & 1462 & 0.5656 \\
\multicolumn{1}{|r|}{\textit{\#4:} SSC} & \multicolumn{4}{l|}{CRFB, TÍTULO IV, CAPÍTULO I, Seção VIII, Subseção III} & 2659 & 0.5655 \\
\multicolumn{1}{|r|}{\textit{\#5:} SEC} & \multicolumn{4}{l|}{CRFB, TÍTULO IV, CAPÍTULO I, Seção V} & 1450 & 0.5602 \\
\hline
Poly-Vector + Blind & 0.6481 & 0.7009 & 0.6662 & 0.0172 & 4428 & 10 \\
\multicolumn{1}{|r|}{\textit{\#1:} I,L} & \multicolumn{4}{l|}{\textbf{urn:lex:br:federal:constituicao:1988-10-05;1988!art52, CRFB, Art. 52.}} & 880 & \textbf{0.7009} \\
\multicolumn{1}{|r|}{\textit{\#2:} LBL} & \multicolumn{4}{l|}{CRFB, Art. 51.} & 239 & 0.6892 \\
\multicolumn{1}{|r|}{\textit{\#3:} I,L} & \multicolumn{4}{l|}{urn:lex:br:federal:constituicao:1988-10-05;1988!art51\_cpt\_inc2, CRFB, Art. 51., caput, Inciso II} & 62 & 0.6704 \\
\multicolumn{1}{|r|}{\textit{\#4:} I,L} & \multicolumn{4}{l|}{urn:lex:br:federal:constituicao:1988-10-05;1988!art51\_cpt, CRFB, Art. 51., caput} & 194 & 0.6671 \\
\multicolumn{1}{|r|}{\textit{\#5:} Blind} & \multicolumn{4}{l|}{\textbf{Chunk \#72} ["idos escritos de informações a Ministros ...  Distrito Federal e dos Municí"]} & 800 & 0.6650 \\
\multicolumn{1}{|r|}{\textit{\#6:} Blind} & \multicolumn{4}{l|}{Chunk \#71 [" renovação de concessão de emissoras de  ... ional de Justiça e do Conselho"]} & 800 & 0.6628 \\
\multicolumn{1}{|r|}{\textit{\#7:} I,L} & \multicolumn{4}{l|}{urn:lex:br:federal:constituicao:1988-10-05;1988!art5\_cpt\_inc52, CRFB, Art. 5º, caput, Inciso LII} & 100 & 0.6552 \\
\multicolumn{1}{|r|}{\textit{\#8:} Blind} & \multicolumn{4}{l|}{Chunk \#80 [" da convocação. § 8º Havendo medidas pro ... ioria relativa de seus membros"]} & 800 & 0.6541 \\
\multicolumn{1}{|r|}{\textit{\#9:} I,L} & \multicolumn{4}{l|}{urn:lex:br:federal:constituicao:1988-10-05;1988!art53\_par1, CRFB, Art. 53., § 1º} & 33 & 0.6487 \\
\multicolumn{1}{|r|}{\textit{\#10:} I,L} & \multicolumn{4}{l|}{urn:lex:br:federal:constituicao:1988-10-05;1988!art53, CRFB, Art. 53.} & 520 & 0.6481 \\
\hline
Poly-Vector + Flat & 0.6083 & 0.7009 & 0.6358 & 0.0285 & 4910 & 21 \\
\multicolumn{1}{|r|}{\textit{\#1:} I,L} & \multicolumn{4}{l|}{\textbf{urn:lex:br:federal:constituicao:1988-10-05;1988!art52, CRFB, Art. 52.}} & 880 & \textbf{0.7009} \\
\multicolumn{1}{|r|}{\textit{\#2:} LBL} & \multicolumn{4}{l|}{\textbf{\textbf{CRFB, Art. 51.}}} & 239 & 0.6892 \\
\multicolumn{1}{|r|}{\textit{\#3:} I,L} & \multicolumn{4}{l|}{urn:lex:br:federal:constituicao:1988-10-05;1988!art51\_cpt\_inc2, CRFB, Art. 51., caput, Inciso II} & 62 & 0.6704 \\
\multicolumn{1}{|r|}{\textit{\#4:} I,L} & \multicolumn{4}{l|}{urn:lex:br:federal:constituicao:1988-10-05;1988!art51\_cpt, CRFB, Art. 51., caput} & 194 & 0.6671 \\
\multicolumn{1}{|r|}{\textit{\#5:} I,L} & \multicolumn{4}{l|}{urn:lex:br:federal:constituicao:1988-10-05;1988!art5\_cpt\_inc52, CRFB, Art. 5º, caput, Inciso LII} & 100 & 0.6552 \\
\multicolumn{1}{|r|}{\textit{\#6:} I,L} & \multicolumn{4}{l|}{urn:lex:br:federal:constituicao:1988-10-05;1988!art53\_par1, CRFB, Art. 53., § 1º} & 33 & 0.6487 \\
\multicolumn{1}{|r|}{\textit{\#7:} I,L} & \multicolumn{4}{l|}{urn:lex:br:federal:constituicao:1988-10-05;1988!art53, CRFB, Art. 53.} & 520 & 0.6481 \\
\multicolumn{1}{|r|}{\textit{\#8:} I,L} & \multicolumn{4}{l|}{urn:lex:br:federal:constituicao:1988-10-05;1988!art50\_par1, CRFB, Art. 50., § 1º} & 59 & 0.6451 \\
\multicolumn{1}{|r|}{\textit{\#9:} I,L} & \multicolumn{4}{l|}{urn:lex:br:federal:constituicao:1988-10-05;1988!art50\_cpt, CRFB, Art. 50., caput} & 107 & 0.6448 \\
\multicolumn{1}{|r|}{\textit{\#10:} URN} & \multicolumn{4}{l|}{urn:lex:br:federal:constituicao:1988-10-05;1988!art50} & 291 & 0.6369 \\
\multicolumn{1}{|r|}{\textit{\#11:} I,L} & \multicolumn{4}{l|}{urn:lex:br:federal:constituicao:1988-10-05;1988!art5\_cpt\_inc51, CRFB, Art. 5º, caput, Inciso LI} & 136 & 0.6321 \\
\multicolumn{1}{|r|}{\textit{\#12:} I,L} & \multicolumn{4}{l|}{urn:lex:br:federal:constituicao:1988-10-05;1988!art95\_par1u\_inc2, CRFB, Art. 95., Parágrafo único., Inciso II} & 41 & 0.6171 \\
\multicolumn{1}{|r|}{\textit{\#13:} I,L} & \multicolumn{4}{l|}{urn:lex:br:federal:constituicao:1988-10-05;1988!art5\_cpt\_inc42, CRFB, Art. 5º, caput, Inciso XLII} & 114 & 0.6143 \\
\multicolumn{1}{|r|}{\textit{\#14:} URN} & \multicolumn{4}{l|}{urn:lex:br:federal:constituicao:1988-10-05;1988!art47} & 77 & 0.6137 \\
\multicolumn{1}{|r|}{\textit{\#15:} I,L} & \multicolumn{4}{l|}{urn:lex:br:federal:constituicao:1988-10-05;1988!art48\_cpt, CRFB, Art. 48., caput} & 446 & 0.6119 \\
\multicolumn{1}{|r|}{\textit{\#16:} URN} & \multicolumn{4}{l|}{urn:lex:br:federal:constituicao:1988-10-05;1988!art152} & 89 & 0.6115 \\
\multicolumn{1}{|r|}{\textit{\#17:} URN} & \multicolumn{4}{l|}{urn:lex:br:federal:constituicao:1988-10-05;1988!art14\_par1\_inc2} & 65 & 0.6091 \\
\multicolumn{1}{|r|}{\textit{\#18:} URN} & \multicolumn{4}{l|}{urn:lex:br:federal:constituicao:1988-10-05;1988!art82} & 76 & 0.6090 \\
\multicolumn{1}{|r|}{\textit{\#19:} I,L} & \multicolumn{4}{l|}{urn:lex:br:federal:constituicao:1988-10-05;1988!art55\_cpt, CRFB, Art. 55., caput} & 146 & 0.6089 \\
\multicolumn{1}{|r|}{\textit{\#20:} URN} & \multicolumn{4}{l|}{urn:lex:br:federal:constituicao:1988-10-05;1988!art65\_par1u} & 15 & 0.6087 \\
\multicolumn{1}{|r|}{\textit{\#21:} I,L} & \multicolumn{4}{l|}{urn:lex:br:federal:constituicao:1988-10-05;1988!art102, CRFB, Art. 102.} & 1220 & 0.6083 \\
\hline
Poly-Vector + Multi-layer & 0.6382 & 0.7009 & 0.6596 & 0.0202 & 6960 & 11 \\
\multicolumn{1}{|r|}{\textit{\#1:} I,L} & \multicolumn{4}{l|}{\textbf{urn:lex:br:federal:constituicao:1988-10-05;1988!art52, CRFB, Art. 52.}} & 880 & \textbf{0.7009} \\
\multicolumn{1}{|r|}{\textit{\#2:} LBL} & \multicolumn{4}{l|}{\textbf{\textbf{CRFB, Art. 51.}}} & 239 & 0.6892 \\
\multicolumn{1}{|r|}{\textit{\#3:} I,L} & \multicolumn{4}{l|}{urn:lex:br:federal:constituicao:1988-10-05;1988!art51\_cpt\_inc2, CRFB, Art. 51., caput, Inciso II} & 62 & 0.6704 \\
\multicolumn{1}{|r|}{\textit{\#4:} I,L} & \multicolumn{4}{l|}{urn:lex:br:federal:constituicao:1988-10-05;1988!art51\_cpt, CRFB, Art. 51., caput} & 194 & 0.6671 \\
\multicolumn{1}{|r|}{\textit{\#5:} I,L} & \multicolumn{4}{l|}{urn:lex:br:federal:constituicao:1988-10-05;1988!art5\_cpt\_inc52, CRFB, Art. 5º, caput, Inciso LII} & 100 & 0.6552 \\
\multicolumn{1}{|r|}{\textit{\#6:} I,L} & \multicolumn{4}{l|}{urn:lex:br:federal:constituicao:1988-10-05;1988!art53\_par1, CRFB, Art. 53., § 1º} & 33 & 0.6487 \\
\multicolumn{1}{|r|}{\textit{\#7:} I,L} & \multicolumn{4}{l|}{urn:lex:br:federal:constituicao:1988-10-05;1988!art53, CRFB, Art. 53.} & 520 & 0.6481 \\
\multicolumn{1}{|r|}{\textit{\#8:} SEC} & \multicolumn{4}{l|}{CRFB, TÍTULO IV, CAPÍTULO I, Seção V} & 1450 & 0.6476 \\
\multicolumn{1}{|r|}{\textit{\#9:} I,L} & \multicolumn{4}{l|}{urn:lex:br:federal:constituicao:1988-10-05;1988!art50\_par1, CRFB, Art. 50., § 1º} & 59 & 0.6451 \\
\multicolumn{1}{|r|}{\textit{\#10:} I,L} & \multicolumn{4}{l|}{urn:lex:br:federal:constituicao:1988-10-05;1988!art50\_cpt, CRFB, Art. 50., caput} & 107 & 0.6448 \\
\multicolumn{1}{|r|}{\textit{\#11:} SEC} & \multicolumn{4}{l|}{CRFB, TÍTULO III, CAPÍTULO VII, Seção I} & 3316 & 0.6382 \\
\hline
Poly-Vector + Multi-layer + Q. Norm. & 0.6199 & 0.6976 & 0.6455 & 0.0253 & 4494 & 17 \\
\multicolumn{1}{|r|}{\textit{\#1:} LBL} & \multicolumn{4}{l|}{\textbf{\textbf{CRFB, Art. 51.}}} & 239 & 0.6976 \\
\multicolumn{1}{|r|}{\textit{\#2:} LBL} & \multicolumn{4}{l|}{\textbf{CRFB, Art. 52.}} & 880 & 0.6877 \\
\multicolumn{1}{|r|}{\textit{\#3:} I,L} & \multicolumn{4}{l|}{urn:lex:br:federal:constituicao:1988-10-05;1988!art52\_par1u, CRFB, Art. 52., Parágrafo único.} & 92 & 0.6790 \\
\multicolumn{1}{|r|}{\textit{\#4:} I,L} & \multicolumn{4}{l|}{urn:lex:br:federal:constituicao:1988-10-05;1988!art52\_cpt, CRFB, Art. 52., caput} & 709 & 0.6732 \\
\multicolumn{1}{|r|}{\textit{\#5:} I,L} & \multicolumn{4}{l|}{urn:lex:br:federal:constituicao:1988-10-05;1988!art51\_cpt\_inc2, CRFB, Art. 51., caput, Inciso II} & 62 & 0.6572 \\
\multicolumn{1}{|r|}{\textit{\#6:} I,L} & \multicolumn{4}{l|}{urn:lex:br:federal:constituicao:1988-10-05;1988!art51\_cpt, CRFB, Art. 51., caput} & 194 & 0.6568 \\
\multicolumn{1}{|r|}{\textit{\#7:} I,L} & \multicolumn{4}{l|}{urn:lex:br:federal:constituicao:1988-10-05;1988!art5\_cpt\_inc52, CRFB, Art. 5º, caput, Inciso LII} & 100 & 0.6465 \\
\multicolumn{1}{|r|}{\textit{\#8:} I,L} & \multicolumn{4}{l|}{urn:lex:br:federal:constituicao:1988-10-05;1988!art50\_par1, CRFB, Art. 50., § 1º} & 59 & 0.6399 \\
\multicolumn{1}{|r|}{\textit{\#9:} I,L} & \multicolumn{4}{l|}{urn:lex:br:federal:constituicao:1988-10-05;1988!art50\_cpt, CRFB, Art. 50., caput} & 107 & 0.6334 \\
\multicolumn{1}{|r|}{\textit{\#10:} LBL} & \multicolumn{4}{l|}{CRFB, Art. 53.} & 520 & 0.6315 \\
\multicolumn{1}{|r|}{\textit{\#11:} I,L} & \multicolumn{4}{l|}{urn:lex:br:federal:constituicao:1988-10-05;1988!art50, CRFB, Art. 50.} & 291 & 0.6289 \\
\multicolumn{1}{|r|}{\textit{\#12:} I,L} & \multicolumn{4}{l|}{urn:lex:br:federal:constituicao:1988-10-05;1988!art53\_par1, CRFB, Art. 53., § 1º} & 33 & 0.6287 \\
\multicolumn{1}{|r|}{\textit{\#13:} I,L} & \multicolumn{4}{l|}{urn:lex:br:federal:constituicao:1988-10-05;1988!art53\_par2, CRFB, Art. 53., § 2º} & 81 & 0.6246 \\
\multicolumn{1}{|r|}{\textit{\#14:} I,L} & \multicolumn{4}{l|}{urn:lex:br:federal:constituicao:1988-10-05;1988!art53\_cpt, CRFB, Art. 53., caput} & 33 & 0.6233 \\
\multicolumn{1}{|r|}{\textit{\#15:} I,L} & \multicolumn{4}{l|}{urn:lex:br:federal:constituicao:1988-10-05;1988!art48\_cpt, CRFB, Art. 48., caput} & 446 & 0.6232 \\
\multicolumn{1}{|r|}{\textit{\#16:} I,L} & \multicolumn{4}{l|}{urn:lex:br:federal:constituicao:1988-10-05;1988!art5\_cpt\_inc51, CRFB, Art. 5º, caput, Inciso LI} & 136 & 0.6217 \\
\multicolumn{1}{|r|}{\textit{\#17:} LBL} & \multicolumn{4}{l|}{CRFB, Art. 48.} & 512 & 0.6199 \\
\hline
\end{longtable}
\clearpage

\end{landscape}

\end{document}